%% file: sample63.tex
\documentclass[twocolumn]{aastex63}

\usepackage{CJK}
\usepackage{amsmath}

\include{affiliations}

\newcommand{\caltech}{Department of Astronomy, California Institute of Technology, Pasadena, CA 91125, USA}

\received{xxx}
\revised{xxx}
\accepted{August 28, 2024}

\shorttitle{HD 118203 System}
\shortauthors{Zhang et al.}

\graphicspath{{./}{figures/}}
\begin{document}
\begin{CJK*}{UTF8}{gbsn}
\title{A Testbed for Tidal Migration: the 3D Architecture of an Eccentric Hot Jupiter HD 118203 b Accompanied by a Possibly Aligned Outer Giant Planet}

\correspondingauthor{Jingwen Zhang}
\email{jingwen7@hawaii.edu}

%\author[0000-0002-0786-7307]{Greg J. Schwarz}

\author[0000-0002-2696-2406]{Jingwen Zhang (张婧雯)}
\altaffiliation{NASA FINESST Fellow}
\affiliation{\UH}

\author[0000-0001-8832-4488]{Daniel Huber}
\affiliation{\UH}
\affiliation{Sydney Institute for Astronomy (SIfA), School of Physics, University of Sydney, NSW 2006, Australia}

\author[0000-0002-3725-3058]{Lauren M. Weiss}
\affiliation{Department of Physics and Astronomy, University of Notre Dame, Notre Dame, IN 46556, USA}

\author[0000-0002-6618-1137]{Jerry W. Xuan}
\altaffiliation{NASA FINESST Fellow}
\affiliation{\caltech}

\author[0000-0002-0040-6815]{Jennifer~A.~Burt}
\affil{\JPL}

\author[0000-0002-8958-0683]{Fei Dai}
\affiliation{\UH}

\author[0000-0003-2657-3889]{Nicholas Saunders}
\altaffiliation{NSF Graduate Research Fellow}
\affiliation{\UH}

\author[0000-0003-0967-2893]{Erik A.\ Petigura}
\affiliation{\UCLA}

\author[0000-0003-3856-3143]{Ryan A. Rubenzahl}
\altaffiliation{NSF Graduate Research Fellow}
\affiliation{\caltech}

\author[0000-0002-4265-047X]{Joshua N. Winn}
\affil{\Princeton}

\author[0000-0002-6937-9034]{Sharon X.~Wang}
\affiliation{Department of Astronomy, Tsinghua University, Beijing 100084, People's Republic of China} 

\author[0000-0002-4290-6826]{Judah Van Zandt}
\affiliation{\UCLA}

\author[0009-0008-9808-0411]{Max Brodheim}
\affiliation{\WMKO}

\author[0000-0002-9879-3904]{Zachary R. Claytor}
\affiliation{Space Telescope Science Institute, 3700 San Martin Drive, Baltimore, MD 21218, USA}
\affiliation{Department of Astronomy, University of Florida, 211 Bryant Space Science Center, Gainesville, FL 32611, USA}

\author{Ian Crossfield}
\affiliation{\Kansas}

\author[0009-0000-3624-1330]{William Deich}
\affil{\UCO}

\author[0000-0003-3504-5316]{Benjamin J.\ Fulton}
\affiliation{\caltech}

\author[0009-0004-4454-6053]{Steven R. Gibson}
\affil{\COO}

\author[0000-0003-1312-9391]{Samuel Halverson}
\affil{\JPL}

\author[0000-0002-7648-9119]{Grant M.\ Hill}
\affiliation{\WMKO}

\author[0000-0002-6153-3076]{Bradford Holden}
\affil{\UCO}

\author[0000-0002-5812-3236]{Aaron Householder}
\affiliation{Department of Earth, Atmospheric and Planetary Sciences, Massachusetts Institute of Technology, Cambridge, MA 02139, USA}

\affil{Kavli Institute for Astrophysics and Space Research, Massachusetts Institute of Technology, Cambridge, MA 02139, USA}

\author[0000-0001-8638-0320]{Andrew W. Howard}
\affiliation{\caltech}

\author[0000-0002-0531-1073]{Howard Isaacson}
\affiliation{\UCB}

\author{Stephen Kaye}
\affiliation{\COO}

\author[0009-0004-0592-1850]{Kyle Lanclos}
\affiliation{\WMKO}

\author[0000-0003-2451-5482]{Russ R. Laher (\begin{CJK*}{UTF8}{bsmi}良主嶺亞\end{CJK*})}
\affiliation{NASA Exoplanet Science Institute/Caltech-IPAC, MC 314-6, 1200 E California Blvd, Pasadena, CA 91125, USA}

\author[0000-0001-8342-7736]{Jack Lubin}
\affiliation{\UCLA}

\author[0009-0008-4293-0341]{Joel Payne}
\affiliation{\WMKO}

\author[0000-0001-8127-5775]{Arpita Roy}
\affiliation{\Schmidt}

\author[0000-0002-4046-987X]{Christian Schwab}
\affil{\Macquarie}

\author[0000-0003-3133-6837]{Abby P.\ Shaum}
\affiliation{\caltech}

\author[0000-0002-6092-8295]{Josh Walawender}
\affiliation{\WMKO}

\author{Edward Wishnow}
\affiliation{\SSL}

\author[0000-0002-4037-3114]{Sherry Yeh}
\affiliation{\WMKO}

%\collaboration{1}{(AAS Journals Data Scientists collaboration)}
%\collaboration{1}{(LaTeX collaboration)}

\begin{abstract}
Characterizing outer companions to hot Jupiters plays a crucial role in deciphering their origins. We present the discovery of a long-period giant planet, HD 118203 c ($m_{c}=11.79^{+0.69}_{-0.63}\ \mathrm{M_{J}}$, $a_{c}=6.28^{+0.10}_{-0.11}$ AU) exterior to a close-in eccentric hot Jupiter HD 118203 b ($P_{b}=6.135\ \mathrm{days}$, $m_{b}=2.14\pm{0.12}\ \mathrm{M_{J}}$, $r_{b}=1.14\pm{0.029}\ \mathrm{R_{J}}$, $e_{b}=0.31\pm{0.007}$) based on twenty-year radial velocities. Using Rossiter-McLaughlin (RM) observations from the Keck Planet Finder (KPF), we measured a low sky-projected spin-orbit angle $\lambda_{b}=-11^{\circ}.7^{+7.6}_{-10.0}$ for HD 118203 b and detected stellar oscillations in the host star, confirming its evolved status. Combining the RM observation with the stellar inclination measurement, we constrained the true spin-orbit angle of HD 118203 b as $\Psi_{b}<33^{\circ}.5\ (2\sigma)$, indicating the orbit normal of the hot Jupiter nearly aligned with the stellar spin axis. Furthermore, by combining radial velocities and Hipparcos-Gaia astrometric acceleration, we constrained the line-of-sight mutual inclination between the hot Jupiter and the outer planet to be $9^{\circ}.8^{+16.2}_{-9.3}$ at $2\sigma$ level. HD 118203 is one of first hot Jupiter systems where both the true spin-orbit angle of the hot Jupiter and the mutual inclination between inner and outer  planets have been determined. Our results are consistent with a system-wide alignment, with low mutual inclinations between the outer giant planet, the inner hot Jupiter, and the host star. This alignment, along with the moderate eccentricity of HD 118203 c, implies that the system may have undergone coplanar high-eccentricity tidal migration. Under this framework, our dynamical analysis suggests an initial semi-major axis of 0.3 to 3.2 AU for the proto-hot Jupiter. 

%We also constrained the line-of-sight stellar obliquity for HD 118203 c to be  $14^{\circ}.7^{+13.3}_{-10.2}$. 

\end{abstract}

\keywords{editorials, notices --- 
miscellaneous --- catalogs --- surveys}

\section{Introduction} \label{sec:intro}

The discovery of Hot Jupiters, defined as giant gaseous planets with short orbital periods ($P<10$ days), upended the traditional planet formation theories based on our Solar System. Various theories have been proposed to explain how these massive gaseous planets end up so close to their stars. The in-situ formation theory suggests that hot Jupiters form directly in their current, close orbits \citep{Bodenheimer2000,Boley2016,Batygin2016}. However, the typical giant planet formation processes -- either core accretion or gravitational instability -- encounter difficulties operating at such close distances \citep{Rafikov2005, Rafikov2006}. Disk-driven migration is another proposed mechanism in which the proto-hot Jupiters initially form in the outer region of the protoplanetary disks and migrate inward due to the torques from the disk \citep{Goldreich1980,Lin1996, Ida2008}.  Alternatively, in the high-eccentricity tidal migration scenario, the orbit of a giant planet that formed beyond the ice line attains a high eccentricity and then shrinks and circularizes through dissipative tidal interactions with the host star \citep{Fabry2007,Fabrycky2009, Naoz2011}. The trigger for tidal migration is the eccentricity excitation of the proto-hot Jupiters, which could be caused by dynamical events such as planet-planet scattering \citep{Rasio1996,Chatterjee2008,Beaug2012, Petrovich2014}, or secular interactions with another planet such as von-Zeipel-Kozai-Lidov cycles \citep{1910AN....183..345V,Kozai1962,Lidov1962}\footnote{The phenomenon is commonly referred to as Kozai-Lidov cycles or Lidov-Kozai cycles, acknowledging that Kozai and Lidov independently discovered it in the 1960s. However, recent investigations into the literature have revealed that Swedish astronomer von Zeipel described a similar theory as early as 1910 \citep{Ito2019}.}.

The spin-orbit angles (also known as stellar obliquities) of stars hosting hot Jupiters are considered a crucial indicator of their formation channel \citep{Winn2010,Albrecht2012,Dawson2018, Rice2022}. Spin-orbit angle is the angle between the star's spin axis and the planet's orbital axis. In scenarios involving tidal migration, where dynamic events cause eccentric orbits, the angular momentum of the proto-hot Jupiters would be altered, leading to high spin-orbit angles. 
Additionally, stellar companions might cause misalignment between protoplanetary disks and their host stars \citep{Batygin2013}. Thus, planets could end up misaligned relative to their host stars even if they migrate through disk-driven migration. However, observations have suggested that the spin-orbit angle distribution may not only reflect the hot Jupiters' origins but also sculpting by tidal realignment. \cite{Schlaufman2010} and \cite{Winn2010}  indicate a significant dichotomy in the spin-orbit angles of stars hosting hot Jupiters based on the star's temperature. They discovered that cool stars hosting hot Jupiters generally show low spin-orbit angles, likely because effective tidal realignment processes erase their initial misalignment. In contrast, hotter stars exhibit a broader range of spin-orbit angles, indicating that they may retain the initial spin-orbit angles from when hot Jupiters first migrated to their close-in orbit. Furthermore, \cite{Rice2022} found that the spin-orbit angles of eccentric hot Jupiters do not exhibit such a trend with stellar temperature, which is consistent with the high-eccentricity migration scenario from their models. 

The presence and properties of companions to hot Jupiters serve as another clue to distinguish different origin theories. In-situ formation could produce hot Jupiters with nearby planets that are not necessarily in orbital resonance \citep{Hansen2013, Boley2016}. In contrast, in the disk migration scenario, hot Jupiters tend to be accompanied by nearby planets in orbital resonance \citep{Malhotra1993, Lee2002}. On the other hand, high-eccentricity tidal migration will destroy the nearby companions during process, but requires a distant companion outside the forming region of the proto-hot Jupiter to trigger the eccentricity excitation. Several surveys have been conducted to search for stellar or planetary companions to hot Jupiters out to tens of AU using radial velocities (RVs) and direct imaging \citep{Endl2014,Knutson2014,Bryan2016,Ngo2016}. \cite{Ngo2016} revealed that fewer than  $16\%$ of hot Jupiters possess stellar companions capable of driving proto-hot Jupiters into highly eccentric orbits, suggesting stellar companions are not typically the instigators of high-eccentricity tidal migration. In contrast, \cite{Knutson2014} reported that $51\pm{10}\%$ of hot Jupiters have an outer giant planet with masses between $1-13\ M_{J}$ and orbital semi-major axes between 1 and 20 AU, which are nearby and massive enough to raise the eccentricities of proto-hot Jupiters. In a recent study, \cite{zink2023} found that outer companions of all five hot Jupiters in their sample are at least three times more massive than the inner hot Jupiters and have an average eccentricity of 0.34. They argue that these features favor coplanar high-eccentricity migration as the dominant mechanism for the formation of hot Jupiters \citep{Petrovich2015}. %\cite{Wu2023} searched for transit timing variations across Kepler data and found that $12\pm{6}\%$ of hot Jupiters have a nearby planetary companion, indicating a quiescent dynamical history of these systems. 

In addition to spin-orbit angles and companion statistics, the mutual inclinations between the hot Jupiters and their outer companions might provide clues about the formation of hot Jupiters. For the secular von-Zeipel-Kozai-Lidov cycles to be effective, there must be an initial mutual inclination of over 40 degrees between the orbits of the two objects. On the other hand, recent studies propose that hot Jupiters originate through coplanar high-eccentricity migration, where the outer planets have eccentric orbits and relatively low mutual inclinations relative to the hot Jupiters \citep{Petrovich2015, ER2023}. \cite{Dawson2014} constrained the mutual inclination between a warm Jupiter and its outer giant planet companion in the Kepler-419 system to be $\sim 9^{\circ}$ using Transit Timing Variations (TTV). By combining radial velocity data with Gaia-Hipparcos astrometric accelerations, \citet{xw2020} found evidence for misalignments between the orbits of an outer giant planet and a smaller inner planet in the $\pi$ Men ($52.3^{\circ}<\Delta I<127.7^{\circ}$) and HAT-P-11 ($63.1^{\circ}<\Delta I<117^{\circ}$) systems (see also \citealt{DeRosa2020, Damasso2020}). However, direct measurements of mutual inclination between transiting planets, their companions and their orientation to the host star have been  limited to only a handful systems.  

 HD 118203 (TOI-1271) is a sub-giant star ($M_*=1.25\pm{0.06}\ M_{\odot}$, $R_* = 2.10\pm{0.06}\ R_{\odot}$) hosting a transiting hot Jupiter with an orbital period of 6.135 days, a mass of  $2.166 ^{+0.074}_{-0.079}\ M_{J}$  and an eccentricity of 0.3 \citep{Silva2006, pepper2020}. In this paper, we first combined Rossiter-McLaughlin observations from the Keck Planet Finder (KPF) and stellar rotation periods derived from Transiting Exoplanet Survey Satellite (TESS) light curves to determine the true spin-orbit angle of HD 118203 b. Furthermore, we present the discovery of HD 118203 c, a long-period giant planet orbiting outside HD 118203 b. We characterized the three-dimensional orbital parameters of this outer planet by combining RVs and  \textit{Hipparcos} and \textit{Gaia} astrometric data. HD 118203 is one of the first systems  where the spin-orbit angle for both the hot Jupiter and the outer giant planet, and the mutual inclination between the hot Jupiter and outer planet have been determined. Recently, \citet{Maciejewski2024} presented an independent detection of HD 118203 c using long-baseline radial velocity data from HARPS-N \citep{Cosentino2012} and the High-Resolution Spectrograph (HRS; \citealt{Tull1998}) at the Hobby-Eberly Telescope \citep{Ramsey1998} \footnote{ \cite{Maciejewski2024}, working in parallel with this work, also discovered HD 118203 c around the same time. The early draft of this work was completed and submitted before the findings of \cite{Maciejewski2024} were made public. However, during the peer-review process, we incorporated  RV data from their paper as soon as it became available to more precisely characterize the system.}. We compare our results with their findings in Section~\ref{sec:3dfitting}.

\section{Observations} \label{sec:obs}

\subsection{ELODIE and HIRES radial velocities} 

We used 56 archival RVs of HD 118203 taken with the ELODIE spectrograph installed on the 1.93m reflector at Observatoire de Haute-Provence in France \citep{Baranne1996} between 2004 May and 2006 June from \cite{Silva2006} and \cite{pepper2020}. We also include 18 RVs from High Accuracy RV Planet Searcher in the northern hemisphere (HARPS-N) and 51 RVs from High-Resolution Spectrograph (HRS) at the 9.2 m Hobby-Eberly Telescope published by \cite{Maciejewski2024}. The HRS observation were taken from  January 2006 and June 2013, and HARP-N data were taken from  December 2012 and August 2015. Furthermore, we collected 26 RV data for HD 118203 from 2010 to 2024 using the High Resolution Echelle Spectrometer (HIRES, \citealt{Vogt1994}) at the Keck I 10 m telescope on Maunakea. The observations are part of a survey aiming to search for substellar/stellar companions to transiting planet hosts with significant \textit{Hipparcos} and \textit{Gaia} astrometric accelerations \citep{zhang2023}. We used the standard California Planet Search (CPS) pipeline described in \cite{Howard2010} to determine RVs. Spectra were obtained with an iodine gas cell in the light path for wavelength calibration. An iodine-free template spectrum bracketed by observations of rapidly rotating B-type stars was used to deconvolve the stellar spectrum from the spectrograph PSF. We then forward-model the spectra taken with the iodine cell using the deconvolved template spectra \citep{Butler1996}. The wavelength scale, the instrumental profile, and the RV in each of the $\sim 700$ segments of 80 pixels were solved simultaneously \citep{Howard2010}. The RVs used in this work are presented in Table~\ref{tab:rvs}.

\input{table_rv}

\subsection{KPF Rossiter-McLaughlin observations}

The Keck Planet Finder \citep[KPF][]{Gibson2016,Gibson2018, Gibson2020} is an echelle spectrometer that was commissioned on the Keck 1 telescope at the W.M. Keck Observatory in March 2023.  KPF covers the wavelength range of 445--870 nm with a resolving power of 98{,}000.
The Rossiter-McLaughlin (RM, \citealt{Rossiter1924, McLaughlin1924} ) observations of HD 118203~b were conducted during one transit on April 09 2023 UT. A total of 76 exposures were obtained, each with an integration time of 180 s and a readout time of 60 s. We achieved a typical SNR of 176 at 550 nm. The observation covered the ingress and egress phases of the transit with additional 30 mins before and after the transit. The observation was interrupted for $\sim1.5$ hours in the middle of the transit due to high humidity. The spectra were reduced with the KPF Data Reduction Pipeline (DRP,\citealt{Gibson2020}), which is publicly available on Github\footnote{\url{https://github.com/Keck-DataReductionPipelines/KPF-Pipeline}}. The KPF DRP performs quadrant stitching, flat-fielding, order tracing, and optimal extraction. KPF's wavelength calibration sources include Th-Ar and U-Ne lamps, a Laser Frequency Comb, a Fabry-P\'erot Etalon, and the Solar Calibrator \citep{KPFSoCal}.  Our observations were taken with simultaneous wavelength calibration sourced from an etalon lamp fed through a dedicated calibration fiber. In addition to simultaneous calibrations, we performed hourly dedicated Etalon calibration exposures, known as `slew-cals,' to monitor the stability of RVs. We used the publicly available template-matching code SpEctrum Radial Velocity Analyser \texttt{serval} \citep{SERVAL} to extract the RVs from our KPF spectra. The KPF RVs used in this work are presented in Table~\ref{tab:rvs}.

\subsection{TESS photometry}
 HD 118203  appears in the \textit{TESS} Asteroseismic Target List \citep{Schofield2019} as a solar-like oscillator to be observed in 2-minute cadence. \textit{TESS} data were available as 2-minute cadence light curves for Sectors 15, 16, 22, 49 and 76 of the  mission, spanning from 2019 Aug 15 to 2024 Mar 26. It was also selected for  \textit{TESS} 20-second cadence observations in Sector 76. The light curves were processed by the Science Processing Operations Center (SPOC) data reduction pipeline \citep{Jenkins2016}. For the transit fitting, we downloaded all five sectors of Pre-search Data Conditioning simple aperture photometry (PDC$\_$SAP, \citealt{Smith2012, Stumpe2012,Stumpe2014})
light curves, and then stitched and normalized them using the \texttt{lightkurve} package \citep{LK2018}. To identify the rotation period, we also used the package Systematics-insensitive Periodogram (SIP, \citealt{Hedges2020}) to download all five sectors of TESS Target Pixel File data and build Simple Aperture Photometry (SAP) light curves of the target.

\subsection{Hipparcos and Gaia astrometric acceleration\footnote{Also known as proper motion anomaly.}}\label{sec:HG}

HD 118203 was identified as a transiting-planet host with significant \textit{Hipparcos} and \textit{Gaia} astrometric acceleration by \cite{zhang2023}, suggesting the existence of an outer companion in the system. We used the astrometric data for HD 118203 from the Hipparcos-Gaia Catalog of Accelerations (HGCA, \citealt{Brandt2021}). The HGCA catalog provides three proper motions in units of $\mathrm{mas}\ \mathrm{yr}^{-1}$: (1) the \textit{Hipparcos} proper motion $\mu_{\rm{H}}$ measured at an epoch near 1991.25; (2) the \textit{Gaia} EDR3 proper motion $\mu_{\rm{G}}$ measured at an epoch near 2016.01; (3) the long-term proper motion $\mu_{\rm{HG}}$ calculated as the difference in positions between \textit{Hipparcos} and \textit{Gaia} divided by the $\sim 25$-year baseline. The long-term proper motion $\mu_{\rm{HG}}$ can be approximately considered as the velocity of stellar linear motion across the sky plane over nearly 25 years. Following \cite{Kervella2019} and \cite{Brandt2021}, we computed the astrometric acceleration by subtracting the long-term proper motion $\mu_{\rm{HG}}$ from the \textit{Hipparcos} or \textit{Gaia} proper motion. Table~\ref{tab:HGCA_obs} presents the astrometric accelerations, which have a signal-to-noise ratio ($\rm{S/N}$) of 3.47 at the \textit{Gaia} epoch, and 2.71 at the \textit{Hipparcos} epoch.

\input{table_hgca}

\section{Host star Properties}\label{sec:starpa}

\subsection{Literature values}
HD 118203 is a sub-giant G0V star at an age of $5.32^{+0.96}_{-0.73}$ Gyr and a distance of $91.811\pm{ 0.236}$ pc \citep{pepper2020}. The star has a mass of $1.25 \pm{0.06}\ M_{\odot}$, a radius of $2.10 \pm{0.06}\ R_{\odot}$, and a metallicity [Fe/H] of $0.223\pm{0.076}$~dex. Its spectroscopically determined projected rotation velocity, $v\sin i_*$, is $5.32\pm{0.5}$~km\ s$^{-1}$ \citep{Luck2017}. \cite{pepper2020} reported a rotation period of $20\pm{5}$ days from TESS light curves of sector 15 and 16. Recently, \cite{Castro2024} reported a period of 6.1 days from \textit{TESS} PDCSAP light curves, which matches with the orbital period of HD 118203 b. They argue it could be resulted from the magnetic interaction between the hot Jupiter and the host star.

\subsection{ High-resolution Spectroscopy}
We obtained iodine-free spectra with Keck/HIRES for HD 118203 to characterize
stellar parameters. We used SpecMatch synthetic methodology (SpecMatch-Synth, \citealt{Petigura_thesis}) to compute stellar properties. We obtained a projected rotation velocity of the star as $5.13\pm{1.0}~\rm{km}\ s^{-1}$. Other parameters are listed in Table~\ref{tab:star-compare}. 

\subsection{Isochrone fitting}
As described in Section~\ref{sec:GPSHO}, we measured asteroseismic oscillations with a frequency at maximum power of $\nu_{\rm{max}}=967.2^{+46.5}_{-51.7}$~$\mu$Hz. We also employed \texttt{isoclassify} \citep{huber2017,berger2020} to derive the stellar parameters with following input observables: (1) the frequency at maximum power $\mathrm{\nu_{max}}$, (2) the spectroscopic stellar effective temperature and metallicity \citep{pepper2020}, (3) the 2MASS $K_{s}$ magnitude \citep{Cutri2003}, and (4) the Gaia DR3 parallax \cite{Gaia}. Our measurement of $\nu_{max}$ helps to break the mass and age degeneracy reported by \cite{pepper2020}. 
We derived a mass of $1.27^{+0.028}_{-0.032}\ M_{\odot}$, a radius of $2.04\pm{0.03}\ R_{\odot}$, and an age of $4.73\pm{0.39}$~Gyr, consistent with the classification of the host star as a sub-giant. The stellar parameters are given in Table~\ref{tab:star-compare}.

\input{table_stellar_params}

%\section{Rossiter-McLaughlin analysis}

\section{spin-orbit angle of HD 118203 \MakeLowercase{b}}

In this paper, we define the spin-orbit angle as the angle between the stellar spin axis and the orbital axis of a planet,  also commonly referred to as stellar obliquity. In single-planet systems, the two definitions are equivalent. However, in multiple-planet systems, stellar obliquity is defined as the angle between the stellar spin axis and the reference vector, typically the total angular momentum of system. Whereas the term spin-orbit angle could be applied individually to each planet in multiple-planet systems.

\subsection{RM effect modelling without oscillations}\label{sec:cRM}

We used the \texttt{allesfitter} \citep{ allesfitter-code,allesfitter-paper} package to determine the sky-projected spin-orbit angle for HD 118203 b. In the first step, we performed a simultaneous global fit of 2-minute cadence light curves from TESS sectors 15, 16, 22, 49, and 76 out-of-transit RVs from ELODIE \citep{Silva2006, pepper2020} and HIRES, and RM measurements from KPF. We allowed 24 free parameters in the fitting (see Table~\ref{tab:rm_result}). The planetary properties and orbital elements of HD 118203 were described by the following parameters: $r_{p}/R_{*}$, $(R_{*}+r_{p})/a_{b}$, $\cos{I_{b}}$, $T_{0,b}$, $P_{b}$, $K_{b}$, $\sqrt{e_{b}}\cos{\omega_{b}}$, and $\sqrt{e_{b}}\sin{\omega_{b}}$, where $R_{*}$ and $r_{p}$ are stellar and planet radii, $\cos{I_{b}}$ is the cosine of the orbital inclination, $T_{0,b}$ is the midtime of transit, $P_{b}$ is the orbital period, $K_{b}$ is the RV semi-amplitude, $e_{b}$ is the eccentricity, and $\omega_{b}$ is the argument of pericenter. The RM effect was modeled with two additional parameters:
$\lambda_{b}$, the projected spin-orbit angle for HD 118203 b, and $v\sin{i_{\star}}$, the projected stellar rotation velocity. In addition, we fitted the transformed quadratic limb darkening coefficients for TESS ($q_{1,\rm{TESS}}$, $q_{2,\rm{TESS}}$) and KPF ($q_{1,\rm{KPF}}$, $q_{2,\rm{KPF}}$). We also included constant zero point offsets for TESS light curves and KPF RVs ($\gamma_{\rm{TESS}}$, $\gamma_{\rm{KPF}}$) and linear slopes for ELODIE and HIRES RVs ($\gamma_{\rm{ELODIE}}$, $\dot{\gamma}_{\rm{ELODIE}}$, $\gamma_{\rm{HIRES}}$, $\dot{\gamma}_{\rm{HIRES}}$). Finally, we added four free parameters to account for the jitter terms of the four instruments ($\ln \sigma_{\rm{TESS}}$, $\ln \sigma_{\rm{ELODIE}}$, $\ln \sigma_{\rm{HIRES}}$, $\ln \sigma_{\rm{KPF}}$). We adopted uniform priors for each of the parameters (see Table~\ref{tab:rm_result}). The initial guesses for the planetary and orbital parameters were adopted from \cite{pepper2020}, while the initial guess for $v\sin{i_{\star}}$ was the spectroscopically determined value from \cite{Luck2017}.
\input{fig1}
\input{fig2}
%and a single ellipse

We performed a dynamic Nested Sampling fit with \texttt{dynesty} \citep{dynest2020} to sample the posterior distributions of all model parameters. Our Nested Sampling analysis used 500 live points and a convergence criterion that required the uncertainty of the Bayesian evidence estimate to be $<0.01$. The best-fit parameters and their $1\sigma$ uncertainties are reported in Table~\ref{tab:rm_result}. The fitting results in a sky-projected spin-orbit angle for HD~118203~b of $\lambda_b=-11_{-13}^{+11}\,^{\circ}$, and a projected stellar rotation speed of $v\sin{i_{\star}}=5.85_{-0.32}^{+0.41}\ \mathrm{km\ s^{-1}}$ (see Appendix~\ref{fig:figureA1} for the best-fit models and Appendix~\ref{fig:figureA2} for the joint posterior distributions of the 24 parameters). Our fitted $v\sin{i_{\star}}$ is consistent with both the spectroscopic measurements ($5.13\pm{1}$~km\, s$^{-1}$) from the HIRES spectra  and the value ($5.32\pm{0.5}$~km\, s$^{-1}$) reported by \citet{Luck2017} .

Additionally, we computed the power spectrum of the KPF RV residuals using a Lomb-Scargle periodogram \citep{Press1989}. The power spectrum reveals a clear pattern of peaks attributable to solar-like oscillations centered around $950\ \mathrm{\mu Hz}$ (see Figure~\ref{fig:figure1}). The observed frequency at the peak is consistent with the expected $\nu_{\rm{max}}$  calculated using the scaling relation $\frac{g}{g_{\odot}}=\frac{\nu_{\rm{max}}}{\nu_{\odot}}\frac{T_{\rm{eff}}}{T_{\rm{eff,\odot}}}$, based on the stellar parameters from \cite{pepper2020}. Because the KPF RM observation only spans about six hours, it is not feasible to distinguish individual oscillation frequencies. On the other hand, we performed Lomb-Scargle periodogram analysis on the combined \textit{TESS} light curves spanning 5 sectors with a 2-minute cadence, as well as on each sector individually. Furthermore, we examined the Lomb-Scargle periodogram of light curves with a 20-second cadence specifically from Sector 76. No oscillation modes were identified in any of these analyses.

\subsection{Gaussian process modelling of the RM effect with oscillations}\label{sec:GPSHO}

Since the stellar oscillation period (17.5~min) is longer than the cadence of the KPF RVs (4~min), we implemented a Gaussian process (GP) model alongside the Rossiter-McLaughlin effect to account for any possible systematic errors in the spin-orbit angle measurement introduced by the asteroseismic oscillations. The oscillations can be modelled with a sum of stochastically driven damped simple harmonic oscillators (SHO), as described by  \cite{Foreman-Mackey2017}. We used a single SHO term to describe the  envelope of the spectrum of oscillations. The power spectrum of a SHO is
\begin{equation}
    S(\omega)=\sqrt{\frac{2}{\pi}}\frac{S_{0}\omega^{4}}{(\omega^{2}-\omega_{\mathrm{max}}^{2})^{2}+\omega^{2}\omega_{\mathrm{max}}^{2}/Q_{0}^{2}}
\end{equation}
where $\omega_{\mathrm{max}}$ is the  angular frequency at the maximum power, which can be converted to frequency at maximum power using $\nu_{max}=\omega_{max}/2\pi$. $Q_{0}$ is the quality factor describing the oscillation damping time-scale, and $S_{0}$ is proportional to the power at $\omega=\omega_{\mathrm{max}}$:$S(\omega_{\mathrm{max}})=\sqrt{2/\pi}S_{0}Q_{0}^{2}$. 
%In practice, a high-quality factor $Q_{0}\gg 1$ effectively models asteroseismic oscillations. Additionally, given that the duration of observation is substantially shorter than the typical decay time of  oscillations in subgiants, which spans several days, a high $Q_{0}$ is also a plausible assumption for our case. In our case, $Q_{0}$ serves to describe the width of the power envelope.

We used 27 free parameters in the Nested Sampling, with the same parameters as described in section~\ref{sec:cRM} and three additional GP parameters ($\ln S_{0}$, $\ln Q_{0}$, and $\ln\omega_{\mathrm{max}}$). We adopted wide uniform priors on the three GP parameters (see Table~\ref{tab:rm_result}). The damping timescale  of sub-giant oscillations is typically a few days, which is significantly longer than our observing baseline \citep{Lund2017}. Therefore, we set a uniform prior of $\ln Q_{0}$ from 1 to 4, consistent with coherent (undamped) oscillations. We used 500 live points, and required the Bayesian evidence estimate to be $<0.01$ for convergence.

Figure~\ref{fig:figure2} shows the results of modeling the RM effect alongside the stellar oscillations. We obtained a sky-projected spin-orbit angle of $\lambda_b = -11^{\circ}.7^{+7.6}_{-10}$ and a projected stellar rotation speed of $v\sin i_* = 5.89^{+0.31}_{-0.26}\ \mathrm{km\ s^{-1}}$, which agrees with the results we obtained when fitting the RM effect without modeling the stellar oscillations (section~\ref{sec:cRM}). The Bayes factor, with a value of $\Delta \log(Z) = 6.19 > 5$, indicates that the RM model with the GP is favored over the RM model without the GP. Despite this, the fitting outcomes are consistent, indicating that the stellar oscillations have a negligible effect on the derived spin-orbit angle. This result aligned with our expectations, given that the amplitude and timescale of the RM signal are both much larger than those of the oscillations. The GP model with an SHO kernel yields a frequency of maximum oscillation power $\nu_{max}=967.2_{-46.5}^{+51.7}\ \mathrm{\mu Hz}$, which corresponds to an oscillation period of $\sim 17 $ minutes. 

\input{table_modelcompare1}

%We also compute the Lomb-scargle periodigram of the KPF RV residuals in the RM fitting with SHO kernel. Figure~\ref{fig:figure1} shows the the SHO kernel effectively models the oscillations and removes periodic patterns from the residuals.

\input{figA4}
\input{fig3}

\subsection{True spin-orbit angle distribution}\label{sec:istar}

The true spin-orbit angle $\Psi_{p}$ of a planet is given by
 \begin{equation}\label{eqn:Psi}
   \cos{\Psi_{p}} = \cos{i_{*}}\cos{i_{p}} + \sin{i_{*}}\sin{i_{p}}\cos{\lambda_{p}},
\end{equation}
where $i_{p}$ is the planet's orbital inclination, $i_{*}$ is the star's spin-axis inclination, and $\lambda_{p}$ is the sky-projected spin-orbit angle. For HD 118203 b, $i_{b}$ is tightly constrained by the TESS transit observations, and $\lambda_{b}$ is constrained by modeling the RM effect. Therefore, it is essential to determine the stellar inclination $i_{*}$ in order to derive the true spin-orbit angle of a planet. When the stellar radius $R_*$ and rotational period $P_{\rm{rot}}$ are known, the equatorial rotational velocities can be calculated as \( v_{eq} = 2\pi R_{*}/P_{\rm{rot}} \). If $R_*$, $P_{\rm{rot}}$ and $v\sin{i_*}$ are independent, $i_*$ can be derived as $\sin^{-1}(\frac{ v\sin{i_*}}{v_{eq}})$. However, \cite{MandW2020M} pointed out that $v_{eq} $ and $v\sin{i_*}$ are not statistically independent, and the consideration of their correlation is needed when determining the probability distribution of $i_*$. Addressing this correlation, \cite{Bowler2023} presents a Bayesian probabilistic framework to constrain the stellar inclination as
 
\begin{equation}{\label{eqn:i}}
P(i_*\mid P_\mathrm{rot}, R_*, v\sin i_*) \propto  \sin i_* \times \frac{e^{- \frac{\big(v \sin i_* - \frac{2\pi R_*}{P_\mathrm{rot}}\sin i_* \big)^2}{2\big(\sigma_{v\sin i_*}^2 + \sigma_{v_\mathrm{eq}}^2 \sin^2 i_* \big)}} }{\sqrt{\sigma_{v\sin i_*}^2 + \sigma_{v_\mathrm{eq}}^2 \sin^2 i_*}},
\end{equation}
where $\sigma_{v\sin i_*}$ and $\sigma_{v_\mathrm{eq}}$ are the uncertainties in the projected velocity $v\sin{i_{*}}$ and the equatorial velocity $v_{eq}$. We approximated $\sigma_{v_\mathrm{eq}}$ as
\begin{equation}{\label{eqn:veqerr}}
\sigma_{v_\mathrm{eq}} =  \frac{2\pi R_*}{P_\mathrm{rot}} \sqrt{\Big(\frac{\sigma_{R_*}}{R_*}\Big)^2 + \Big(\frac{\sigma_{P_\mathrm{rot}}}{P_\mathrm{rot}}\Big)^2 }.
\end{equation}
 
 We applied a Systematics-insensitive Periodogram (SIP, \citealt{Hedges2020}) to stitch and detrend systematics in the TESS light curves from sectors 15, 16, 22, 49, and 76. This process enabled us to generate a Lomb-Scargle periodogram. We identified a rotation period $P_\mathrm{rot}$ of $23.1\pm{3.8}$ days (see Figure~\ref{fig:figureA4}). We derived the uncertainty from the width of the periodogram peak, adding an extra $10\%$ to account for  surface differential rotation \citep{Epstein2014, Claytor2022}. Our result is consistent with the rotation period of $20\pm{5}$ days reported by \cite{pepper2020}. We adopted the stellar projected rotational velocity $v\sin i_*$ of $5.32\pm{0.5}$ km s$^{-1}$ reported from \cite{Luck2017} and stellar radius $R_{*}$of $2.04\pm{0.03}\ R_{\odot}$ (see Section~\ref{sec:starpa} for details). Figure~\ref{fig:figure3} b shows the resulting distribution of the stellar inclination which ranges from $0^{\circ}$ to $180^{\circ}$ to accommodate the possibility that the stellar axis  points either toward ($<90^{\circ}$) or away($>90^{\circ}$) from the observer. We obtained a stellar inclination of $i_{\star}=89^{\circ}.9^{+13.7}_{-13.8}$ at $1\sigma$ confidence, suggesting that the stellar rotation axis is orthogonal to the line of sight. %The findings align with the calculation in \cite{pepper2020}, where a rotational period of 22 days was estimated based on the $v\sin i_*$ and $R_{*}$, assuming that the axis of stellar rotation is perpendicular to our line of sight. 

By combining this information with the sky-projected spin-orbit angle from our RM measurements with Eqn.~\ref{eqn:Psi}, we derived the cosine term of  the true spin-orbit angle $\cos{\Psi_{b}}$, which is strongly peaked at 1 (see Figure~\ref{fig:figure3} c). This is consistent with expectations as we have projected spin-orbit angle $\lambda_b$ is close to 0 deg and stellar inclination peaks at 90 deg. Figure~\ref{fig:figure3} d shows the distribution of the true spin-orbit angle $\Psi_b$ for HD 118203 b after transforming from cosine terms to angles. However, this distribution shows a low probability near $\Psi_b=0$. This does not necessarily imply that the true spin-orbit angle is away from zero, but rather reflects the effect of transforming from cosine terms to angles. Therefore, we believe it is more appropriate to provide an upper limit for $\Psi_b$, which is $\Psi_{b}<33^{\circ}.5 $ at $2\sigma$ significance level. The low spin-orbit angle indicates HD 118203 b is nearly aligned with the host star.

\section{The Discovery of an outer long-period planet}\label{sec:dist}
\input{fig11}
\input{fig4}

\cite{Silva2006} and \cite{pepper2020} both reported an RV trend in addition to HD 118203 b's Keplerian signal across two years of ELODIE observations. The Gemini-North/Alopeke speckle imaging ruled out the possibility of a stellar companion within 1.25 arcseconds. Therefore, the RV trend could possibly arise from either a massive planet or a brown dwarf companion.  Recently, \citet{Maciejewski2024} used long-baseline RV data from HARPS and HRS to detect the outer planet, HD 118203 c. In this work, we combine archival data with our new HIRES RVs, which extended the observation baseline to nearly 20 years, to further refine the orbit of the outer planet.

%The HIRES observations presented here extended the observation baseline to nearly 20 years

\subsection{Lomb-Scargle Periodigram}
%\subsection{Comparing single-planet and two planet models }\label{sec:51}
%Figure~\ref{fig:figure4} presents the radial velocities of HD 118203 taken using ELODIE, Keck/HIRES, HRS, and HARP-N. We first performed a single-planet Keplerian orbital fit to the RVs using \textit{RadVel} \citep{Fulton2018}. We obtained orbital parameters and masses for HD 118203 b that are consistent with those of \cite{pepper2020}. However, the RV residuals reveal a noticeable curvature,  indicating an additional long-period planet. Furthermore, the single-planet MCMC fit resulted in a HIRES jitter term $\sigma_{\rm{HIRES}}$ of $30\ m\ s^{-1}$. This value is significantly greater than the typical HIRES uncertainties ($\sim3\ m\ s^{-1}$) and the stellar jitter due to oscillations ($\sim2\ m\ s^{-1}$), implying that the single-planet model is inadequate.

Figure~\ref{fig:figure11} (a) displays the radial velocity (RV) periodogram in which the vertical axis shows the Bayesian Information Criterion (BIC) difference, $\Delta \rm{BIC}$, when comparing single-planet and two-planet models. These calculations were performed using the \texttt{RVsearch} package, as detailed in \cite{Rosenthal2021}. Specifically, \texttt{RVsearch} establishes an orbital period grid ranging from two days to five times the span of observational data to search for periodic signals. The algorithm iteratively fits a sinusoid at each fixed period within this grid and computes $\Delta \rm{BIC}$ for each point. The algorithm calculates an empirical false alarm probability (FAP) by fitting a linear model to a log-scale histogram of periodogram power values. From this, it extrapolates a detection threshold in terms of $\Delta \rm{BIC}$ corresponding to an empirical FAP of  $0.1\%$, indicating that only  $0.1\%$ of periodogram values are anticipated to exceed this threshold (see \cite{Howard2016} for details). We calculated two $\rm{\Delta\ BIC}$ periodograms: the first one using all available RVs and the second one only using HIRES RVs. The two periodograms both reveal pronounced peaks at $\sim4979$ days surpassing the detection threshold (a FAP of $0.1\%$). The lower peak observed at shorter periods above FAP is attributable to the harmonics of the period. The peak disappears if we remove the signal at $\sim4979$ days. To investigate whether stellar activity could account for the long-period signal, we also conducted a periodogram analysis of the \textit{S}-values. We computed \textit{S}-values for HIRES data by measuring the core flux of Calcium H $\&$ K lines. Note that we only used the \textit{S}-values from the HIRES observations. We do not see the peak at 4979 days shown in the RV periodogram (see Figure~\ref{fig:figure4} d).  Moreover, HD 118203 shows a Hipparcos-Gaia astrometric acceleration with $>3\sigma$ significance, which can not be accounted by the stellar activity, and suggests the star undergoes orbital motion due to gravitational interaction with a long-period giant planet.  %Therefore, the RV reversal is consistent with the discovery of a long-period giant planet (hereafter HD 118203 c). Furthermore, the semi-amplitude shown in the RV residuals is at least  $120\ m\ s^{-1}$, which is much larger than activity induced RV variability seen in similar stars. \cite{Isaacson2010} measured the chromospheric activity and RV jitter of nearly 2600 main-sequence and sub-giantstars. For sub-giants similar to HD 118203 in their sample, the typical root mean square(rms) in measured RVs are around $4\sim15\ m\ s^{-1}$.

\subsection{RV-only Keplerian Fit}

We first performed a two-planet Keplerian orbital fit to the RVs using \texttt{RadVel} \citep{Fulton2018}. The model includes five orbital elements for each planet $K$, $P$, $T_{conj}$, $\sqrt{e}sin\omega$,$\sqrt{e}cos\omega$, where $K$ is the RV semi-amplitude, $P$ is the orbital period, $T_{conj}$ is the time of conjunction, $e$ is the eccentricity, and $\omega$ is the argument of pericenter. For HD 118203 b, we used Gaussian priors informed by \textit{TESS} observations \citep{pepper2020} for the orbital period $P_b$ and conjunction time $T_{conj,b}$.  We set bounds on $0<e<1$, $K>0$ for all planets.

\input{rvonlytab}

We performed the Markov Chain Monte Carlo (MCMC) exploration with \texttt{emcee} \citep{FM2013} to estimate parameter credible levels.  Our MCMC analysis used 50 walkers and ran for $1.5 \times 10^6$ steps per walker, achieving a maximum Gelman-Rubin(GR) statistic of 1.004. Figure~\ref{fig:figure4} shows the best fit Keplerian solution.  We derived a minimum mass for HD 118203 c from RV amplitudes as $m_{c}\sin{c}=11.70^{+0.87}_{-0.83} \ \rm M_{J}$, semi-major axis as $a_c = 6.32^{+0.17}_{-0.18}$ AU and an eccentricity of $e_c=0.268^{+0.028}_{-0.026}$. The derived planetary parameters are given in Table~\ref{tab:rvparams}. 

\input{fig5}

\subsection{Three-dimensional orbit fitting  }\label{sec:3dfitting}

We then characterized the three-dimensional orbit of the outer companion by conducting a joint fit to the Hipparcos-Gaia astrometric accelerations and RV time series, following the methodology outlined in \cite{xw2020}. The long-baseline RVs over 20 years provide crucial information about the period, eccentricity, and companion mass. Meanwhile, the two astrometric accelerations measured at the Hipparcos and Gaia epochs, with a baseline around 25 years, offer the essential constraint on the inclination. We used the parallel-tempering Markov chain Monte Carlo (PT-MCMC) ensemble sampler in \texttt{emcee} \citep{Foreman-Mackey2017}. We use 40 different temperatures to sample the parameter space, and our results are taken from the `coldest' chain, which corresponds to the original, unmodified likelihood function. For each of the 40 temperatures, we use 100 walkers to sample the 18-parameter model. Two of these parameters define the mass of the host star $M_{\star}$ and the parallax of the system $\varpi$. Seven parameters define the orbits and mass of the outer companion: the true mass of companion $m_{c}$, the cosine  of its orbital inclination $\cos{I_{c}}$, the orbital period $P_{c}$, the longitude of the ascending node $\Omega_{c}$, the eccentricity $e_{c}$, and the argument of periastron $\omega_{c}$ (fitted as $\sqrt{e_{c}}\cos{\omega_{c}}$ and $\sqrt{e_{c}}\sin{\omega_{c}}$)\footnote{Note that $\omega_{c}$ is the argument of periastron of the {\it planet}, not the star. The argument of periastron of the planet is offset by $\pi$ from that of the star.}, and the epoch of periastron $\tau_{c}$ at a reference epoch ($t_{\rm{ref}}=2451544.5$ JD). $\tau$ is a dimensionless quantity ranging from 0 to 1, calculated relative to a reference epoch $t_{\mathrm{ref}}$, the time of pariastron $t_{\mathrm{p}}$ and the planet orbital period $P$ : $\tau = (t_{p}-t_{\mathrm{ref}})/P$ \citep{Blunt2020}. Determining the bounds of $\tau$ is straightforward, even if the orbital period $P$ is uncertain for planets with long orbital periods. Five parameters are for the orbit of the inner hot Jupiter HD 118203 b: RV semi-amplitude $K_{b}$, orbital period $P_{b}$, $\sqrt{e_{b}}\cos{\omega_{b}}$, $\sqrt{e_{c}}\sin{\omega_{b}}$ and mean longitude at a reference epoch $\tau_{b} $. Given that the orbital period of HD 118203 b (6.13 days) is significantly shorter than the observation periods of both \textit{Hipparcos} and \textit{Gaia}, its astrometric signal would be canceled out. Therefore, we did not consider it in the astrometric model and set its inclination to be $I_{b}=88^{\circ}.48$, which is well-determined by the transiting observations. Finally, we include four parameters to account for the RV zero points ($\gamma_{\rm{ELODIE}}$, $\gamma_{\rm{HIRES}}$, $\gamma_{\rm{HRS}}$, $\gamma_{\rm{harps}}$) and jitter terms ($\sigma_{\rm{ELODIE}}$, $\sigma_{\rm{HIRES}}$, $\sigma_{\rm{HRS}}$, $\sigma_{\rm{harps}}$) for each instrument. The final likelihood used in our MCMC is given by

%Compared to a regular single MCMC analysis, PT-MCMC enhances efficiency in exploring various areas of the parameter space by running multiple MCMCs at different `temperatures,' corresponding to different modified likelihoods for the posterior. Therefore, PT-MCMC can help to prevent our fit from becoming confined to local minima.

\begin{equation}
    \ln\mathcal{ L} = -\frac{1}{2}(\chi^2_{\Delta \mu}+ \chi^2_{RV}),
\label{eq:loglike}
\end{equation}

with

\begin{equation}\label{eqn:chi2_rv}
    \begin{split}
    \chi^2_{RV} =\sum_{j}^{N_{\rm{inst}}} \sum_{i}^{N_{\rm{RV}}}[\frac{(\mathcal{M}[{\rm RV}_i] - {\rm RV}_i + {\rm RV}_{\rm{offset,j}})^2}{\sigma_i^2 + \sigma_{\rm{jit,j}}^2}  \\
     +  \ln{2\pi (\sigma_i^2 + \sigma_{\rm{jit,j}}^2)}],
    \end{split}
\end{equation}

\begin{equation}\label{eqn:PMa}
    \begin{split}
    \chi^2_{\Delta \mu} = \sum_{j}^{G, H} \left[\left(\frac{\mathcal{M}[\Delta \mu_{j,\alpha}] - \Delta \mu_{j,\alpha}}{\sigma[\Delta \mu_{j,\alpha}]}\right)^{2}\right . \\
    \left.+ \left(\frac{\mathcal{M}[\Delta \mu_{j,\delta}] - \Delta \mu_{j,\delta}}{\sigma[\Delta \mu_{j,\delta}]}\right)^{2}\right]
    \end{split}
\end{equation}

Equation~\ref{eqn:chi2_rv} presents the chi-square of RV fitting $\chi^2_{RV}$, where $\mathcal{M}[{\rm RV}_i]$, ${\rm RV}_i$, and $\sigma_i$ are the model, data, and uncertainties of the radial velocity at time $i$. ${\rm RV}_{\rm zero,j}$ and $\sigma_{\rm jit,j}$ are the RV zero-points and jitter terms for the instrument $j$. The RV model is given by the third component of Eq. 4 in \cite{xw2020}. Specifically, we consider a two-planet model to fit RV data.

\input{table_3d}
\input{fig6}
Equation~\ref{eqn:PMa} calculates the chi-square of astrometric fitting $\chi^2_{\Delta \mu}$ by aggregating data across \textit{Hipparcos} and \textit{Gaia} epochs in the Right Ascension (RA,$\alpha$) and Declination(DEC,$\delta$) dimensions. $\Delta \mu$ and $\sigma[\Delta \mu]$ are the astrometric accelerations and uncertainties listed in Table~\ref{tab:HGCA_obs}. $\mathcal{M}[\Delta \mu]$ is the astrometric model that depends on the orbital elements of the outer companion, including $I_{c}$, $M_{c}$, $\Omega_{c}$ (for details, see \citealt{xw2020}). We neglect HD 118203 b and only consider the outer companion in the astrometric model. We also take into account two important systematic corrections in the astrometric model. First, \textit{Hipparcos} and \textit{Gaia} do not provide truly instantaneous proper motions, but averaged values over multiple measurements during the Hipparcos ($\delta_{\rm{H}}=1227$ days) and  Gaia DR3 ($\delta_{\rm{GDR3}}=1002$ days) observation windows. Consequently, the astrometric acceleration from a companion whose orbital periods are comparable to or shorter than the time spans of 1227 days or 1002 days will experience a smearing effect throughout the observation period. Furthermore, even targets with longer periods will suffer some amount of smearing. To account for the smearing effect, we compute the tangential velocities at every individual observation time. The \textit{Hipparcos} observation times can be found from the Hipparcos Epoch Photometry Annex \citep{vanLeeuwen1997} and the \textit{Gaia} DR3 observation times can be downloaded in \textit{Gaia} Observation Forecast Tool \footnote{https://gaia.esac.esa.int/gost/} within the {\it Gaia} DR3 and {\it Hipparcos} observing periods, and then average over them to mimic the smearing effect. Second, the star was in different orbital phases during the \textit{Hipparcos} and \textit{Gaia} epochs, and the star's orbital motion contributed to the long-term mean motion vector $\mu_{HG}$. To account for this effect, we modeled the orbital positions of the star at each observation time of {\it Hipparcos} and {\it Gaia DR3} and averaged these positions to obtain the mean orbital positions at the two epochs. 

Our PT-MCMC analysis stabilized in the mean and root mean square (rms) of the posteriors of each of the model parameters after $1\times 10^{5}$ steps. We saved every 100th step of our chains and discarded the first $75\%$ of the chain as the burn-in portion. Figure~\ref{fig:figure5} shows the astrometric models and data. Table~\ref{tab:3d} presents the posterior distributions of our fitted parameters, as well as some derived parameters. 

Figure~\ref{fig:figure6} presents the joint posterior distributions for the orbital inclination, mass and orbital period of  HD 118203 c (see Appendix~\ref{fig:figureA5}  for additional parameters). Our results show that HD 118203 c is a giant planet with a mass of $m_{c}=11.79^{+0.69}_{-0.63}\ \mathrm{M_{J}}$ 
semi-major axis of $a_{c}=6.28^{+0.10}_{-0.11}$ AU and eccentricity $e_{c}=0.26^{+0.03}_{-0.02}$. We obtained the orbital inclination of the outer giant planet as $I_{c}=93^{\circ}.7^{+11.9}_{-13.8}$, suggesting that HD 118203 c orbits the star in an almost edge-on orbit. Our results are consistent with the values reported from \cite{Maciejewski2024} within $1\sigma$.

%The two peaks of the $I_{c}$ distribution add up to $180^{\circ}$. This bi-modal nature likely stems from a prograde-retrograde degeneracy, where two orbits lie in the same plane but exhibit opposite directions of orbital motion. The current data quality is insufficient to resolve the orbital direction. However, solutions with $I_{c} > 90^{\circ}$ are more favored. If we only consider solutions with $I_{c}>90^{\circ}$, the orbital inclination of HD 118203 c is $I_{c}=106.7^{+13.8}_{-10.6}$ deg at $1\sigma$ confidence and $I_{c}=106.7^{+32.1}_{-15.6}$ deg at $2\sigma$ confidence. On the other hand, for solutions $I_{c} < 90^{\circ}$, we get $I_{c}=74.1^{+11.0}_{-15.7}$ deg at $1\sigma$ confidence and $I_{c}=74.1^{+15.1}_{-33.7}$ deg at $2\sigma$ confidence. Note that the solutions with $I_{c} > 90^{\circ}$ do not necessarily imply that the outer planet's orbit is retrograde, as the direction of the star's rotation—whether clockwise or counterclockwise—is not known. Based on the prograde orbit of HD 118203 b observed from RM observation in Section~\ref{sec:GPSHO}, it is more probable that HD 118203 c also orbits in a prograde direction.

\input{fig8}
\input{fig9}

\subsection{Constraints on the planet-planet mutual inclination and spin-orbit angle for HD 118203 c}

Figure~\ref{fig:figure8} a$\&$b depict the fundamental orbital geometry illustrating the mutual inclination between two planets, defined as the angle between the their orbital axes. In the diagrams, the z-axis corresponds to the line of sight, and is perpendicular to the x-y sky plane. The orientation of a planet's orbit is determined by two parameters: the orbital inclination $I$, which represents the angle between a planet's orbital axis and the line of sight (z-axis), and the longitude of ascending node $\Omega$ defined as the angle of the line of nodes from a reference direction (celestial north). In principle, the mutual inclination between two planets can be calculated by measuring their inclination and longitude of ascending node \citep{Fabrycky2009}:  
\begin{equation}
    \cos{\Delta I_{12}} = \cos{I_1}\cos{I_2} + \sin{I_1}\sin{I_2}\cos{(\Omega_1 - \Omega_2)}
    \label{eq:Imut}
\end{equation} 
where $I_i$ and $\Omega_i$ is the inclination and longitude of ascending node for planet $i$. Typically, using transit and radial velocity data, we can infer the inclination of a transiting planet, but not its longitude of the ascending node. In such scenarios, the orbital axis of a transiting planet with an inclination close to $90^{\circ}$ might be oriented in any direction within the sky plane.  Similarly, for a non-transiting planet with an inclination away from $90^{\circ}$, its orbital axis could lie on the surface of a cone, which is either towards ($I_2 < 90^{\circ}$) or away from ($I_2 > 90^{\circ}$) the observer. %Furthermore, when two orbits have an inclination sum equal to $180^{\circ}$, as seen in $P_1$ and $P^{\prime}_1$ in Figure~\ref{fig:figure8}b, their orbital axes align along the same line but in opposite directions. This indicates that both orbits occupy the same plane but the orbit directions are opposite. This phenomenon is known as the retrograde-prograde degeneracy.  
Therefore, if we only have information about the orbital inclinations of two planets and the longitude of ascending nodes for one or both planets is unknown, we cannot determine the true mutual inclinations. Instead, we can only derive the mutual inclinations as projected along the line of sight.

%a broader range of possibilities exists for the true mutual inclinations.

Figure~\ref{fig:figure9}~a presents a comparison of the orbital inclinations of HD 118203 b, c and stellar inclination using polar coordinates. By calculating the absolute difference between $I_b$ and $I_c$,  we can obtain their mutual inclination in the line of sight direction, as shown in Figure~\ref{fig:figure9}~b. The resulting distribution is compressed towards zero, yielding a line of sight $\Delta I_{bc}$ as 
$9^{\circ}.8^{+8.6}_{-6.7}$ at $1\sigma$ and 
$9^{\circ}.8^{+16.2}_{-9.3}$ at $2\sigma$. The peak of this distribution, where the probability is highest, is at $1.5^\circ$. The measurement is consistent with the scenario of a low mutual inclination between HD 118203  b and c. Similarly, we can also constrain the spin-orbit angle of HD 118203 c in line of sight by calculating the absolute difference between stellar inclination and $I_c$.  Figure~\ref{fig:figure9}~c illustrates that the line of sight $\Psi_c$ is $14^{\circ}.7^{+13.3}_{-10.2}$ at the $1\sigma$ confidence level and $14^{\circ}.7^{+26.4}_{-13.9}$ at the $2\sigma$ level. The most probable value is $4^\circ.5$. Our findings indicate that HD 118203 c is aligned with the host star at the $2\sigma$ level.  While the line of sight measurements of $\Delta I_{bc}$ and $\Psi_c$ only provide minimum values because the angles in the sky plane are not determined, the similar inclinations observed suggest a strong possibility of alignment across the entire system. For instance, if the $\cos{i}$ values for the inclinations of planets b and c and the stellar inclination were to be randomly selected from a uniform distribution between -1 and 1, the likelihood that all of these bodies would have inclinations within the range of $70^{\circ}$to $110^{\circ}$ is merely $3.4\%$.

%To constrain the mutual inclination between HD 118203 b $\&$c, we randomly sampled $I_{c}$ and $\Omega_{c}$ from the chains of our MCMC fitting and $I_{b}$ from a normal distribution $\mathcal{N}(88^{\circ}.48,0^{\circ}.7)$ which is well-determined from the TESS transiting observations. For the  unknown $\Omega_{b}$, we randomly sample from a uniform distribution between 0 and $2\pi$. Figure~\ref{fig:figure9} shows the resulting $\Delta I_{bc}$ distribution is bi-modal and symmetric about $90^{\circ}$ due to the prograde-retrograde degeneracy. The broad distribution can also be attributed to our lack of knowledge about $\Omega_{b}$, as well as the wide distribution of $I_c$. The mutual inclination is $41^{\circ}<\Delta I_{bc}<129^{\circ}$ at $1\sigma$ level and $16^{\circ}<\Delta I_{bc}<164^{\circ}$ at $2\sigma$ level. However, we acknowledge that assuming $\Omega_{b}$ to be uniformly distributed from 0 to $2\pi$ may not accurately reflect reality. For instance, as depicted in Figure~\ref{fig:figure9}, there's a high probability that two orbits are orthogonal to each other ($\Delta I_{bc}$ close to $90{^\circ}$), which may be uncommon in nature.% This discrepancy could be attributed to the uniform assumption regarding $\Omega_{b}$. 

In summary, our analysis reveals that the inner hot Jupiter, HD 118203 b, is nearly aligned with the host star's spin axis. Our measurements are also consistent with the outer giant planet, HD 118203 c, exhibiting a low mutual inclination relative to both planet b and the host star spin axis.

%\section{Discussion}
\section{Formation Scenarios for HD 118203 \MakeLowercase{b}}

\subsection{Disk-driven migration}

Numerous studies in the past decades have explored the eccentricities of hot Jupiters within the context of disk-driven migration, e.g. \citealt{Goldreich2003,Bitsch2013,Duffell2015}. These studies generally agree that interactions between the planets and protoplanetary disks would lead to damping of their eccentricities ( $e<0.1$).
However, recent studies have found that the hot Jupiter could have eccentricity up to 0.4 if they migrate inside low-density cavities in the proto-planetary disks \citep{Debras2021,  Li2023,Romanova2023, Romanova2024MNRAS}. Therefore，we can not rule out the possibility of disk-drive migration for HD 118203 b given the system-wide alignment of the system. But it is an open question why HD 118203 c  remains at an outer orbit of $\sim 6.23$ AU, if  HD 118203 b moved inward through disk-driven migration in the similar condition.

%While the outer giant planet could potentially increase the eccentricity of the hot Jupiter following disk-driven migration, the hot Jupiter would experience apsidal precession due to general relativity and tidal forces at its close-in orbit. This precession could counteract the gravitational perturbations induced by the outer giant planet. \cite{Fabrycky2010} present an approximate criterion for the precession caused by the outer planet to overcome the general-relativistic precession utilizing the masses and orbital periods of both the inner and outer planets:
%\begin{equation}
%    \left(\frac{m_{out}}{M_{J}}\right)\left(\frac{37\ \rm{day}}{P_{out}}\right)^{2} \geq \left(\frac{M_{\odot}}{M_*}\right)^{5/3}\left(\frac{3\ \rm{day}}{P_{in}}\right)^{8/3}
%\end{equation}

%With a mass of $m_{c}=11.79^{+0.69}_{-0.63}\ \mathrm{M_{J}}$ and an orbital period of $\sim 5103$ days, HD 118203 c does not meet the threshold required to excite the eccentricity of HD 118203 b in its current close-in orbit.

\input{fig10}
\subsection{Traditional high-eccentricity tidal migration vs.\ Coplanar high-eccentricity tidal migration}
The hierarchical star-planet-planet configuration in HD 118203 system fits the high-eccentricity tidal migration scenario for explaining the origin of the hot Jupiter. Tidal migration involves two steps: first, diminishing the orbital angular momentum of the proto-hot Jupiter by exciting it into a highly eccentric orbit, and second, tidally dissipating the orbital energy of the planet through the interaction with the host star. We will examine how our study of HD 118203 system sheds light on each of these steps below.

%Furthermore, disk migration alone cannot fully explain the misalignment between two planets, nor the misalignment of the outer planet relative to the star we observed in HD 118203 system. In single systems, disk migration usually remain the alignment between the star and planets \cite{}. \cite{Batygin2013} proposed that primordial misalignment could be a natural consequence of disk migration in binary systems. However, speckle imaging \citep{pepper2020} and RV measurements have rule out the presence of close stellar companions to HD 118203.
 
%\subsection{High-eccentricity Tidal migration}

\subsubsection{Eccentricity excitation}\label{sec:EE}
In the first step, gravitational interactions with a third body extract orbital angular momentum from the proto-hot Jupiter and trigger the excitation of its orbital eccentricity, although the specific mechanism responsible for this process is still unknown. 
 
\textit{Secular perturbation} is a plausible mechanism in which a proto-hot Jupiter swaps angular momentum with a third object gradually, on timescales much longer than the orbital periods. Unlike short-term perturbations causing noticeable changes in an orbit, secular perturbations manifest over a timescale of many orbits, often affecting the orbit's shape and orientation gradually and cumulatively. Von-Zeipel-Kozai-Lidov cycles \citep{1910AN....183..345V, Kozai1962, Lidov1962} are a specific type of secular perturbation that can occur in hierarchical triple systems. Traditionally, the von Zeipel-Kozai-Lidov mechanism describes the evolution of the inner orbit under secular perturbation from a circular outer orbit, which can be analyzed using a perturbing function truncated to quadrupole order in the semi-major axis ratio ($a_1/a_2$). This approximation provides a well-defined analytical solution for the periodic exchange between the orbital eccentricity and inclination of the inner orbit. For von-Zeipel-Kozai-Lidov cycles to occur, there must be an initial misalignment of at least $40^{\circ}$ between the two objects. 

However, our findings that the outer planet has an eccentricity of $e_{c}=0.26$, imply that we need to take into account the effects of the non-circular outer orbit. \cite{Naoz2011} explored a more generalized mechanism in which the outer orbit can be eccentric, called the eccentric Kozai-Lidov (EKL) mechanism. They found that considering an eccentric outer orbit requires an expansion of the perturbing function to octupole order, which is proportional to $(a_1/a_2)^{3}$. In the octupole level of approximation,  EKL mechanism can drive the eccentricity of the inner orbit to extremely high levels, and can even induce a complete reversal in the orientation of the inner orbit, flipping it from prograde to retrograde. The EKL mechanism is also applicable across a broader range of parameters, including scenarios with low initial mutual inclinations between the two objects. \cite{Li2014b} demonstrated that 
when the two planets have eccentric orbits, the inner planet's eccentricity can grow
to large values even when the orbits are
initially close to being coplanar.  \cite{Petrovich2015} investigated the consequence of this eccentricity excitation in the context of tides and found that subsequent high-eccentricity migration kept the two planets nearly coplanar. According to their study, hot Jupiters formed via this coplanar high-eccentricity migration (CHEM) are likely to have distant ( $a \gtrsim 5 \rm{AU}$) and massive (three times more massive than the hot Jupiter) companions with relatively low mutual inclinations ( $\leq 20^\circ$) and moderate eccentricities (0.2-0.5). Our measurements for HD 118203 c, with properties of $a_c = 6.28^{+0.10}_{-0.11}\ \rm{AU}$, $m_c = 11.79^{+0.69}_{-0.63}\ M_{J}$, and $e_c = 0.26^{+0.03}_{-0.02}$, align well with these predictions. Furthermore, our analysis indicates that $\approx 95\%$  values from the posterior distribution of the line-of-sight  $\Delta I_{bc}$  are below $25^\circ$, consistent with the outer giant planet having a low mutual inclination relative to the inner hot Jupiter. 

In reality, there are additional effects beyond secular perturbations from the outer companion that can influence the evolution of inner orbit. For example, general relativity (GR) effect from the host star could induce apsidal precession of the inner orbit to the opposite direction than that induced by the outer giant planet. If GR precession takes place at a faster rate, either quadrupole-level von Zeipel-Kozai-Lidov cycles or octupole-level EKL would be suppressed. Therefore, we need to compare the apsidal precession rate induced by GR effect and the precession rate induced by the outer giant planet. Here, we used the well-determined analytic solution for the quadrupole-level von Zeipel-Kozai-Lidov cycles \citep{Naoz2013} to provide a rough estimate of the relevant time-scales. 

\begin{equation}
    \tau_{\rm{quad}}= P_{b}\frac{m_{b}+M_{\star}}{m_{c}}(\frac{a_{c}}{a_{b}})^{3}(1-e_{c})^{3/2}
\end{equation}

In the other hand, the precession rate due to GR is given by \cite{Fabry2007}, 
\begin{equation}\label{eq:t_gr}
     \tau_{\rm{GR}}  = \frac{a_b c^2 (1-e_b^2)}{3 G (M_\star+m_b)} P_{b} 
\end{equation}
 where $c$ is the speed of light, and $G$ is the gravitational constant. In the top panel of Figure~\ref{fig:figure10}, HD~118203 c is located to the right of the orange line, indicating the general relativistic (GR) precession at the current close-in orbit of HD 118203 b is significantly faster than the precession induced by the distant secular perturber. Therefore, the hot Jupiter has decoupled from the outer perturber as tides shrank its semi-major axis to current close-in orbit. However, the companions to the left of blue lines are massive and nearby enough to counter the GR effect at the initial orbital distances of the proto-hot Jupiters. The exact boundary depends on the initial orbital distances of the proto-hot Jupiter. Our results suggest that HD 118203 c is able of triggering the high-eccentricity migration as long as the proto-hot Jupiter starts from a position beyond 0.3 AU.

%We obtained a timescale $\tau_{\rm{KL}}$ as $4.2\times 10^{6}$ yr.
 %We obtained $\tau_{\rm{GR}}$ for HD 118203 b as $2.8\times 10^{4}$ yrs. 
 %The timescale of apsidal precession caused by tidal torques on the inner planet from the central star can be estimated as \citep{Pu2019}
%\begin{equation}\label{eq:t_tides}
% \tau_{\rm{tides}}  =\frac{2}{15k_{2}}(\frac{m_{b}}{M_{\star}})^{5}(\frac{a_b}{R_b})^{5}P_b
%\end{equation}
%where $k_2$ is the planet love number for which we adopted $k_2=0.565$ measured from Jupiter \citep{Durante2020}. The tidal timescale is $\tau_{\rm{tides}}=2.4\times 10^{5}$ yrs. Therefore, we find $\tau_{\rm{GR}}<\tau_{\rm{tides}}<\tau_{\rm{KL}}$, which implies that GR precession and tidal precession are able to counteract and suppress von-Zeipel-Kozai-Lidov oscillations at the present close-in orbit of the hot Jupiter HD 118203 b.
%Furthermore, since the mass of HD 118203 b ($m_{b}=2.3M_{\mathrm{J}}$) is comparable to that of the outer planet ($m_{b}=8.3 M_{\mathrm{J}}$), the inner orbit might also exert a torque on the outer orbit during EKL cycles. However, future N-body simulations are necessary to determine if this interaction can lead to the observed misalignment between the orbit of the outer planet and the stellar spin axis ($ \Psi_{c} >32^{\circ}.6^{+18}_{-14}$ at $1\sigma$ level.

\textit{Secular chaos} can also excite eccentricities of innermost planets in multi-planet systems with  two or more eccentric and inclined giant planets \citep{Wu2011, Teyssandier2019}. Due to the planet-planet interaction, the innermost planet may become very eccentric and/or inclined compared to other planets over time. When the inner planets reach a sufficiently large eccentricity, it would undergo the tidal migration.   Theoretical studies suggest that such scenarios could account for the clustering of hot Jupiter orbital periods around 3 days and the observed spin-orbit misalignment \citep{Wu2011, Teyssandier2019}.

\textit{Planet-planet scattering} is another mechanism to create high eccentricity. In systems where multiple giant planets are tightly packed, their gravitational interactions can destabilize their orbits, causing them to cross paths and have close encounters \citep{Rasio1996,Chatterjee2008,Beaug2012, Petrovich2014}. These encounters can lead to a range of outcomes, including the ejection of one or more planets from the system, collisions between planets or with the central star, and significant alterations in the eccentricity and inclination of planetary orbits. The last scenario could produce planets with highly eccentric orbits that can subsequently undergo tidal orbital circularization. Numerical simulations have found that planet-planet scattering would also alter the orientation of planet orbits and increase their orbital inclination \citep{Chatterjee2011,Li2021}. Possibly, the HD 118203 system initially contained more than two giant planets. A series of close encounters among these planets may have led to the ejection of at least one planet, and the eccentricity excitation of HD 118203 b followed by tidal migration. However, it is highly unlikely that HD 118203 b reached its close-in orbit solely through planet-planet scattering, as the system's total orbital energy is conserved. For instance, a Jupiter-mass planet would need to eject 100 other planets of similar mass to shrink its semi-major axis by a factor of 100 \citep{Dawson2018}.
%It is highly unlikely that HD 118203 b can arrive it close-in orbit solely via planet-planet scattering, as this process would require a giant planet to eject 100 other planets of its own mass to reduce its semi-major axis by a factor of 100 \citep{Dawson2018}. 

\input{fig7}

\subsubsection{Tidal dissipation and realignment }

Once the proto-hot Jupiter's orbit becomes sufficiently eccentric, tidal dissipation in the planet initiates its inward migration. Specifically, during each close approach to the host star, tides raised on the planet by the star convert a small fraction of the planet's orbital energy into heat, resulting in the gradual shrinking and circularization of the planet's orbit. During the tidal dissipation process, the planet's angular momentum remains constant. Therefore, we can reverse the trajectory of the hot Jupiter's semi-major axis $a(t)$ and eccentricity $e(t)$ over time as $a_{\rm{final}} = a(t)[1-e(t)^2] = \rm{constant}$, where $ a_{\rm{final}} $ is the final semi-major axis of the hot Jupiter when it's fully circularized \citep{Dawson2014}. The bottom panel of Figure~\ref{fig:figure10} shows that HD~118203 b may have experienced high-eccentricity tidal migration along constant angular momentum tracks. Given its current semi-major axis of $a = 0.07$ AU and eccentricity of $e = 0.32 $, assuming an initial eccentricity of 0.99 would imply an upper limit for the initial semi-major axis of 3.2 AU. In Section~\ref{sec:EE}, we have shown that the initial semi-major axis needs to exceed 0.3 AU for the outer companion to overcome GR effects. Therefore, the initial position of the proto-hot Jupiter could range from 0.3 to 3.2 AU. 

Upon reaching its close-in orbit, the hot Jupiter might have been misaligned relative to the host star due to tilting throughout the tidal migration. However, previous studies have found that hot Jupiters orbiting cool stars with temperatures below the Kraft break (6250 K, \citealt{kraft1967}) tend to be aligned with their host stars, while hot stars show a range of spin-orbi angles \cite{Winn2010, Schlaufman2010}. To interpret this, \cite{Winn2010} proposed that cool stars realign with the hot Jupiters' orbits due to tidal dissipation in their convective zones while hot stars cannot realign because of their thinner convective zones. \cite{lai2012} suggested that in the presence of a convective envelope,  the dissipation of inertial waves would cause efficient spin–orbit alignment for cool star systems. Recently, \cite{Zanazzi2024} suggested that the gravity mode pulsations in the radiative interior of cool stars can lock the hot Jupiters' orbits and  efficiently reduce the system’s spin-orbit misalignment. Our analysis reveals that HD 118203 b has a low spin-orbit angle of $<33^{\circ}.5\ (95\%)$, indicating the hot Jupiter orbits its host star in a nearly aligned orbit. This alignment is consistent with the above trend as HD 118203 is a cool star ($T_{\rm{eff}}=5742^{+93}_{-95}\ K$). The stellar evolutionary models for HD 118203 show that it is at the sub-giant evolutionary stage and its effective temperature has remained below the Kraft Break (6250K) throughout its entire life (see Appendix~\ref{fig:figureA8}). It is possible that the orbit of HD 110233 b has undergone realignment process. As a sub-giant system, our result is also consistent with those of \cite{Saunders2024}, who report on the aligned orbits of three hot Jupiters orbiting sub-giant stars and suggest an increased efficiency in reducing the spin-orbit angle  after a star leaves the main sequence. 

On the other hand, the orbital eccentricity of HD 118203 b is currently 0.32. If we assume that the hot Jupiter was misaligned with its host star upon arriving the close-in orbit, the damping of spin-orbit angle  would occur more rapidly than the damping of eccentricity. Based on the analytical model from \cite{lai2012}, the timescale of spin-orbit angle damping depends on tidal quality factor of the star $Q_*$, which ranges over multiple orders of magnitude for sub-giants. With an age of $4.73\pm{0.39}\  \mathrm{Gyr}$ determined by our isochrone model in Section~\ref{sec:starpa}, we assume an upper limit for the timescale of spin-orbit angle damping ($\tau_{\lambda}<4.73\pm{0.39} \  \mathrm{Gyr}$) and constrain the tidal quality factor of the star $Q_* < (2.5\pm{0.2})
 \times 10^{5}$. On the other hand, we constrain a lower limit for the timescale of eccentricity damping ($\tau_{e}>4.73\pm{0.39} \  \mathrm{Gyr}$). If we only consider the eccentricity damping driven by tidal dissipation within the planet, the timescale $\tau_{e}$ could be derived with  the planet's effective tidal quality factor $Q_{p}$, planet's love number $k_{2}$, star mass $M_{*}$ and the orbital parameters of the hot Jupiter\citep{Murray1999,Rice2022}. If we adopted a $k_{2}=0.565$ and used the known stellar and planetary parameters, we can obtain a constraint on the planet's tidal quality factor $Q_{p}>(1.135\pm{0.095}) \times 10^{5}$.

Alternatively, the coplanar high-eccentricity migration provides a plausible explanation for the aligned yet eccentric orbits of HD 118203 b. A hot Jupiter undergoing coplanar high-eccentricity migration would maintain a low spin-orbit angle throughout the entire process. Recent observations of warm Jupiters that have high eccentricities and aligned orbits support this scenario \citep{ER2023,Hu2024,Rubenzahl2024}.

\subsection{Comparison to other systems}
Figure~\ref{fig:figure7} compares the dynamical architecture of HD 118203 b to other exoplanet systems with hot Jupiters ($P<10 \ \rm{days}$). We can see that HD 118203 b exhibits an eccentric but aligned orbit, a trait shared by many other hot Jupiters. This raises an intriguing question: Why are the orbits of these planets aligned with their host star yet not fully circularized? One possibility is that they experience spin-orbit angle damping more rapidly than eccentricity damping. Alternatively, these planets might have a low spin-orbit angle upon reaching their current close-in orbit, potentially resulting from a formation pathway such as coplanar high-eccentricity migration. Future Gaia Data Release 4 is expected to offer high-precision epoch astrometry, which will improve our ability to constrain the mutual inclination between hot Jupiters and their distant giant companions and thereby investigate their formation \citep{EspinozaRetamal2023}. 

Figure~\ref{fig:figure7} reveals that most hot Jupiters with eccentricity, either with low or large spin-orbit angle, usually orbit around cool stars. However, 
the sample only includes six systems with eccentric hot Jupiters orbiting host stars hotter than 6250 K, making it difficult  to draw any statistical conclusions. The lack of eccentric hot Jupiters around hot stars  could be a result of detection bias because hot stars typically rotate rapidly, making it challenging to obtain reliable RVs. However, if the trend is true, that might suggest that the hot stars would facilitate the circularization of the hot Jupiter orbits. 

Although dozens of hot Jupiters have been discovered with outer giant planets, direct measurements of mutual inclination between these planets are still scarce. Here, we compare HD 118203 with other systems with constrained mutual inclinations between inner close-in planets and outer giant planets.

\begin{itemize}

\item $\pi$ men c is a transiting super Earth ($\rm{r_p\sim 2\ R_{\oplus}}$,  $\rm{m_{p}\sim 4.5\ M_{\oplus}}$) with an orbital period of 6.26 days \citep{Huang2018,Gandolfi2018} accompanied by an outer giant planet $\pi$ men b at $\sim 3.2$ AU \citep{Jones2002, Butler2006}. While $\pi$ men c doesn't fall into the category of hot Jupiters, it has been found to be significantly misaligned relative to the outer planet \citep{xw2020, DeRosa2020, Damasso2020} and the system serves as a comparison to HD 118203. Furthermore, based on RM observations with ESPRESSO, \cite{Kunovac2021} measured that $\pi$ men c is moderately misaligned relative to the host star' spin axis ($\lambda = -24^{\circ}\pm{4.1}$). The N-body simulations show that the inclined outer giant planet may titled the orbits of the inner planet and cause the observed spin-orbit misalignment \cite{xw2020}.

\item HAT-P 11 b is a transiting planet with a mass of $0.09\ M_J$ and an orbital period of 4.88 days.  Unlike the aligned orbit of HD 118203 b, HAT-P-11 b has a large sky-projected spin-orbit angle ($\sim 100^{\circ}$, \citealt{Winn2010b,Hirano2011}) although its host star has an effective temperature below Kraft break. \cite{Yee2018} discovered a outer giant planet at $\sim 4 $ AU and with a eccentricity of 0.6. \cite{xw2020, An2024} combined RVs and Hipparcos-Gaia astrometric acceleration to determine the mutual inclination between HAT-P-11 b and the outer giant planet to be at least $50^{\circ}$. This is consistent with previous studies that the nodal precession driven by an inclined outer companions could cause the spin-orbit misalignment of inner planets \citep{huber2013,Pu2019,Zhang2021}. The N-body simulation shows that HAT-P-11 b has undergone nodal precession driven by the outer giant planet HAT-P-11 c, which surpassed the tidal realignment forces exerted by the star \citep{Yee2018,xw2020,Lu2024}. To interpret the difference between HD 118203 with HAT-P 11, we have estimated the nodal precession rate of HD 118203 b driven by HD 118203 c using the analytical model from \cite{Lai2018}. It shows that  HD 118203 c is not massive or nearby enough to overcome the tidal realignment forces exerted by the star, potentially explaining why we see HD 118203 b aligns with the host star, but HAT-P-11 b does not.
    
\item  Kepler-419 b is eccentric warm Jupiter with a mass of $2.5\ M_J$ and eccentricity of 0.83 at  0.37 AU. \cite{Dawson2014} used RVs and TTV observations to discover a non-transiting Kepler-419 c at 1.68 AU wth a mass of $7.3\ M_J$. They also constrained that mutual inclination between the two planets as low as $\sim 9^{\circ}$ from TTV observations. The system serves as an example of undergoing co-planar high eccentricity migration. 

\item  WASP-148 b is a hot Jupiter ($\rm{P=8.8}$ days, $\rm{M=0.29\ M_{J}}$) with a nearby warm Jupiter companion ( $\rm{P=34.5}$ days, $\rm{M=0.4\ M_{J}}$ \citealt{Hernard2020}). The TTV observation show that the mutual inclination between the two planets to be below $35^{\circ}$. \cite{Wang2022} found that WASP-148 b is aligned with the host stars with a sky-projected spin-orbit angle of $\sim8^{\circ}$.  

\item  HAT-P 2 b\citep{Bakos2007} is an eccentric massive hot Jupiter ($\rm{P=5.63}$ days, $M=9.04\pm{0.50}\ M_{J}$, $e=0.52\pm{0.01}$) with a low sky-projected skin-orbit angle ($\lambda \sim 9^{\circ}$, \citealt{Winn2007, Albrecht2012}). \cite{deBeurs2023} discovered a long-period giant planet HAT-P 2 c and measured its orbital inclination of $90^{\circ}\pm{16}$. The results suggest that HAT-P 2 c may also have a low mutual inclination relative to the transiting hot Jupiter HAT-P 2 b. HAT-P 2 system exhibits a similar architecture as HD 118203.

\end{itemize}

\section{Conclusions}

We measured the true spin-orbit angle of a hot Jupiter HD 118203 b and presented the discovery of a long-period giant
planet HD 118203 c in the same system. HD118203 is one of first systems for which the true spin-orbit angle of the host star relative to the hot Jupiter, and the mutual inclination between the hot Jupiter and an outer giant planet, have both been determined.
Our main conclusions are as follows:

\begin{itemize}
    \item HD 118203 is a sub-giant ($M_{*}=1.27^{+0.028}_{-0.032}\ M_{\odot}$, $R_{*}=2.04\pm{0.03}\ R_{\odot}$, $T_{\rm{eff}}=5742^{+93}_{-95}\ K$) hosting a hot Jupiter HD 118203 b with a moderate eccentricity ($P_{b}=6.135$ days, $e_{b}=0.31\pm{0.007}$, $m_{b}=2.14\pm{0.12}\ M_{\rm{J}}$, $r_{b}=1.14\pm{0.029}\ R_{\rm{J}}$). We detected stellar oscillations using KPF RVs and measured the frequency at maximum power to be $\mu_{\rm{max}}=967.2^{+51.7}_{-46.5}\ \mu Hz$, which confirms that HD 118203 is a sub-giant. 
    
    \item We measured the sky-projected spin-orbit angle of HD 118203 b to be $\lambda_{b}=-11.7^{+7.6}_{-10.0}$ deg through RM observations using KPF. We measured the stellar rotational period as $23.1\pm{3.8}$ days using TESS light curves and constrained the stellar inclination to be $i_{*}=89^{\circ}.9^{+13.7}_{-13.8}$. We obtained the true spin-orbit angle of HD 118203 b as $\Psi_{b}<33^{\circ}.5\ (2\sigma) $. Our results reveal that HD 118203 is nearly aligned with the stellar spin axis.  
    
    \item HD 118203 c is a long-period giant planet ($m_{c}=11.79^{+0.69}_{-0.63}\ \mathrm{M_{J}}$, $a_{c}=6.28^{+0.10}_{-0.11}$ AU, $e_{c}=0.26^{+0.03}_{-0.02}$) with moderate eccentricity outside the hot Jupiter. We constrained the orbital inclination of HD 118203 c ($I_{c}=93^{\circ}.7^{+11.9}_{-13.8}$ ) by combing RV and Hipparcos-Gaia astrometric accelerations.  The line-of-sight mutual inclination between HD 118203 b $\&$ c is $9^{\circ}.8^{+8.6}_{-6.7}$ at $1\sigma$ and 
    $9^{\circ}.8^{+16.2}_{-9.3}$ at $2\sigma$
    confidence level. The line-of-sight spin-orbit angle for HD 118203 c is  $14^{\circ}.7^{+13.3}_{-10.2}$ at the $1\sigma$ confidence level and $14^{\circ}.7^{+26.4}_{-13.9}$ at the $2\sigma$ level confidence level. Our measurements demonstrate that the orbit of HD 118203 c is most likely aligned with both the host star spin axis and the orbit of the inner hot Jupiter. 
    
    %$41^{\circ}<\Delta I<139^{\circ}$ at $1\sigma$ confidence level, and $16^{\circ}<\Delta I<164^{\circ}$ at $2\sigma$ confidence level. The stellar obliquity for HD 118203 c is  $43^{\circ}<\Psi_c <137^{\circ}$ at $1\sigma$ confidence level, and $20^{\circ}<\Psi_c<159^{\circ}$ at $2\sigma$ confidence level.$16^{\circ}.7^{+14.6}_{-10.9}$$17^{\circ}.8^{+18.6}_{-12.4}$

    \item The architecture of the HD 118203 system (star, hot Jupiter, distant giant planet) is compatible with the high-eccentricity tidal migration scenario as a possible explanation for the origin of the hot Jupiter. The observed low mutual inclination between HD 118203 b and c, along with the moderate eccentricity of HD 118203 c, suggests that the system underwent eccentric Kozai-Lidov cycles (EKL) or another coplanar high eccentricity migration scenario.
    %von Zeipel-Kozai-Lidov cycles or planet-planet scattering events in the past.
    \item The low spin-orbit angle of HD 118203 b indicates that either the hot Jupiter has always been aligned with the host star, for example through co-planar high-eccentricity migration, or that tidal dissipation in the host star has realigned the stellar spin and hot Jupiter's orbit. If tidal realignment did occur, the moderate eccentricity of HD 118203 b suggests that spin-orbit angle damping was more rapid than eccentricity damping. 

    \item Based on the constant angular momentum evolution during high-eccentricity migration and the requirement of companion secular perturbation to counteract general relativity effects from the host star, we constrain the initial position of the proto-hot Jupiter to be between 0.3 and 3.2 AU.
    
    %\item The epoch astrometry that is expected to be available in Gaia DR4 will help to better constrain the orbital solution of HD 118203 c, including the orbital inclination and mutual inclination between the two planets. This data will also facilitate the characterization of the three-dimensional orbital architecture of more systems similar to HD 118203, paving the way for statistical studies to understand the origin channels of hot Jupiters. 
\end{itemize}

\acknowledgments
The authors wish to recognize and acknowledge the very significant cultural role and reverence that the summit of Mauna Kea has always had within the indigenous Hawaiian community. We are most fortunate to have the opportunity to conduct observations from this mountain.

J.Z. would like to thank Maximilian N. G$\ddot{u}$nther for  helpful suggestion and advice regarding the use of package \texttt{allesfitter}. J.Z. and D.H. acknowledge support from the Alfred P. Sloan Foundation, the National Aeronautics and Space Administration (80NSSC22K0303).  D.H. also acknowledges support from the Alfred P. Sloan Foundation, and the Australian Research Council (FT200100871).  J.Z. and L.M.W. acknowledge support from NASA-Keck Key Stragetic Mission Support Grant No. 80NSSC19K1475.  L.M.W. also acknowledges support from the NASA Exoplanet Research Program through grant 80NSSC23K0269.  This research was carried out, in part, at the Jet Propulsion Laboratory and the California Institute of Technology under a contract with the National Aeronautics and Space Administration and funded through the President's and Director's Research \& Development Fund Program. This research has made use of the NASA Exoplanet Archive and ExoFOP, which are operated by the California Institute of Technology, under contract with the National Aeronautics and Space Administration under the Exoplanet Exploration Program.

\vspace{5mm}
\facilities{Keck(HIRES), Keck(KPF), ELODIE, TESS, Gaia, Hipparcos}


\software{\texttt{allesfitter} \citep{allesfitter-code,allesfitter-paper} \texttt{RadVel} \citep{Fulton2018} \texttt{TESS-SIP}\citep{Hedges2020}}


\newpage
\counterwithin{figure}{section}

\appendix

\section{Supplementary figures}
\input{figA1}
\input{figA2}

\newpage

\input{figA3}

\newpage

\newpage

\input{figA5}

\newpage

\input{figA8}

%% dr
%/sample63
\newpage
\bibliography{sample63}{}
\bibliographystyle{aasjournal}


\end{CJK*}
\end{document}

%% file: affiliations.tex
\newcommand{\Macquarie}{School of Mathematical and Physical Sciences, Macquarie University, Balaclava Road, North Ryde, NSW 2109, Australia}

\newcommand{\JPL}{Jet Propulsion Laboratory, California Institute of Technology, 4800 Oak Grove Drive, Pasadena, CA 91109, USA}

\newcommand{\Princeton}{Department of Astrophysical Sciences, Princeton University, 4 Ivy Lane, Princeton, NJ 08540, USA}

\newcommand{\UCO}{UC Observatories, University of California, Santa Cruz, CA 95064, USA}

\newcommand{\WMKO}{W.\ M.\ Keck Observatory, 65-1120 Mamalahoa Hwy, Kamuela, HI 96743, USA}
\newcommand{\SSL}{Space Sciences Laboratory, University of California, Berkeley, CA 94720, USA}
\newcommand{\UH}{Institute for Astronomy, University of Hawai‘i, 2680 Woodlawn Drive, Honolulu, HI 96822, USA}
\newcommand{\UCB}{Department of Astronomy, 501 Campbell Hall, University of California, Berkeley, CA 94720, USA}
\newcommand{\UCLA}{Department of Physics \& Astronomy, University of California Los Angeles, Los Angeles, CA 90095, USA}

\newcommand{\COO}{Caltech Optical Observatories, California Institute of Technology, Pasadena, CA 91125, USA}

\newcommand{\Kansas}{Department of Physics and Astronomy, University of Kansas, Lawrence, KS, USA}

\newcommand{\Schmidt}{Astrophysics \& Space Institute, Schmidt Sciences, New York, NY 10011, USA}

%% file: table_rv.tex
\begin{deluxetable}{cccc}

\tablecaption{HD 118203 RVs}\label{tab:rvs}
\tablehead{\colhead{Time} & \colhead{RV} & \colhead{$\sigma_{\mathrm{RV}}$} & \colhead{Inst} \\ 
\colhead{(BJD - 2450000)} & \colhead{(m/s)} & \colhead{(m/s)} & \colhead{} } 
\startdata
3151.4302000002936 & -29262.81 & 13.40 & ELODIE\\
3153.423499999568 & -29230.81 & 14.47 & ELODIE\\
3155.3946000002325 & -29639.81 & 10.79	 & ELODIE\\
... & ... & ... &  ... \\
5370.815355828032 & -126.58 & 0.86 & HIRES\\
5371.83661200013 & 44.42 & 1.24 & HIRES\\
8885.015345999971 & -327.16 & 1.58 & HIRES\\
8906.171260000207 & 38.02 & 1.41 & HIRES\\
... & ... & ... &  ... \\
10043.80571338 & 42.93 & 0.96 & KPF \\
10043.808489768 & 43.79 & 0.96 & KPF \\
10043.811272037 & 47.59 & 1.05 & KPF \\
... & ... & ... &  ... \\
\enddata
\tablecomments{Times are in BJD - 2450000.0. The RV uncertainties do not include RV jitter. ELODIE RVs are from \cite{pepper2020}, HIRES and KPF RVs are from this work. All RV data utilized in this paper, including those sourced from the literature, are available in a machine-readable format. }

\end{deluxetable}

%% file: table_hgca.tex
\begin{table}
\setlength{\tabcolsep}{4pt}
\centering
\caption{Hipparcos-Gaia astrometric acceleration in declination and right ascension for HD 118203 from \cite{Brandt2021}. The $\Delta \mu_{\delta *}$ components have the $\cos{\delta}$ factor included. $\sigma[\Delta\mu]$ represent the uncertainties. We assume that the uncertainties on the proper motions and $\mu_{\rm{HG}}$ are independent and add them in quadrature to calculate these uncertainties (see \citealt{zhang2023} for details).}\label{tab:HGCA_obs}
\begin{tabular}{cccccc}
\hline
\hline
 Data & $\Delta\mu_{\alpha}$ & $\sigma[\Delta\mu_{\alpha}]$ & $\Delta\mu_{\delta *}$ & $\sigma[\Delta\mu_{\delta *}]$ & S/N\\
  epoch & \multicolumn{2}{c}{$\mathrm{mas}\ \mathrm{yr}^{-1}$} & \multicolumn{2}{c}{$\mathrm{mas}\ \mathrm{yr}^{-1}$} & \\
\hline
 {\it Gaia} & -0.1038& 0.0377 &0.0905 &0.0462 & 3.47 \\
 {\it Hipparcos} & 0.842 & 0.663 & -1.628 &  0.884 & 2.71 \\
\hline
\end{tabular}
\end{table}

%% file: table_stellar_params.tex
\begin{deluxetable}{lccr}
\tablecaption{Stellar Parameters of HD 118203}
\label{tab:star-compare}
\tablehead{\colhead{Parameters }&\colhead{SpecMatch} &\colhead{Isochrone} &\colhead{Reference}  \\
\colhead{ } & \colhead{Synth} & \colhead{Fitting} & \colhead{ Value}  }
\startdata
$T_{\rm eff}\ (\rm{K})$ &  $5802\pm{100}$ & $5742^{+93}_{-95}$    & $5683^{+84}_{-85}$ (a)   \\
$M_{\star}$ ($M_{\odot}$)& $1.32\pm{0.09}$ & $1.27^{+0.028}_{-0.032}$ & $1.25\pm{0.06}$ (a)   \\
$R_{\star}$ ($R_{\odot}$)& $2.08\pm{0.05}$ & $2.04\pm{0.03}$ & $2.10\pm{0.06}$ (a) \\
$\rho_{\star}$ (cgs)& $0.206\pm{0.02}$ & $0.21\pm{0.01}$ & $0.19\pm{0.01}$ (a)  \\
log \textit{g} (cgs)& $3.93\pm{0.1}$ & $3.92\pm{0.02}$ & $3.89\pm{0.02}$(a)   \\
$\rm [Fe/H]$ (dex)&  $0.27\pm{0.06}$  & $0.287^{+0.069}_{-0.111}$ &$0.223\pm0.076$ (a)    \\
$V_{mag}$&  --  & -- &$8.135\pm0.03$ (a)    \\
Age (Gyr)& -- & $4.73\pm{0.39}$ & $5.12^{+0.64}_{-0.61}$   (a)   \\
$v\sin i_*$ ($km\ s^{-1}$)&    $5.13\pm{1}$ &-- &  $5.32\pm{0.5}$   (b)  \\
$\varpi$ (mas)&-- & -- & $10.86\pm{0.018}$   (c)    \\
$\nu_{\rm{max}}$ ($\rm{\mu Hz}$)& --  & $967.2_{-51.7}^{+46.5}$   &  --    \\
$\mathrm{P_{rot}}$ (days)$^{*}$ & --  & $23.1\pm{3.8}$ & $20\pm{5}$ (a)     \\
$i_*$ (deg)$^{*}$&  -- & $89.9_{-13.8}^{+13.7}$  &  --     \\
\enddata
\tablecomments{ (a). \citet{pepper2020}; (b). \citet{Luck2017}; The reported velocity ($7\ \mathrm{km\ s^{-1}}$) in \citet{Luck2017} represents a combination of rotation and macroturbulence velocities, despite being referred to as Vr. After deducting the macroturbulence velocities, the $v\sin i_{*}$ stands at $5.32\ \mathrm{km\ s^{-1}}$. (c). \citet{Gaia}. * $\mathrm{P_{rot}}$ is measured from TESS light curves described in Section~\ref{sec:istar}. $i_{*}$ is constrained with spectroscopic $v\sin i_*$ and stellar rotation period described in Section~\ref{sec:istar}. } 
\end{deluxetable}

%% file: fig1.tex
\begin{figure}
    \centering
    \includegraphics[width=\linewidth]{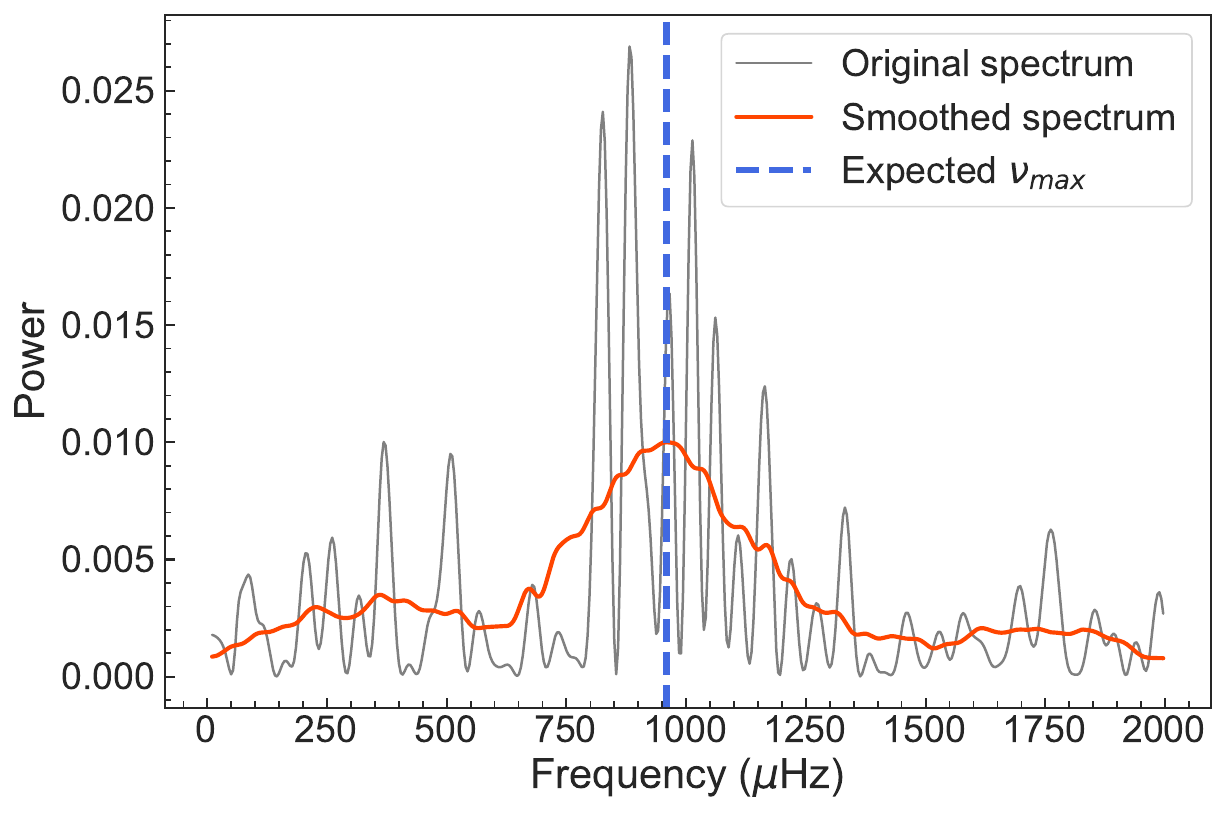}
    \caption{Lomb-Scargle periodograms of the KPF RV residuals from  RM fitting without oscillation. The solid grey and orange lines represent the original and smoothed periodograms (filter width=$340 \rm{\ \mu Hz}$). The blue dashed line indicates the predicted frequency at maximum power, $\nu_{\rm{max}}$, derived using scaling relations $\frac{g}{g_{\odot}}=\frac{\nu_{\rm{max}}}{\nu_{\odot}}\frac{T_{\rm{eff}}}{T_{\rm{eff,\odot}}}$. The values for the star's surface gravity $g$ and effective temperature $T_{\rm{eff}}$ were taken from \cite{pepper2020}. }
    \label{fig:figure1}
\end{figure}

%% file: fig2.tex
\begin{figure*}
    \centering
    \includegraphics[width=0.8\linewidth]{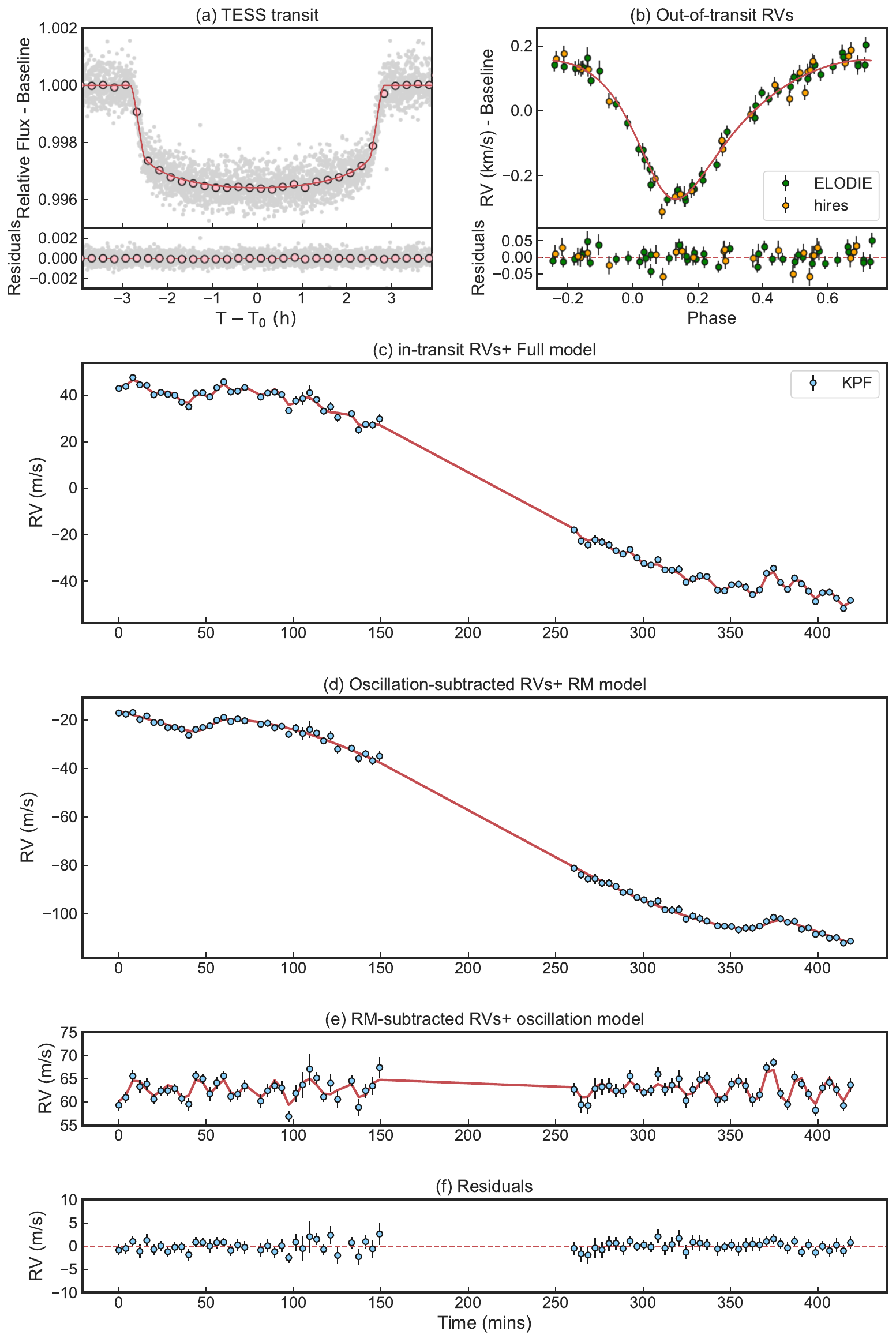}
    \caption{Panel a: TESS SPOC light curves phase-folded to the best-fit orbital period of HD 118203 b. The pink points present the TESS data binned every 15 minutes, and the red line shows the best-fit transit model. Panel b: The out-of-transit RVs from ELODIE (green) and HIRES (orange) phase-folded to the best-fit orbital period of HD 118203 b.  The red line depicts the best-fit RV model. The linear baseline has been removed. Panel c: the in-transit RVs from KPF in blue and the combined RM and GP model in red. Panel d: KPF data and RM model after removing the contribution from  oscillations. Panel e: KPF data and SHO GP model after removing RM effect.  Panel f: the residuals of the fit.}
    \label{fig:figure2}
\end{figure*}

%% file: table_modelcompare1.tex
\newpage
\begin{scriptsize}
\begin{longtable*}[!t]{p{6.8cm}cccc}
\caption{Priors and posterior distributions of RM modelling for HD 118203 b}\label{tab:rm_result}\\
\hline
 & \multicolumn{2}{c}{RM without oscillation}  &\multicolumn{2}{c}{RM+ oscillation}\\
   Parameter (Unit) &  Prior&  Value   & Prior &  Value \\
\hline
\hline
\endfirsthead
\hline
\endhead
\hline
\endfoot
\multicolumn{5}{l}{\textbf{Planetary Parameters}}\\[3pt]
 Planet-to-star radius ratio $r_b / R_\star$ &     $\mathcal{U}(0.01,0.1)$   &  $0.05568_{-0.00021}^{+0.00024}$  & $\mathcal{U}(0.01,0.1)$ & $0.05573_{-0.00021}^{+0.00025}$ \\
 Sum of radii over semi-major axis $(R_\star + r_b) / a_b$ & $\mathcal{U}(0.1,0.3)$   &   $0.1443_{-0.0016}^{+0.0024}$    & $\mathcal{U}(0.1, 0.3)$ & $0.1445_{-0.0016}^{+0.0025}$ \\
 Cosine of the orbital inclination $\cos{I_b}$& $\mathcal{U}(0,0.2)$   &  $0.022_{-0.014}^{+0.018}$    & $\mathcal{U}(0,0.2)$ &$0.027_{-0.013}^{+0.016}$ \\
 Orbital period $P_b$ (days)  & $\mathcal{U}(6.12 ,6.15)$   & $6.1349787\pm0.000002$    & $\mathcal{U}(6.12, 6.15)$ &$6.1349779\pm0.000002$\\
  Transit epoch - 2459000 $T_{0;b}$ (BJD) & $\mathcal{U}(1711, 1714)$   &  $-262.80110\pm0.0008^{*}$    & $\mathcal{U}(1711, 1714)$ &$-268.93581\pm0.0008^{*}$\\
  RV semi-amplitude $K_b$ ($\mathrm{km\ s^{-1}}$) & $\mathcal{U}(0.1, 0.3)$   & $0.2165\pm0.0031$    & $\mathcal{U}(0.1, 0.3)$ & $0.2169\pm0.0031$\\   
  Eccentricity term $\sqrt{e_b} \cos{\omega_b}$& $\mathcal{U}(-0.7,0.7)$ & $-0.508\pm0.012$    & $\mathcal{U}(-0.7,0.7)$ & $-0.510\pm0.012$\\
  Eccentricity term $\sqrt{e_b} \sin{\omega_b}$& $\mathcal{U}(-0.7,0.7)$ &$0.235\pm0.016$   & $\mathcal{U}(-0.7,0.7)$ &$0.232\pm0.016$\\
  \hline
  \multicolumn{5}{l}{\textbf{Rossiter-McLaughlin Parameters}}\\[3pt]
  Sky-projected \textbf{spin-orbit angle} $\lambda_{b}$ (deg)  & $\mathcal{U}(-180,180)$& $-11_{-13}^{+11}$   & $\mathcal{U}(-180,180)$ &$-11.7_{-10.}^{+7.6}$\\
 Projected rotational velocity $v\sin{i_*} $ ($\mathrm{km\ s^{-1}}$)& $\mathcal{U}(1,10)$&  $5.85_{-0.32}^{+0.41}$  & $\mathcal{U}(1,10)$ &$5.89_{-0.26}^{+0.31}$\\
  GP SHO parameter $lnS_{0}\ (\mathrm{km^{2}s^{-2}})$& --& --   & $\mathcal{U}(-35,15)$ &$-20.47_{-0.78}^{+0.54}$\\
  GP SHO parameter $lnQ_{0}$ & --& --  & $\mathcal{U}(1,4)$ &$1.70_{-0.46}^{+0.70}$\\
  GP SHO parameter $ln\omega_{\rm{max}}\ (\mathrm{d^{-1}})$& --& --   & $\mathcal{U}(5,7)$ & $6.266\pm0.055$\\
  \hline
  \multicolumn{5}{l}{\textbf{Other Parameters}}\\[3pt]
  Transformed limb darkening $q_\mathrm{1; TESS}$& $\mathcal{U}(0,1)$ & $0.248_{-0.040}^{+0.043}$     & $\mathcal{U}(0,1)$ &$0.250\pm0.038$\\
  Transformed limb darkening $q_\mathrm{2; TESS}$& $\mathcal{U}(0,1)$ & $0.341_{-0.069}^{+0.076}$    & $\mathcal{U}(0,1)$ &$0.333_{-0.061}^{+0.071}$\\
  Transformed limb darkening $q_\mathrm{1; KPF}$& $\mathcal{U}(0,1)$ & $0.69_{-0.11}^{+0.14}$    & $\mathcal{U}(0,1)$ & $0.658_{-0.079}^{+0.11}$ \\
  Transformed limb darkening $q_\mathrm{2; KPF}$& $\mathcal{U}(0,1)$ & $0.81_{-0.18}^{+0.13}$    & $\mathcal{U}(0,1)$ & $0.864_{-0.15}^{+0.095}$ \\
  Jitter term  $\ln{\sigma_\mathrm{TESS}}$ ($\mathrm{km\ s^{-1}}$) & $\mathcal{U}(-8,-5)$ & $-7.697\pm0.013$    & $\mathcal{U}(-8,-5)$ & $-7.697\pm0.012$\\ 
  Jitter term  $\ln{\sigma_\mathrm{ELODIE}}$ ($\mathrm{km\ s^{-1}}$)& $\mathcal{U}(-10,-1)$ & $-4.24\pm0.20$   & $\mathcal{U}(-10,-1)$ &$-4.24_{-0.21}^{+0.19}$\\
  Jitter term  $\ln{\sigma_\mathrm{HIRES}}$ ($\mathrm{km\ s^{-1}}$)& $\mathcal{U}(-10,-1)$ &  $-3.63_{-0.14}^{+0.15}$   & $\mathcal{U}(-10,-1)$ &$-3.64_{-0.14}^{+0.16}$\\
  Jitter term  $\ln{\sigma_\mathrm{KPF}}$ ($\mathrm{km\ s^{-1}}$)& $\mathcal{U}(-10,-1)$ & $-6.19\pm0.11$  & $\mathcal{U}(-10,-1)$ &$-7.32_{-1.5}^{+0.61}$\\
  Baseline offset $\gamma_\mathrm{TESS}$& $\mathcal{U}(-0.003, 0.003)$ &  $0.00023\pm0.000015$  & $\mathcal{U}(-0.003, 0.003)$ &$0.00023\pm0.000015$\\
  Baseline offset $\gamma_\mathrm{ELODIE}$ ($\mathrm{km\ s^{-1}}$) & $\mathcal{U}(-50,50)$ & $-29.3671\pm0.0057$      & $\mathcal{U}(-50,50)$ &$-29.3670\pm0.0057$\\
  Baseline slope $\dot{\gamma}_\mathrm{ELODIE}$ ($\mathrm{km\ s^{-1}d^{-1}}$) & $\mathcal{U}( -0.1,0.1)$ & $0.0549\pm0.0098$    & $\mathcal{U}( -0.1 ,0.1)$ &$0.0554\pm0.0095$\\
  Baseline offset  $\gamma_\mathrm{HIRES}$ ($\mathrm{km\ s^{-1}}$) & $\mathcal{U}(-50,50)$ & $-0.030\pm0.015$     & $\mathcal{U}(-50,50)$ &$-0.029\pm0.015$\\
  Baseline slope $\dot{\gamma}_\mathrm{HIRES}$ ($\mathrm{km\ s^{-1}d^{-1}}$) & $\mathcal{U}( -0.1, 0.1)$ &  $0.021\pm0.017$    & $\mathcal{U}( -0.1, 0.1)$ &$0.020\pm0.018$\\
  Baseline offset $\gamma_\mathrm{KPF}$ ($\mathrm{km\ s^{-1}}$) & $\mathcal{U}(-50,50)$ & $0.0624\pm0.0024$     & $\mathcal{U}(-50,50)$ &$0.0628\pm0.0025$\\
  \hline
  \multicolumn{5}{l}{\textbf{Derived Parameters}}\\[3pt] 
  Planet mass $m_{b}$ ($\mathrm{M_{J}}$)  &  --& $2.13\pm0.12$  & --&$2.14\pm0.12$  \\
  Planet radius $r_{b}$ ($\mathrm{R_{J}}$)  &  --& $1.139\pm0.028$  & --&$1.140\pm0.029$ \\
  Semi-major axis $a_b$ (AU) &  --& $0.0714\pm0.002$  & --&$0.0713\pm0.002$  \\
  Inclination  $I_\mathrm{b}$ (deg)  &  --& $88.71_{-1.02}^{+0.79}$  & --&$88.48_{-0.93}^{+0.73}$  \\ 
  Eccentricity  $e_\mathrm{b}$  &  --& $0.3135\pm0.0078$ &  --&$0.3136\pm0.0075$\\ 
  Argument of periastron  $w_\mathrm{b}$ (deg)  &  --& $155.2_{-2.0}^{+1.9}$ & --&$155.5_{-2.0}^{+1.9}$ \\ 
  Impact parameter  $b_\mathrm{tra;b}$  &  --&  $0.131_{-0.080}^{+0.10}$ & --&$0.155_{-0.074}^{+0.090}$ \\ 
  Total transit duration  $T_\mathrm{tot;b}$ (h)  &  --& $5.640_{-0.014}^{+0.015}$  & --&$5.643_{-0.014}^{+0.016}$ \\ 
  Full-transit duration  $T_\mathrm{full;b}$ (h)  &  --& $5.028_{-0.016}^{+0.014}$   & --&$5.026_{-0.017}^{+0.014}$\\ 
  Planet Equilibrium temperature $T_\mathrm{eq;b}$ (K) &  --& $1360\pm23$ &  --&$1361_{-22}^{+23}$\\ 
  Planet density $\rho_\mathrm{b}$ (cgs)  &  --& $1.78_{-0.18}^{+0.21}$ &  --& $1.79_{-0.18}^{+0.21}$\\ 
  Planet surface gravity  $g_\mathrm{b}$ (cgs)  &  --&$4198_{-168}^{+155}$ & --&$4192_{-179}^{+150}$\\ 
  Limb darkening $u_\mathrm{1; TESS}$  &  --& $0.339\pm0.049$ &  --&$0.333\pm0.046$\\ 
  Limb darkening $u_\mathrm{2; TESS}$  &  --& $0.159\pm0.087$  &  --& $0.166\pm0.078$\\ 
  Limb darkening $u_\mathrm{1; KPF}$  &  --&$1.33_{-0.23}^{+0.16}$ &  --&$1.39_{-0.17}^{+0.11}$\\ 
  Limb darkening $u_\mathrm{2; KPF}$  &  --& $-0.51_{-0.18}^{+0.29}$  &  --& $-0.59_{-0.13}^{+0.21}$\\ 
  Jitter term $\sigma_\mathrm{ELODIE}(m\ s^{-1})$  &  --& $14.38_{-2.6}^{+3.0}$  &  --& $14.4_{-2.7}^{+3,0}$\\ 
  Jitter term $\sigma_\mathrm{HIRES}(m\ s^{-1})^{\dagger}$  &  --& $26.5_{-3.3}^{+4.3}$  &  --& $26.4_{-3.4}^{+4.4}$\\ 
  Jitter term $\sigma_\mathrm{KPF}(m\ s^{-1})$  &  --& $2.0_{-0.21}^{+0.23}$  &  --& $0.66_{-0.51}^{+0.55}$\\ 
  Host density from all orbits $\rho_\mathrm{\star}$ (cgs)  &  --&  $0.1968_{-0.0095}^{+0.0068}$ &  --&$0.1960_{-0.0099}^{+0.0067}$ \\ 
  Frequency at maximum power $\nu_{\rm{max}}$ ($\mu \rm{Hz}$)  &  --& -- &  --&$967.2_{-46.5}^{+51.7}$\\ 
\hline
\multicolumn{5}{l}{\textbf{Model Comparison}}\\[3pt] 
Bayesian evidence $\rm{log(Z)}$  &  --& $20089.45\pm{0.42}$ &  --&$20095.64\pm{0.43}$\\
\hline
\tablecomments{The epoch has been shifted by -322 periods to put the epoch in the middle of the dataset. $\dagger$ We only consider a single-planet Keplerian model in the global fitting of RM effect. Therefore, the excess of HIRES jitter is attributed to the presence of the outer planet (detail see section~\ref{sec:dist}).}
\end{longtable*}
\end{scriptsize}

%% file: figA4.tex
\begin{figure}
    \centering
    \includegraphics[width=\linewidth]{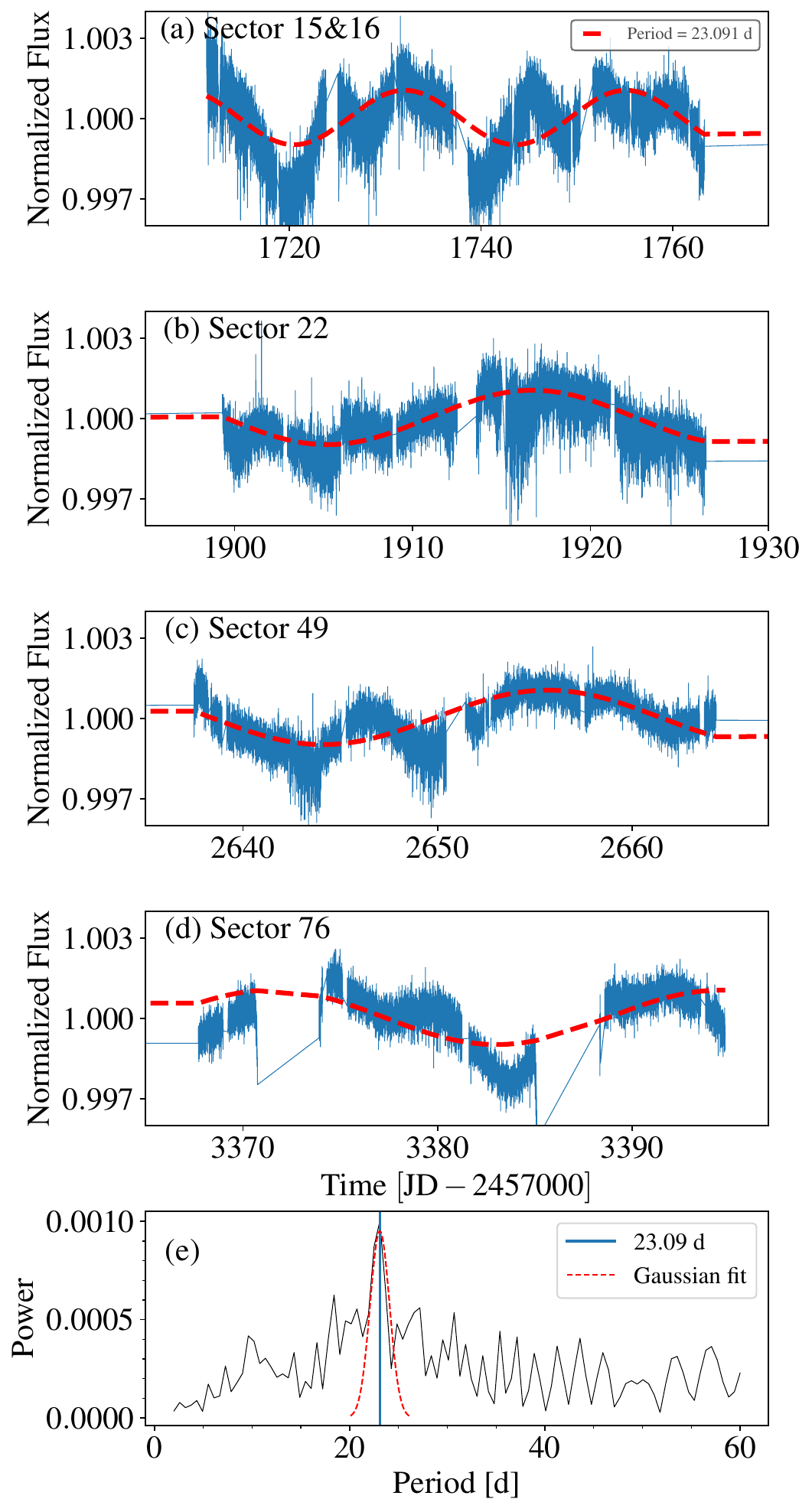}
    \caption{Panel a-d: TESS light curves of HD 118203 obtained in Sector 15, 16, 22, 49 and 76 with the modulation model at a period of 23.091 days. Panel e: the Lomb-Scargle periodogram of the detrended TESS lightcurves of HD 118203 using SIP \citep{Hedges2020}. The highest peak in the periodogram is located at 23.091 days. The red dashed line shows the Gaussian fit to the peak to determine the uncertainty of rotation period.  }
    \label{fig:figureA4}
\end{figure}

%% file: fig3.tex
\begin{figure*}
    \centering
    \includegraphics[width=\linewidth]{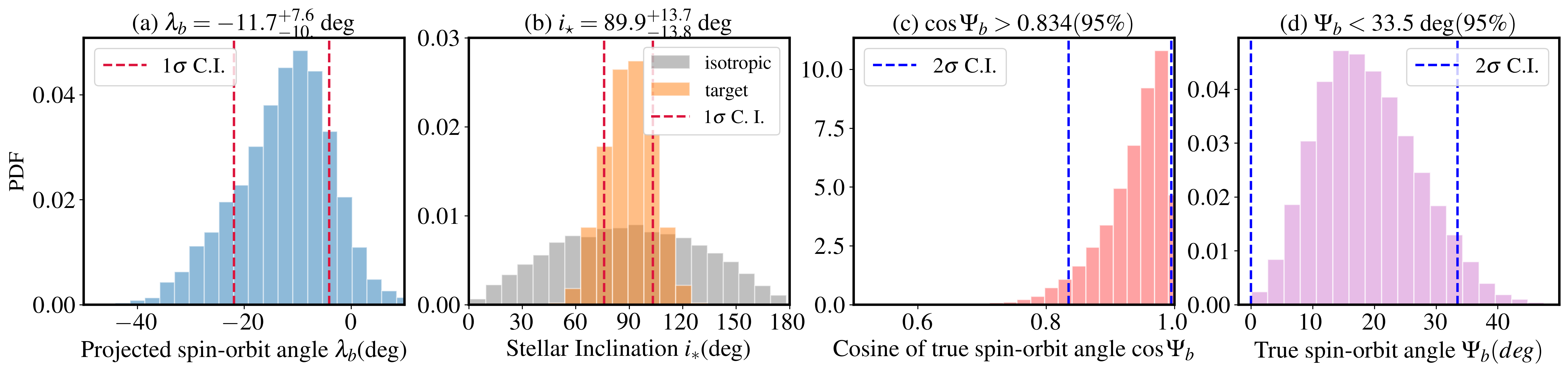}
    \caption{(a): Posterior distribution of sky-projected spin-orbit angle for HD 118203 b, obtained from RM effect modeling with stellar oscillations. (b): Posterior distribution of stellar inclination (in orange) versus an isotropic distribution. The stellar inclination ranges from $0^{\circ}$ to $180^{\circ}$ to accounts for scenarios where the stellar axis may point either toward ($<90^{\circ}$) or away ($>90^{\circ}$) from the observer. The red dashed lines correspond to $1\sigma$ significance. (c): Posterior distribution of cosine of true spin-orbit angle for HD 118203 b. (d): True spin-orbit angle distribution for HD 118203 b. The blue dashed lines correspond to $2\sigma$ significance. }  
    \label{fig:figure3}
\end{figure*}

%% file: fig11.tex
\begin{figure}
    \centering
    \includegraphics[width=\linewidth]{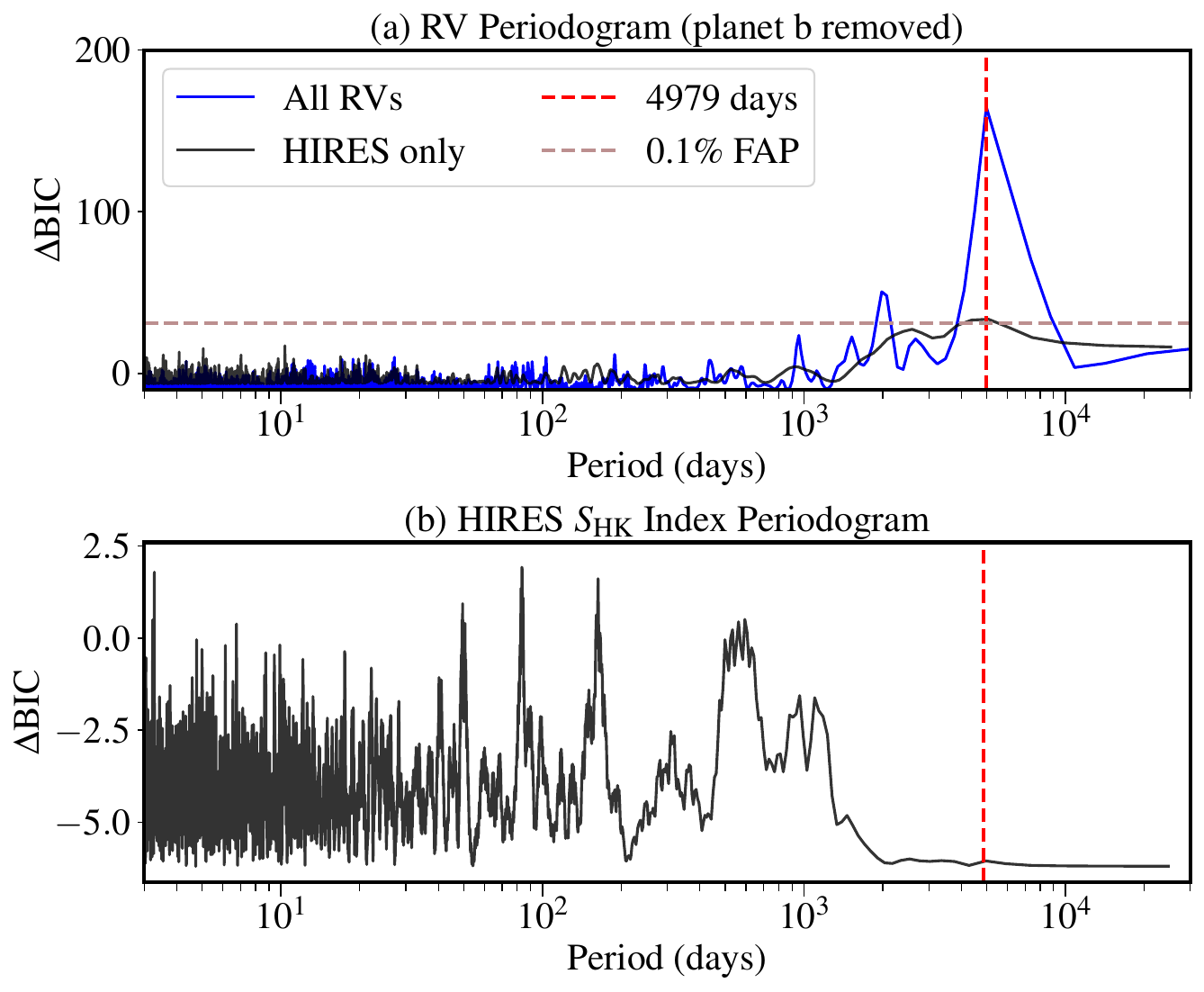}
    \caption{ a: The periodogram of RV residuals computed using \textit{RVsearch}. The signal from the hot Jupiter HD 118203 b has been subtracted. The blue line corresponds to the periodogram using all RVs. The black line corresponds to the periodogram using only HIRES RVs. b: The periodogram of HIRES $S_{\rm{HK}}$ index.} 
    \label{fig:figure11}
\end{figure}

%% file: fig4.tex
\begin{figure}
    \centering
    \includegraphics[width=\linewidth]{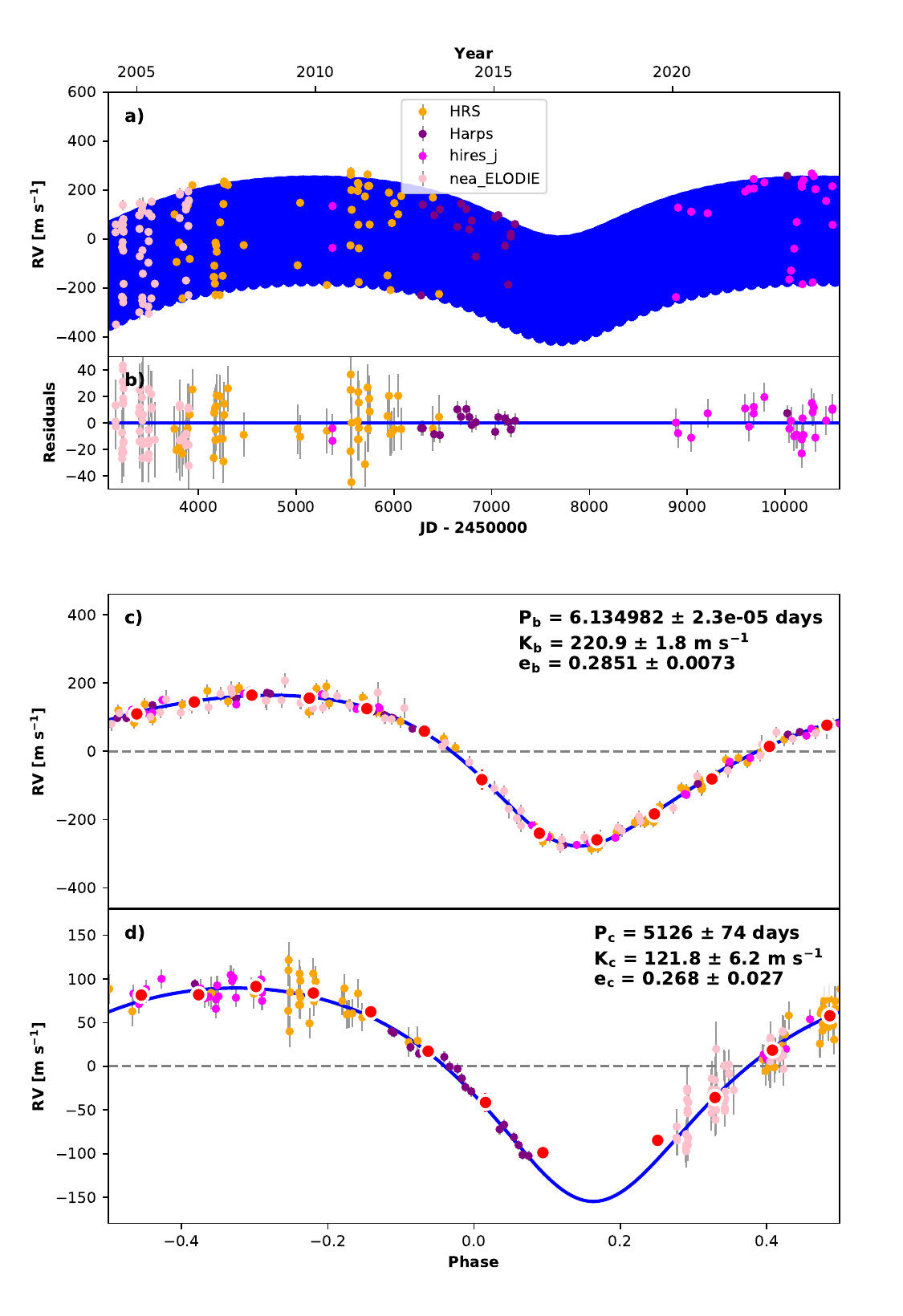}
    \caption{ Best-fit 2-planet Keplerian orbital model for HD 118203 a: HD 118203 RVs with errors (black) and their best fit model (blue) as a function of time. b: the residuals. c$\sim$d: RV data and models for each planet phase-folded at the best-fit orbital period with all other planets' signals removed. } 
    \label{fig:figure4}
\end{figure}

%% file: rvonlytab.tex
\begin{deluxetable}{lrrr}
\tablecaption{ MCMC Posteriors for RV-only fitting}\label{tab:rvparams}
\tablehead{
  \colhead{Parameter} & 
  \colhead{Credible Interval} & 
  \colhead{Max Likelihood} & 
  \colhead{Units}
}
\startdata
\multicolumn{4}{c}{\bf{Fitted parameters}} \\[1pt]
\hline
  $P_{b}$ & $6.134982^{+2.2e-05}_{-2.3e-05}$ & $6.134983$ & days \\
  $T\rm{conj}_{b}$ & $2459129.795\pm 0.022$ & $2459129.794$ & JD \\
  $T\rm{peri}_{b}$ & $2459130.44^{+0.035}_{-0.037}$ & $2459130.444$ & JD \\
  $e_{b}$ & $0.2851^{+0.0074}_{-0.0072}$ & $0.2841$ &  \\
  $\omega_{b,*}^{\dagger}$ & $2.697^{+0.032}_{-0.034}$ & $2.702$ & radians \\
  $K_{b}$ & $220.9\pm 1.8$ & $221.0$ & m s$^{-1}$ \\
  $P_{c}$ & $5126^{+76}_{-72}$ & $5123$ & days \\
  $T\rm{conj}_{c}$ & $2456857^{+54}_{-51}$ & $2456856$ & JD \\
  $T\rm{peri}_{c}$ & $2457654^{+68}_{-71}$ & $2457662$ & JD \\
  $e_{c}$ & $0.268^{+0.028}_{-0.026}$ & $0.267$ &  \\
  $\omega_{c,*}^{\dagger}$ & $3.066^{+0.07}_{-0.073}$ & $3.078$ & radians \\
  $K_{c}$ & $121.8^{+6.6}_{-5.7}$ & $122.1$ & m s$^{-1}$ \\
\hline
\multicolumn{4}{c}{\bf{Other parameters}} \\[1pt]
\hline
  $\gamma_{\rm ELODIE}$ & $\equiv-29290.0306$ & $\equiv-29290.0306$ & m s$-1$ \\
  $\gamma_{\rm HIRES}$ & $\equiv-95.6072$ & $\equiv-95.6072$ & m s$-1$ \\
  $\gamma_{\rm HARPS}$ & $\equiv-29180.777$ & $\equiv-29180.777$ & m s$-1$ \\
  $\gamma_{\rm HRS}$ & $\equiv-22.9028$ & $\equiv-22.9028$ & m s$-1$ \\
  $\dot{\gamma}$ & $\equiv0.0$ & $\equiv0.0$ & m s$^{-1}$ d$^{-1}$ \\
  $\ddot{\gamma}$ & $\equiv0.0$ & $\equiv0.0$ & m s$^{-1}$ d$^{-2}$ \\
  $\sigma_{\rm ELODIE}$ & $14.0^{+2.8}_{-2.5}$ & $13.3$ & $\rm m\ s^{-1}$ \\
  $\sigma_{\rm HIRES}$ & $11.7^{+2.2}_{-1.6}$ & $10.7$ & $\rm m\ s^{-1}$ \\
  $\sigma_{\rm HARPS}$ & $6.9^{+2.1}_{-1.5}$ & $5.6$ & $\rm m\ s^{-1}$ \\
  $\sigma_{\rm HRS}$ & $15.2^{+2.4}_{-2.0}$ & $14.4$ & $\rm m\ s^{-1}$ \\
\hline
\multicolumn{4}{c}{\bf{Derived parameters}} \\[1pt]
\hline
  $M_b$ & $2.24\pm 0.12$ & $2.23$ & M$_{\rm Jup}$ \\
  $a_b$ & $0.0711^{+0.0018}_{-0.0019}$ & $0.0709$ &  AU \\
  $M_c\sin i$ & $11.70^{+0.87}_{-0.83}$ & $11.64$ & M$_{\rm Jup}$ \\
  $a_c$ & $6.32^{+0.17}_{-0.18}$ & $6.31$ &  AU \\
\hline
\enddata
\tablenotetext{}{$\dagger$ The default output from \textit{RadVel} is the argument of periastron for the star. In the \textit{RadVel} coordinate system, the +Z direction points away from the observer. }
\label{tab:params}
\end{deluxetable}

%% file: fig5.tex
\begin{figure*}
    \centering
\includegraphics[width=\linewidth]{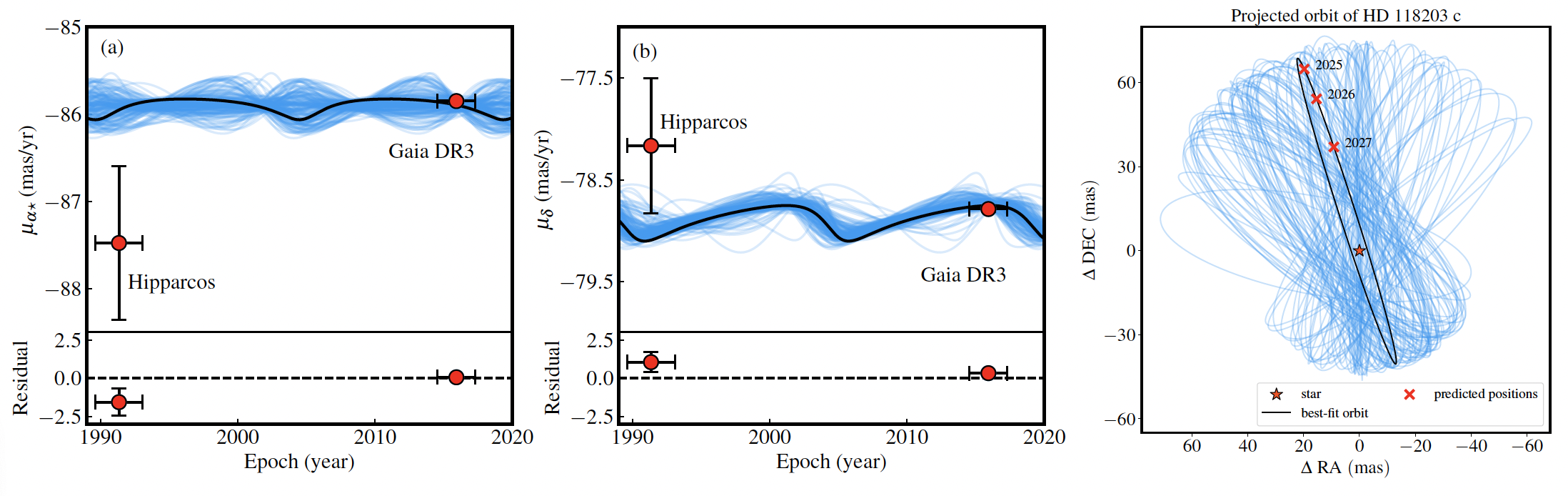}
    \caption{ Three-dimensional orbit characterization of the outer giant planet around HD 118203 using RVs and \textit{Hipparcos}- \textit{Gaia} astrometric acceleration.  (a) $\&$ (b): Observed and fitted Hipparcos and Gaia proper motion of HD118203 in right ascension and declination. (c): astrometric orbits of HD 118203 c projected on the sky plane from our model. In all of above panels, the thicker black lines represent the best-fit orbit in the MCMC chain while the other 100 blue lines represent random draws from the chain.}  
    \label{fig:figure5}
\end{figure*}

%% file: table_3d.tex
\begin{deluxetable*}{lcc}[t!]
\tablecaption{MCMC Posteriors for combined fitting of RV and Hipparcos-Gaia astrometric acceleration$^{a}$}\label{tab:3d}
\setlength{\tabcolsep}{0.10in}
\tablewidth{0pt}
\tablehead{
\colhead{Parameter (Unit)}              &
\colhead{Median $\pm$1$\sigma$} &
\colhead{Prior}                 }
\startdata
\multicolumn{3}{c}{Fitted parameters} \\[1pt]
\hline
\multicolumn{3}{c}{} \\[-5pt]
\multicolumn{3}{l}{\textbf{HD 118203~b}}\\[3pt]
RV semi-amplitude of HD118203~b $K_{b}$ ($\mathrm{m\,s^{-1}}$)                                      & $220.5\pm1.6$                      & $\mathcal{N}(217.7 ,5)$                                                \\[3pt]
Orbital periods $P_{b}$ (days)                                                     & $6.13499\pm{0.000025}$                & $\mathcal{N}(6.13 ,0.1)$                                                   \\[3pt]
Eccentricity term$\sqrt{e}_{b}\sin{\omega_{b}}$                                                      & $-0.48\pm{0.01}$            & no prior                                                            \\[3pt]
Eccentricity term$\sqrt{e_{b}}\cos{\omega_{b}}$                                                      & $0.23\pm{0.02}$              & no prior                                                           \\[3pt]
Epoch of periastron at 2451544.5 JD,  $\tau_{\rm b}$ & $0.51\pm{0.01}$                         & no prior       \\[3pt]
\multicolumn{3}{l}{\textbf{HD 118203~c}}\\[3pt]
Planet mass $m_{\rm c}$ ($\mathrm{M_{J}}$)                                       & $11.79^{+0.69}_{-0.63}$                           & no prior                                                 \\[3pt]
Orbital periods $P_{c}$ (days)                                                     & $5103_{-77}^{+76}$                &  no prior                                                    \\[3pt]
Cosine term of inclination $\cos{I_{c}}$                                                   & $-0.06_{-0.20}^{+0.24}$              &  no prior                              \\[3pt]
Longitude of ascending node  $\Omega_{c}^{b}$  (deg)                                &  $10_{+55}^{-43}$                &  no prior                                                             \\[3pt]
Eccentricity term$\sqrt{e}_{c}\sin{\omega_{c}}$                                                    & $0.04^{+0.04}_{-0.03}$               &  no prior                              \\[3pt]
Eccentricity term$\sqrt{e_{c}}\cos{\omega_{c}}$                                          
            &  $-0.51\pm{0.03}$              &  no prior                                                            \\[3pt]
Epoch of periastron at 2451544.5 JD, $\tau_{\rm c}$ & $0.19\pm{0.02}$                       &  no prior         \\[3pt]
\multicolumn{3}{l}{\textbf{Others}}\\[3pt]
Host-star mass $M_{*}$ ($\mathrm{M_{\odot}}$)                                       & $1.27\pm0.05$                      & $\mathcal{N}(1.27 ,0.09)$                                                    \\[3pt]
Parallax $\varpi$ (mas)                                                              & $10.86\pm0.02$                          & $\mathcal{N}(\varpi_{\rm DR3} ,\sigma[\varpi_{\rm DR3}])^{c}$ \\[3pt]
RV zero point $\gamma_{\rm{ELODIE}}$($\mathrm{m\,s^{-1}}$ )                                      &$29378.5^{+4.0}_{-3.9}$                     & no prior                                                            \\[3pt]
RV zero point $\gamma_{\rm{HIRES}}$($\mathrm{m\,s^{-1}}$ )                                   &       $220.9^{+4.5}_{-4.2}$                       & no prior                                                            \\[3pt]
RV zero point $\gamma_{\rm{HARPS}}$($\mathrm{m\,s^{-1}}$ )                                      &$29180^{+3.6}_{-3.5}$                     & no prior                                                            \\[3pt]
RV zero point $\gamma_{\rm{HRS}}$($\mathrm{m\,s^{-1}}$ )                                      &       $110.6^{+7.7}_{-7.9}$                       & no prior                                                            \\[3pt]
RV jitter term $\sigma_{\rm{ELODIE}}$($\mathrm{m\,s^{-1}}$)                                      &     $14.8^{+2.2}_{-1.9}$                       & $\mathcal{U}(0,25)$                                                           \\[3pt]
RV jitter term $\sigma_{\rm{HIRES}}$($\mathrm{m\,s^{-1}}$)                                      & $6.5^{+1.8}_{-1.3}$                            & $\mathcal{U}(0,25)$                                                           \\[3pt]
RV jitter term $\sigma_{\rm{HARPS}}$($\mathrm{m\,s^{-1}}$)                                      &     $11.5^{+2.0}_{-1.6}$                       & $\mathcal{U}(0,25)$                                                           \\[3pt]
RV jitter term $\sigma_{\rm{HRS}}$($\mathrm{m\,s^{-1}}$)                                      & $13.4^{+2.7}_{-2.5}$                            & $\mathcal{U}(0,25)$                                                           \\[3pt]
\hline
\multicolumn{3}{c}{} \\[-5pt]
\multicolumn{3}{c}{Derived parameters }\\[1pt]
\hline
\multicolumn{3}{c}{} \\[-5pt]
$\rm{Inclination}$ $I_{c}$ (deg)                                   & $93.7_{-13.8}^{+11.9}$                & no prior                                                            \\[3pt]
Semi-major axis $a_{c}$ (AU)                                                     & $6.28^{+0.10}_{-0.11}$                             & no prior                                                            \\[3pt]
Eccentricity $e_{c}$                                                            & $0.26^{+0.03}_{-0.02}$                 & $\mathcal{U}(0,0.99)$                                                            \\[3pt]
Argument of periastron $\omega_{c}$ (deg)                                   & $175\pm{4}$                & no prior                                                            \\[3pt]
Time of periastron $T_{p,c}$ (JD) & $2452527_{-76}^{+73}$    & no prior                                                            \\[3pt]
\enddata
\tablecomments{ (a).The $\chi^{2}$ of RVs is 148 for 151 measurements.  (b). The distribution of  $\Omega_c$ is centered  around $0^{\circ}$, which breaks down into two halves. We shifted values of  $\Omega_c$ greater than $\pi$ by $-2\pi$ to combine the distribution. (c).$\varpi_{\rm DR3}$ and $\sigma[\varpi_{\rm DR3}]$ present the parallax and parallax uncertainty of HD 118203 from Gaia DR3 observations. (d). The HIRES, HRS RVs are median-subtracted, whereas the ELODIE and HARPS RVs are not. (e). we reported the argument of periastron for planets from our 3D fitting here. In our coordinate system +Z points toward the observer, which is the opposite of the RV convention in \textit{RadVel}. }
\end{deluxetable*}

%% file: fig6.tex
\begin{figure}
    \centering
    \includegraphics[width=\linewidth]{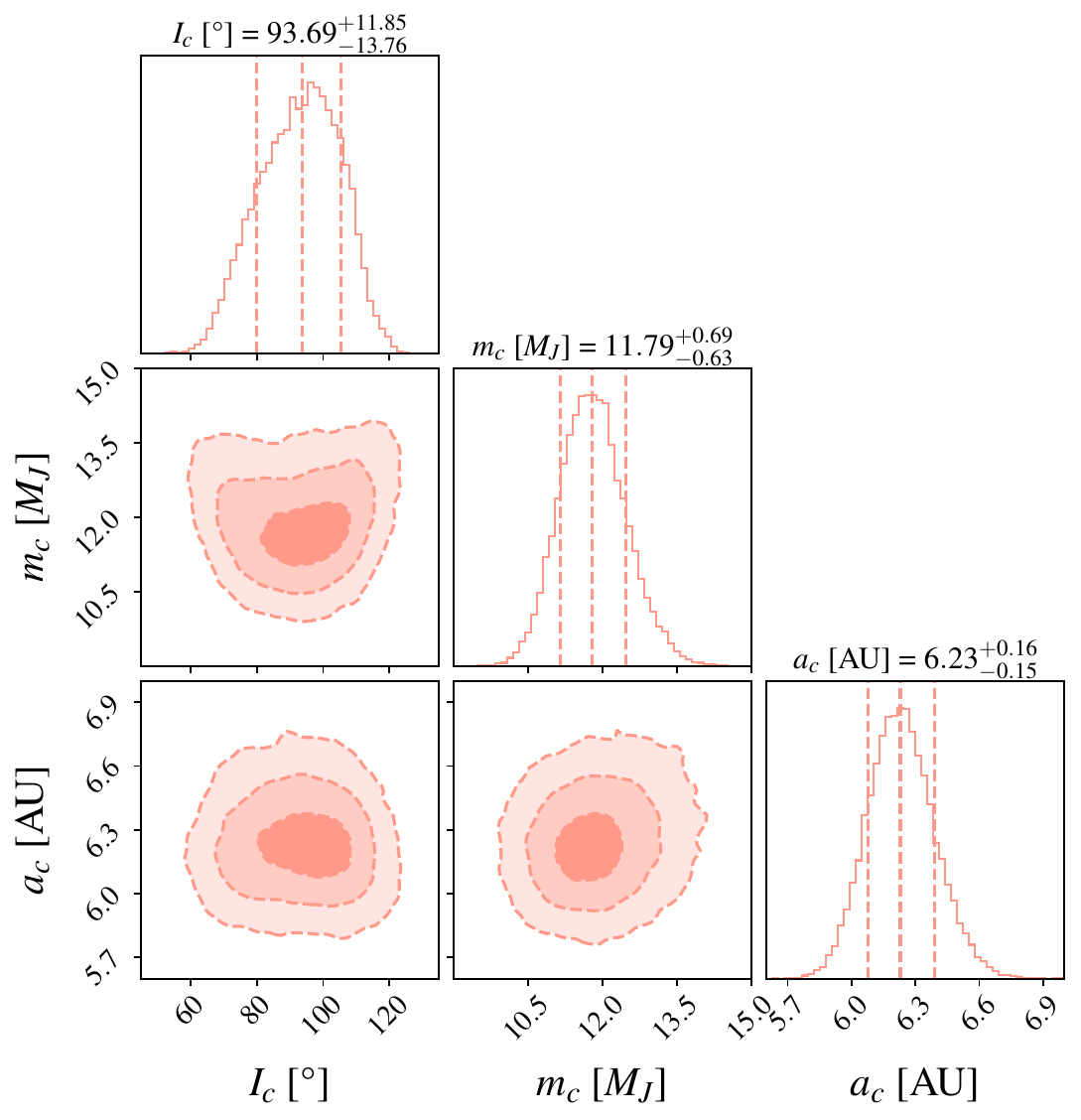}
    \caption{ The joint posterior distributions for orbital inclination $I_c$, mass $m_c$, and semi-major axis $a_c$ of HD 118203 c from combined fitting of RV and Hipparcos-Gaia astrometric acceleration. Moving outward, the dashed lines on the corner plots correspond to $1\sigma$, $2\sigma$ and $3\sigma$ contours. The orange dashed lines in the 1D distribution denote mean values with $1\sigma $  interval. } 
    \label{fig:figure6}
\end{figure}

%% file: fig8.tex
\begin{figure*}
    \centering
    \includegraphics[width=0.8\linewidth]{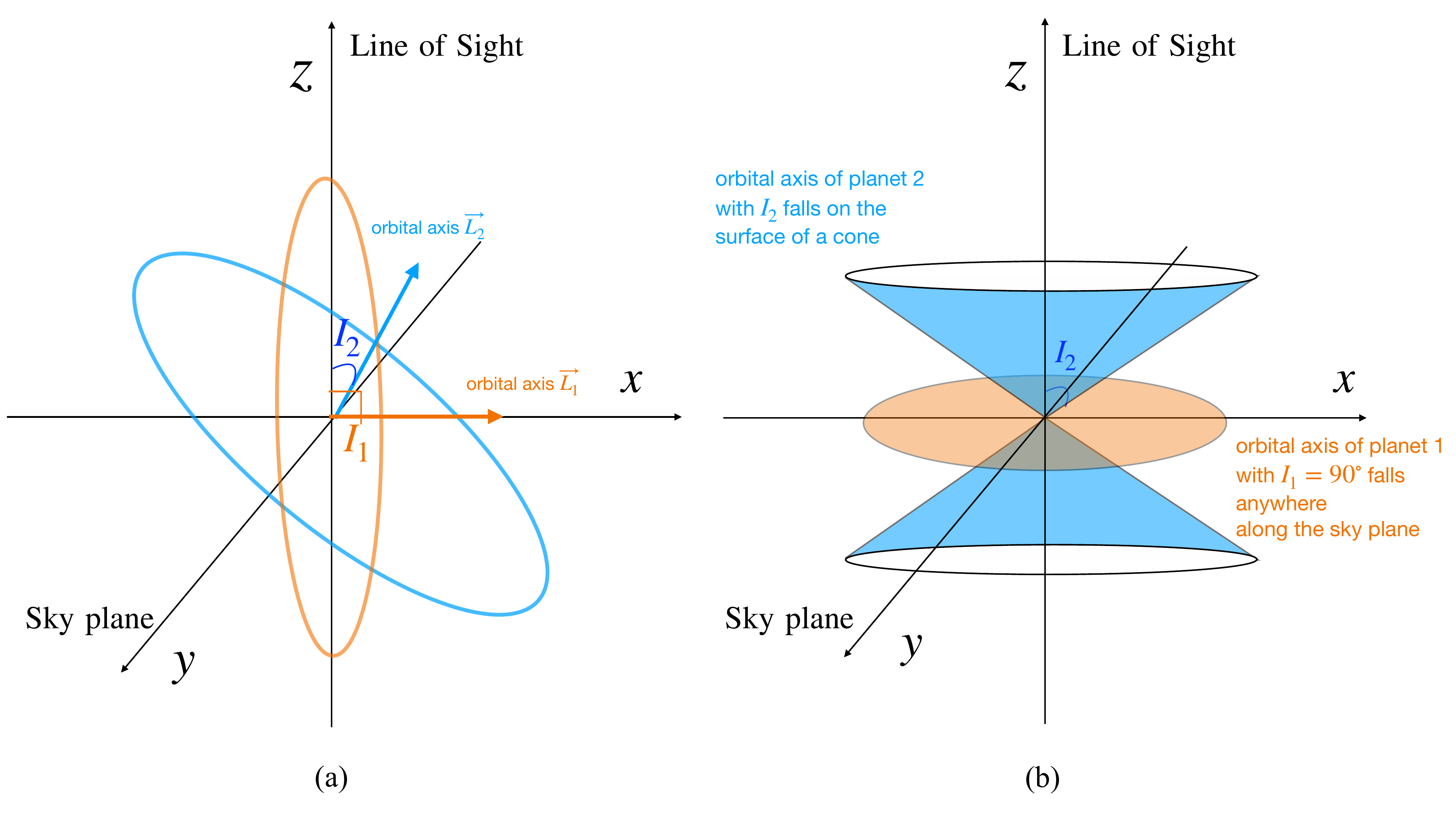}
    \caption{ Panel a: sketch showing the orbital geometry for the mutual inclination between HD 118203 b$\&$c. The observer is looking down from the z-axis. The x-y plane is the plane of the sky, and the star is centered on the origin. The orbital inclination of a planet is defined as the angle between the planet orbital axis and line of sight direction (z-axis). The mutual inclination is the angle between the orbital axes of the two planets.  Panel b: For a transiting planet with orbital inclination $I_1= 90^{\circ}$ without knowing the  longitude of acceding node, its orbital axis could fall anywhere along the sky plane (orange). For a planet with an orbital inclination $I_2\ne 90^{\circ}$ without knowing the  longitude of acceding node, its orbital axis falls on the surface of a cone (blue) with a vertex angle equals to the orbital inclination (toward observer, $I_2<90^{\circ}$) or $\pi-$ orbital inclination (away from observer, $I_2>90^{\circ}$). } 
    \label{fig:figure8}
\end{figure*}

%% file: fig9.tex
\begin{figure*}
    \centering
    \includegraphics[width=0.9\linewidth]{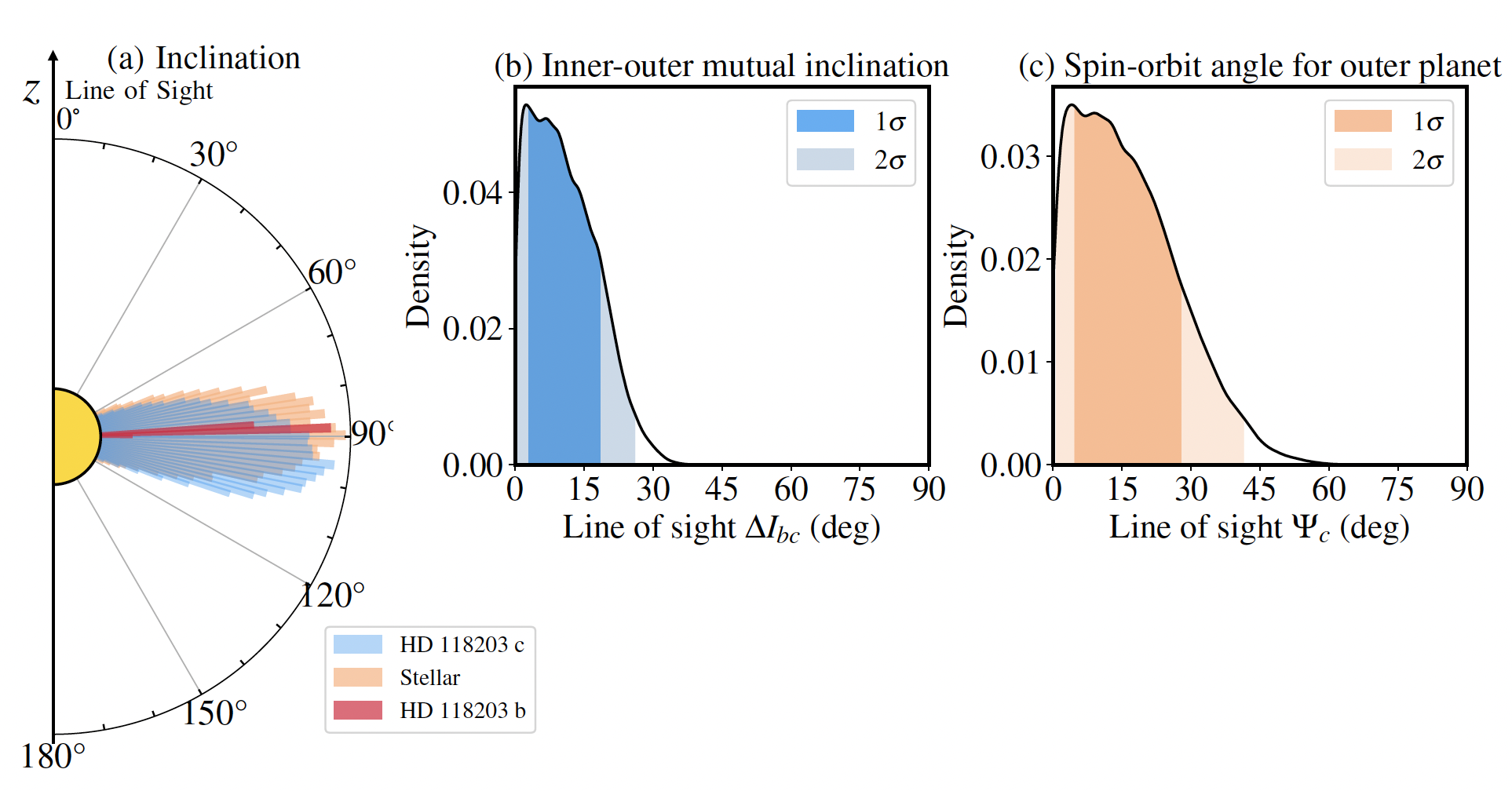}
    \caption{(a) Orbital inclination of HD 118203 c ($I_c$, blue) , compared with that of HD 118203 b ($I_b$, red)  and the line-of-sight stellar inclination ($i_*$, orange). Zero degrees  correspond to the line of sight. The stellar inclination ranges from $0^{\circ}$ to $180^{\circ}$ to accounts for scenarios where the stellar axis may point either toward ($<90^{\circ}$) or away ($>90^{\circ}$) from the observer. (b): Distribution of HD 118203 b$\&$c mutual inclination projected in line of sight direction. (c): Distribution of spin-orbit angle of HD 118203 c projected in line of sight direction.  } 
    \label{fig:figure9}
\end{figure*}

%% file: fig10.tex
\begin{figure*}
    \centering
    \includegraphics[width=0.9\linewidth]{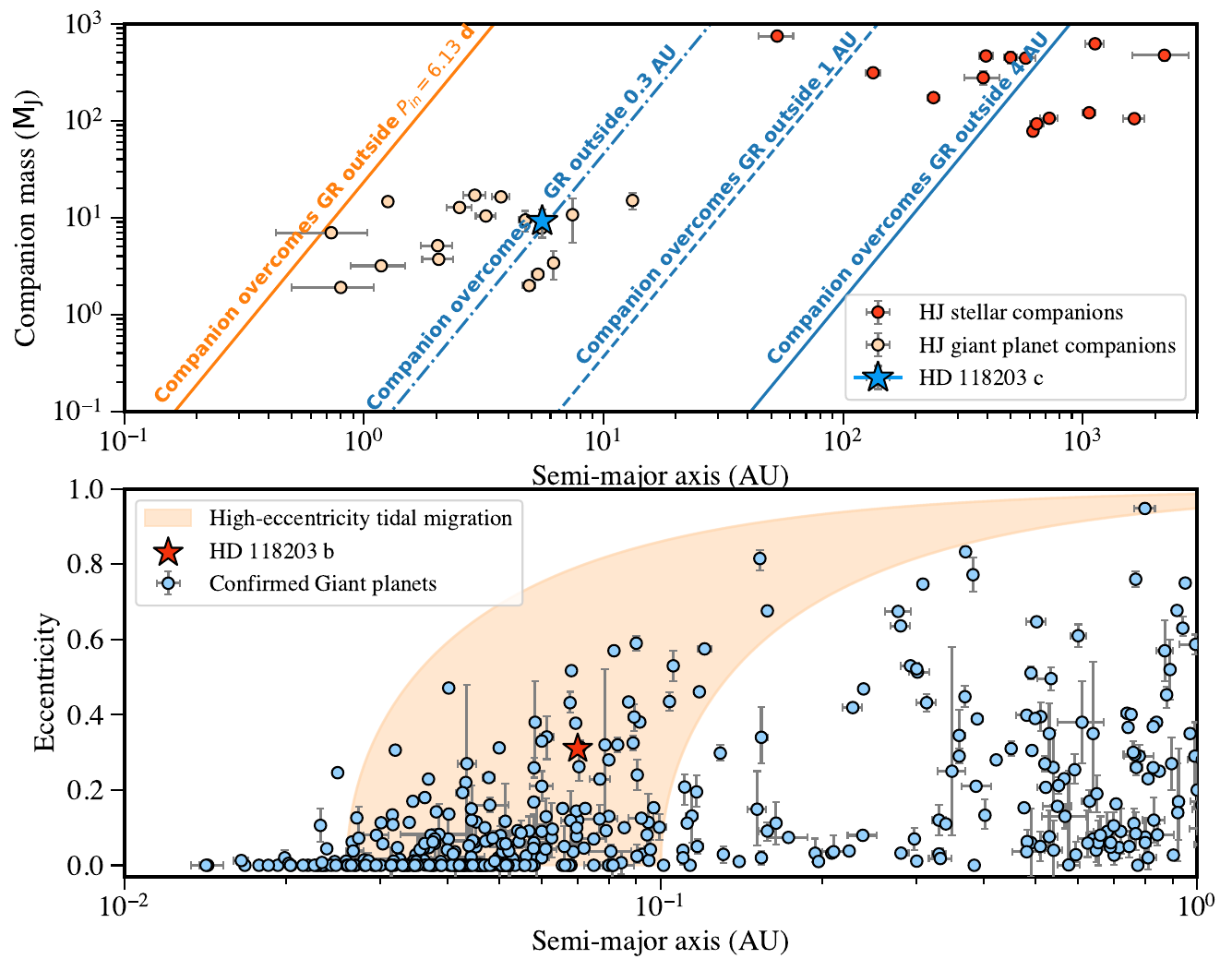}
    \caption{Top: Mass versus semi-major axis of detected companions to hot Jupiters. Red circles are stellar companions from \cite{Ngo2016}, and orange circles are giant planet companions from \cite{Knutson2014, Bryan2016, Rosenthal2021}. The orange line represents the limit for a companion to overcoming GR precession at a close-in orbit of 6.13 days. The blue lines represent the limit for a companion overcoming GR precession at 0.3 AU, 1 AU and 4 AU, respectively. Bottom:  Eccentricity versus semi-major axis of confirmed planets with mass larger tha $0.8 \mathrm{M_J}$ within 1 AU, based on data from the NASA Exoplanet Archive as of June 2024. The orange region indicates the trajectory associated with high-eccentricity tidal migration, where tidal dissipation within the planet reduces its orbital energy while keeping constant angular momentum. The final semi-major axis range is set between 0.02 and 0.1 AU, representing the upper and lower limits. }
    \label{fig:figure10}
\end{figure*}

%% file: fig7.tex
\begin{figure}
    \centering
    \includegraphics[width=\linewidth]{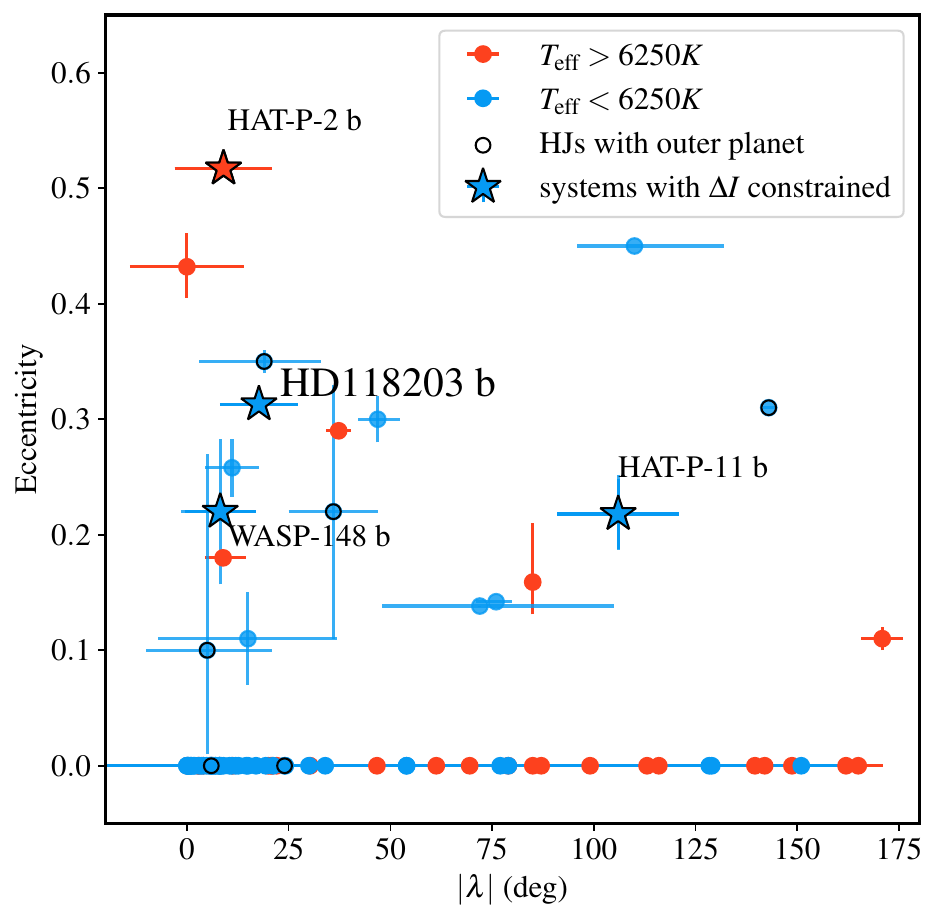}
    \caption{Eccentricity vs. spin-orbit angle of HD 118203 b in comparison to other hot Jupiters ($P<10\ \rm{days}$) from \cite{Rice2022}.  HD 118203 b is one of few systems where the mutual inclinations between the hot Jupiters and outer planets have been directly measured. The colors denote the effective temperature of the host star. }  
    \label{fig:figure7}
\end{figure}

%% file: figA1.tex
\begin{figure*}[b]
    \centering
    \includegraphics[width=0.6\linewidth]{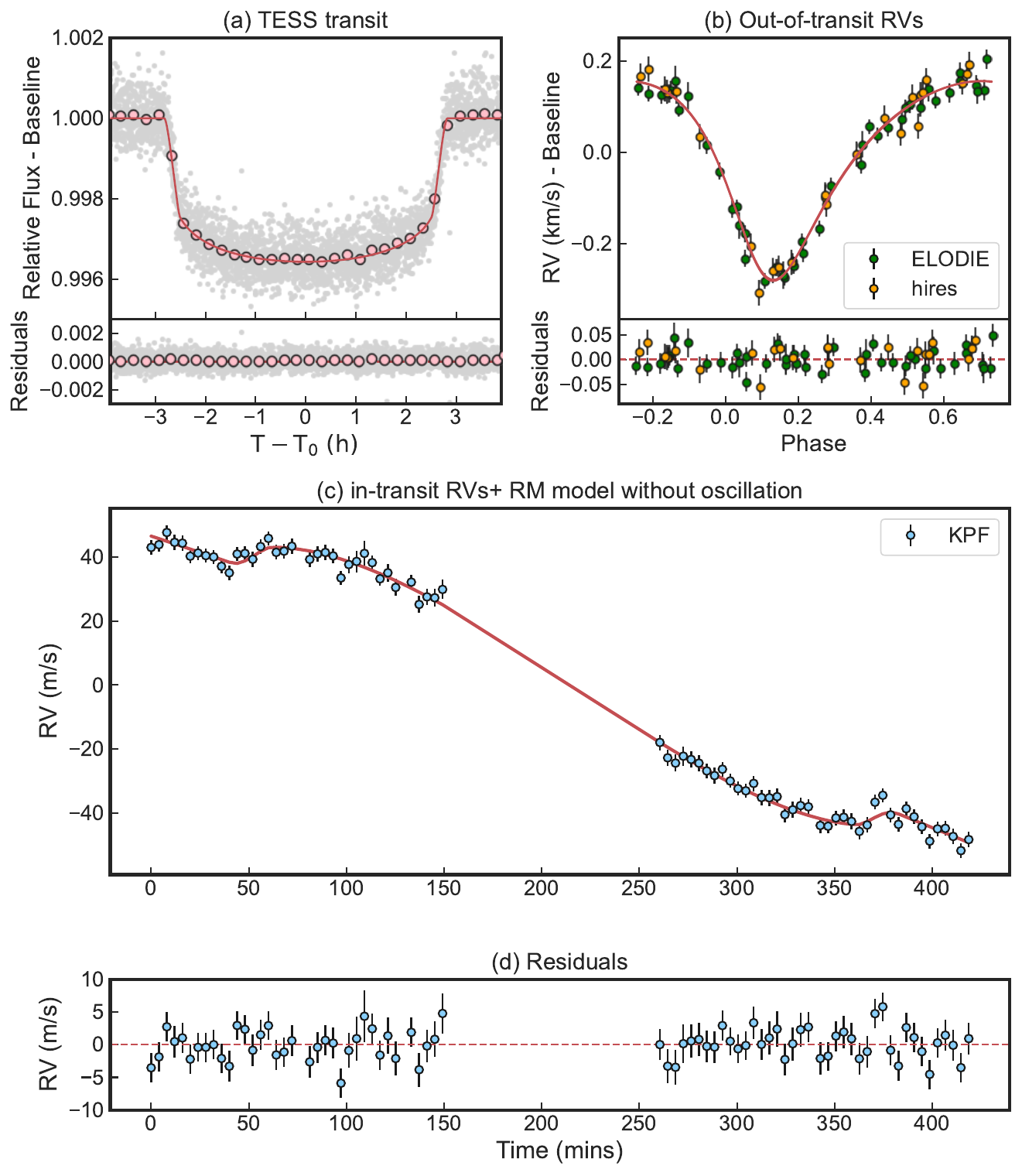}
    \caption{The Rossiter-McLaughlin (RM) model for HD 118203 b without GP for oscillation, integrating a global fitting of photometric data, non-transit radial velocity (RV) measurements, and RM data during transit. Panel a: TESS SPOC light curves phase-folded to the best-fit orbital period of HD 118203 b. The pink points present the TESS data binned every 15 minutes, and the red line shows the best-fit transit model. Panel b: The out-of-transit RVs from ELODIE (green) and HIRES (orange) phase-folded to the best-fit orbital period of HD 118203 b.  The red line depicts the best-fit RV model. Panel c: the in-transit RVs from KPF (blue) and best-fit RM model (red). Panel d: The residuals of the fitting. The residuals show periodic patterns that could come from the oscillation of the star. }
    \label{fig:figureA1}
\end{figure*}

%% file: figA2.tex
\begin{figure*}
    \centering
    \includegraphics[width=1.1\linewidth]{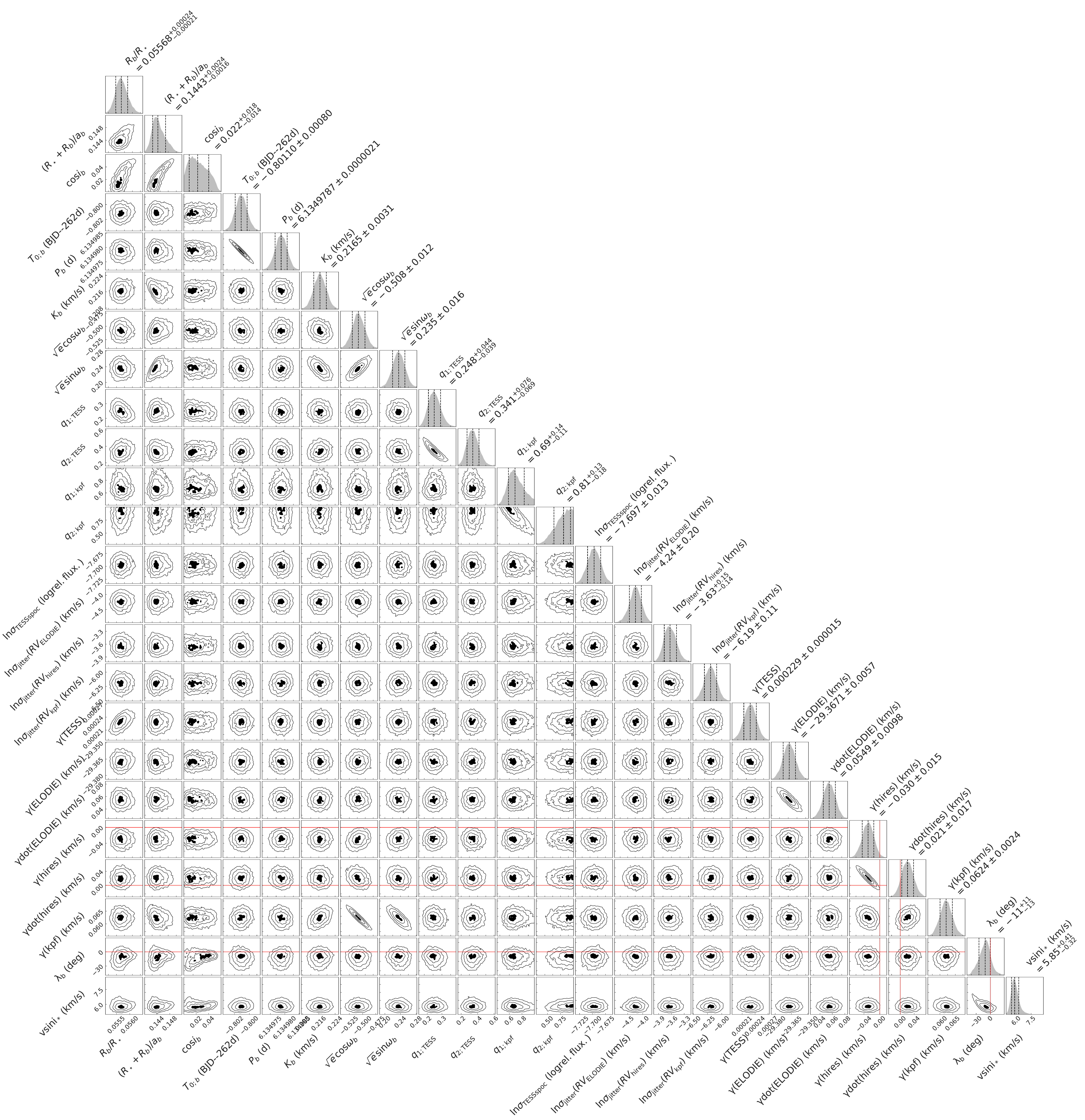}
    \caption{Joint posterior distributions for 24 parameters used in the Nested Sampling fitting of the RM effect without oscillation. The values and histogram distributions of all parameters are shown, along with 1 $\sigma$ uncertainties.}
    \label{fig:figureA2}
\end{figure*}

%% file: figA3.tex
\begin{figure*}
    \centering
    \includegraphics[width=1.1\linewidth]{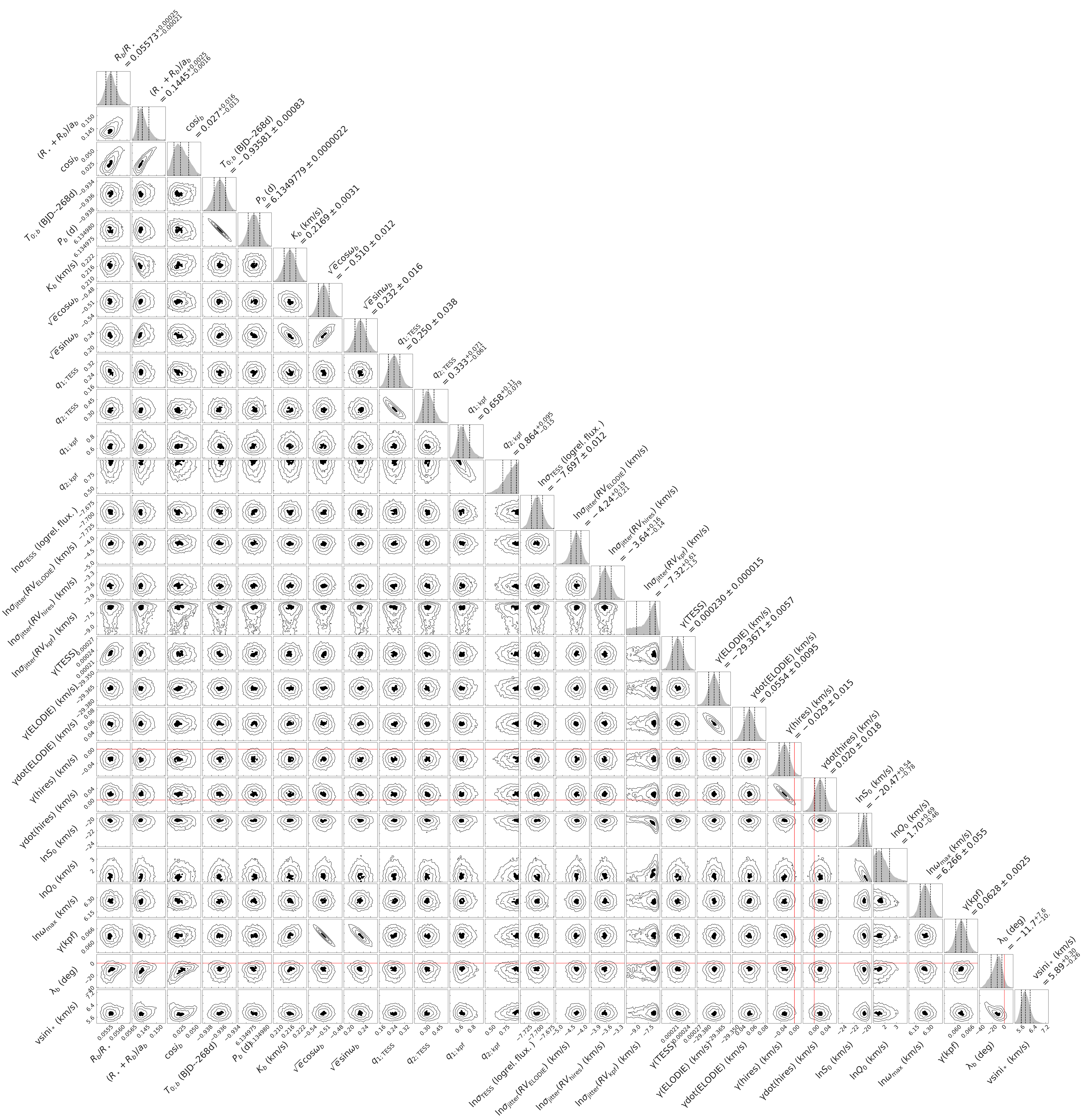}
    \caption{Joint posterior distributions for 27 parameters used in the Nested Sampling fitting of the RM effect with SHO GP to account for stellar oscillation. The values and histogram distributions of all parameters are shown, along with 1 $\sigma$ uncertainties.}
    \label{fig:figureA3}
\end{figure*}

%% file: figA5.tex
\begin{figure}
    \centering
    \includegraphics[width=1.05\linewidth]{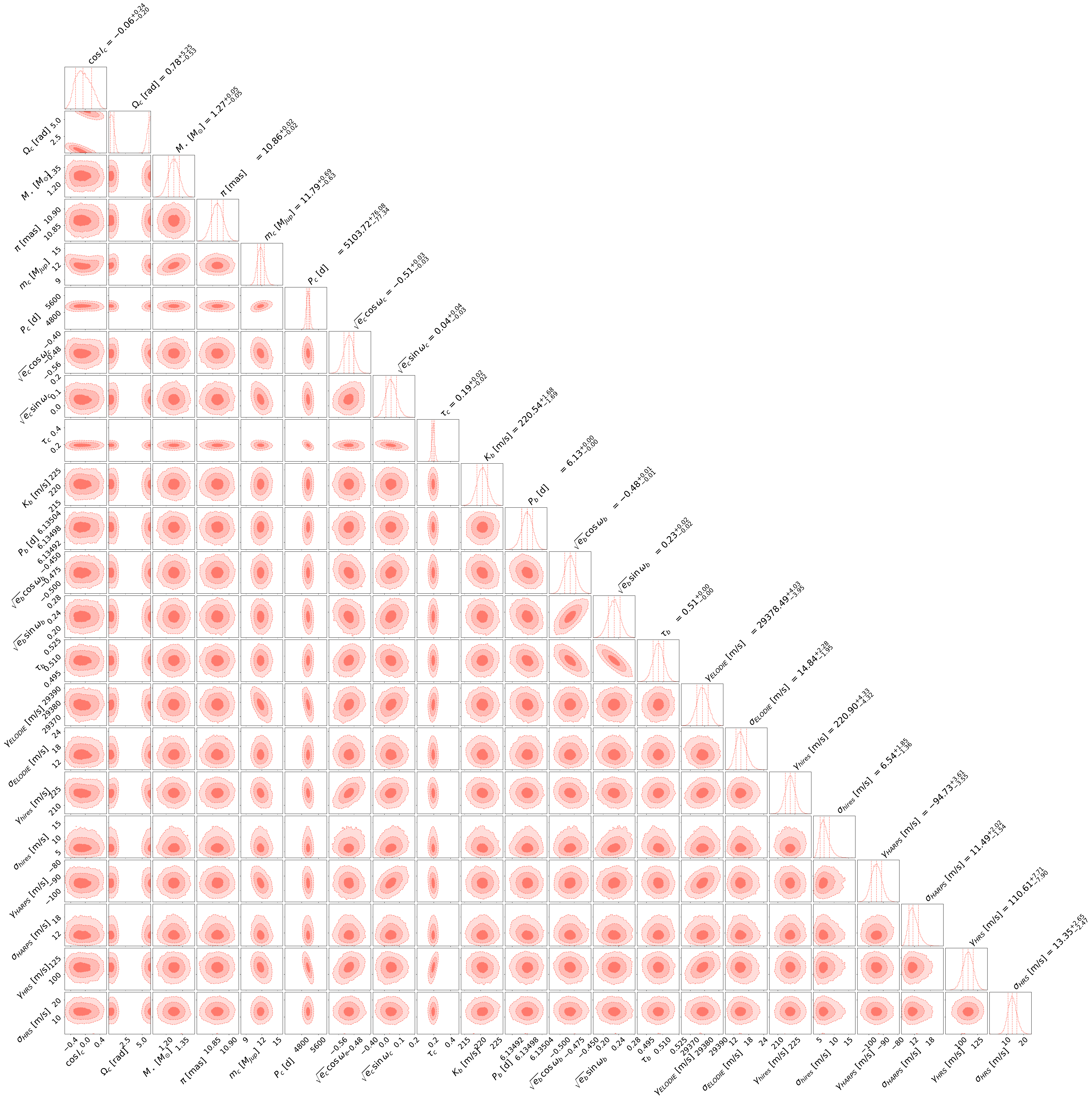}
\caption{Joint posterior distributions for 18 parameters used in the MCMC fitting for the 3D orbits of the outer planet HD 118203 c. The values and histogram distributions of all parameters are shown, along with 1 $\sigma$ uncertainties.}
    \label{fig:figureA5}
\end{figure}

%% file: figA8.tex
\begin{figure}
    \centering
    \includegraphics[width=1.05\linewidth]{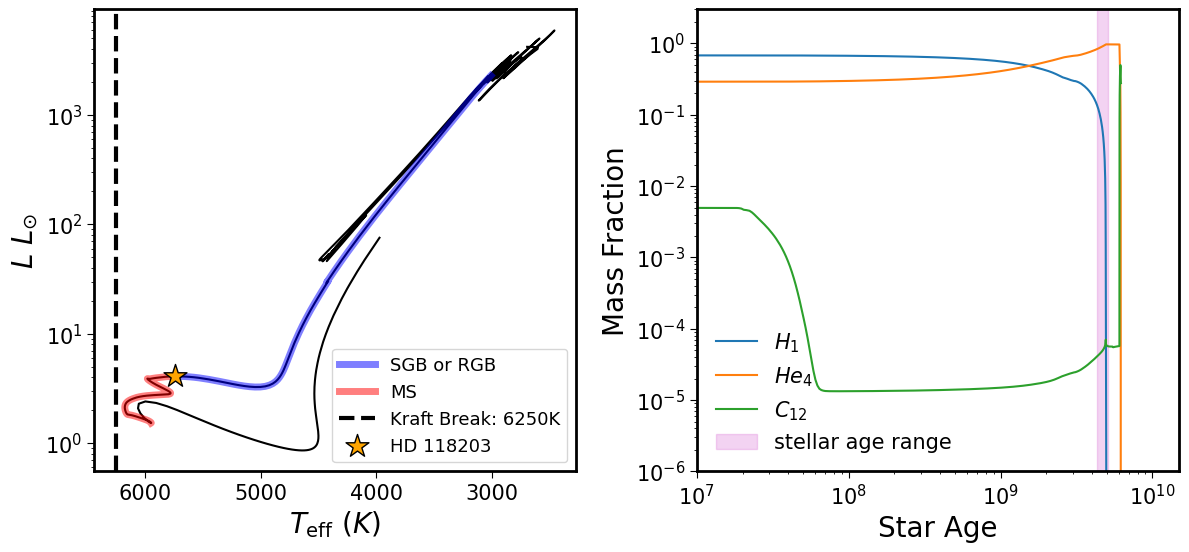}
\caption{Left: the stellar evolution models fit to HD 118203, and
traces back to their ZAMS location. Right: the mass fraction of hydrogen, helium and carbon in the core of the star from the evolution models. At the estimate stellar age, the hydrogen has just been used up and the star evolved to sub-giant. }
    \label{fig:figureA8}
\end{figure}

%% file: sample63.bbl
\begin{thebibliography}{}
\expandafter\ifx\csname natexlab\endcsname\relax\def\natexlab#1{#1}\fi
\providecommand{\url}[1]{\href{#1}{#1}}
\providecommand{\dodoi}[1]{doi:~\href{http://doi.org/#1}{\nolinkurl{#1}}}
\providecommand{\doeprint}[1]{\href{http://ascl.net/#1}{\nolinkurl{http://ascl.net/#1}}}
\providecommand{\doarXiv}[1]{\href{https://arxiv.org/abs/#1}{\nolinkurl{https://arxiv.org/abs/#1}}}

\bibitem[{{Albrecht} {et~al.}(2012){Albrecht}, {Winn}, {Johnson}, {Howard}, {Marcy}, {Butler}, {Arriagada}, {Crane}, {Shectman}, {Thompson}, {Hirano}, {Bakos}, \& {Hartman}}]{Albrecht2012}
{Albrecht}, S., {Winn}, J.~N., {Johnson}, J.~A., {et~al.} 2012, \apj, 757, 18, \dodoi{10.1088/0004-637X/757/1/18}

\bibitem[{{An} {et~al.}(2024){An}, {Lu}, {Brandt}, {Brandt}, \& {Li}}]{An2024}
{An}, Q., {Lu}, T., {Brandt}, G.~M., {Brandt}, T.~D., \& {Li}, G. 2024, arXiv e-prints, arXiv:2405.19510, \dodoi{10.48550/arXiv.2405.19510}

\bibitem[{{Bakos} {et~al.}(2007){Bakos}, {Kov{\'a}cs}, {Torres}, {Fischer}, {Latham}, {Noyes}, {Sasselov}, {Mazeh}, {Shporer}, {Butler}, {Stefanik}, {Fern{\'a}ndez}, {Sozzetti}, {P{\'a}l}, {Johnson}, {Marcy}, {Winn}, {Sip{\H{o}}cz}, {L{\'a}z{\'a}r}, {Papp}, \& {S{\'a}ri}}]{Bakos2007}
{Bakos}, G.~{\'A}., {Kov{\'a}cs}, G., {Torres}, G., {et~al.} 2007, \apj, 670, 826, \dodoi{10.1086/521866}

\bibitem[{{Baranne} {et~al.}(1996){Baranne}, {Queloz}, {Mayor}, {Adrianzyk}, {Knispel}, {Kohler}, {Lacroix}, {Meunier}, {Rimbaud}, \& {Vin}}]{Baranne1996}
{Baranne}, A., {Queloz}, D., {Mayor}, M., {et~al.} 1996, \aaps, 119, 373

\bibitem[{{Batygin}(2013)}]{Batygin2013}
{Batygin}, K. 2013, arXiv e-prints, arXiv:1304.5166, \dodoi{10.48550/arXiv.1304.5166}

\bibitem[{{Batygin} {et~al.}(2016){Batygin}, {Bodenheimer}, \& {Laughlin}}]{Batygin2016}
{Batygin}, K., {Bodenheimer}, P.~H., \& {Laughlin}, G.~P. 2016, \apj, 829, 114, \dodoi{10.3847/0004-637X/829/2/114}

\bibitem[{{Beaug{\'e}} \& {Nesvorn{\'y}}(2012)}]{Beaug2012}
{Beaug{\'e}}, C., \& {Nesvorn{\'y}}, D. 2012, \apj, 751, 119, \dodoi{10.1088/0004-637X/751/2/119}

\bibitem[{{Berger} {et~al.}(2020){Berger}, {Huber}, {van Saders}, {Gaidos}, {Tayar}, \& {Kraus}}]{berger2020}
{Berger}, T.~A., {Huber}, D., {van Saders}, J.~L., {et~al.} 2020, \aj, 159, 280, \dodoi{10.3847/1538-3881/159/6/280}

\bibitem[{{Bitsch} {et~al.}(2013){Bitsch}, {Crida}, {Libert}, \& {Lega}}]{Bitsch2013}
{Bitsch}, B., {Crida}, A., {Libert}, A.~S., \& {Lega}, E. 2013, \aap, 555, A124, \dodoi{10.1051/0004-6361/201220310}

\bibitem[{{Blunt} {et~al.}(2020){Blunt}, {Wang}, {Angelo}, {Ngo}, {Cody}, {De Rosa}, {Graham}, {Hirsch}, {Nagpal}, {Nielsen}, {Pearce}, {Rice}, \& {Tejada}}]{Blunt2020}
{Blunt}, S., {Wang}, J.~J., {Angelo}, I., {et~al.} 2020, \aj, 159, 89, \dodoi{10.3847/1538-3881/ab6663}

\bibitem[{{Bodenheimer} {et~al.}(2000){Bodenheimer}, {Hubickyj}, \& {Lissauer}}]{Bodenheimer2000}
{Bodenheimer}, P., {Hubickyj}, O., \& {Lissauer}, J.~J. 2000, \icarus, 143, 2, \dodoi{10.1006/icar.1999.6246}

\bibitem[{{Boley} {et~al.}(2016){Boley}, {Granados Contreras}, \& {Gladman}}]{Boley2016}
{Boley}, A.~C., {Granados Contreras}, A.~P., \& {Gladman}, B. 2016, \apjl, 817, L17, \dodoi{10.3847/2041-8205/817/2/L17}

\bibitem[{{Bowler} {et~al.}(2023){Bowler}, {Tran}, {Zhang}, {Morgan}, {Ashok}, {Blunt}, {Bryan}, {Evans}, {Franson}, {Huber}, {Nagpal}, {Wu}, \& {Zhou}}]{Bowler2023}
{Bowler}, B.~P., {Tran}, Q.~H., {Zhang}, Z., {et~al.} 2023, \aj, 165, 164, \dodoi{10.3847/1538-3881/acbd34}

\bibitem[{{Brandt}(2021)}]{Brandt2021}
{Brandt}, T.~D. 2021, \apjs, 254, 42, \dodoi{10.3847/1538-4365/abf93c}

\bibitem[{{Bryan} {et~al.}(2016){Bryan}, {Knutson}, {Howard}, {Ngo}, {Batygin}, {Crepp}, {Fulton}, {Hinkley}, {Isaacson}, {Johnson}, {Marcy}, \& {Wright}}]{Bryan2016}
{Bryan}, M.~L., {Knutson}, H.~A., {Howard}, A.~W., {et~al.} 2016, \apj, 821, 89, \dodoi{10.3847/0004-637X/821/2/89}

\bibitem[{{Butler} {et~al.}(1996){Butler}, {Marcy}, {Williams}, {McCarthy}, {Dosanjh}, \& {Vogt}}]{Butler1996}
{Butler}, R.~P., {Marcy}, G.~W., {Williams}, E., {et~al.} 1996, \pasp, 108, 500, \dodoi{10.1086/133755}

\bibitem[{{Butler} {et~al.}(2006){Butler}, {Wright}, {Marcy}, {Fischer}, {Vogt}, {Tinney}, {Jones}, {Carter}, {Johnson}, {McCarthy}, \& {Penny}}]{Butler2006}
{Butler}, R.~P., {Wright}, J.~T., {Marcy}, G.~W., {et~al.} 2006, \apj, 646, 505, \dodoi{10.1086/504701}

\bibitem[{{Castro-Gonz{\'a}lez} {et~al.}(2024){Castro-Gonz{\'a}lez}, {Lillo-Box}, {Correia}, {Santos}, {Barrado}, {Morales-Calder{\'o}n}, \& {Shkolnik}}]{Castro2024}
{Castro-Gonz{\'a}lez}, A., {Lillo-Box}, J., {Correia}, A.~C.~M., {et~al.} 2024, \aap, 684, A160, \dodoi{10.1051/0004-6361/202348722}

\bibitem[{{Chatterjee} {et~al.}(2008){Chatterjee}, {Ford}, {Matsumura}, \& {Rasio}}]{Chatterjee2008}
{Chatterjee}, S., {Ford}, E.~B., {Matsumura}, S., \& {Rasio}, F.~A. 2008, \apj, 686, 580, \dodoi{10.1086/590227}

\bibitem[{{Chatterjee} {et~al.}(2011){Chatterjee}, {Ford}, \& {Rasio}}]{Chatterjee2011}
{Chatterjee}, S., {Ford}, E.~B., \& {Rasio}, F.~A. 2011, in The Astrophysics of Planetary Systems: Formation, Structure, and Dynamical Evolution, ed. A.~{Sozzetti}, M.~G. {Lattanzi}, \& A.~P. {Boss}, Vol. 276, 225--229, \dodoi{10.1017/S1743921311020229}

\bibitem[{{Claytor} {et~al.}(2022){Claytor}, {van Saders}, {Llama}, {Sadowski}, {Quach}, \& {Avallone}}]{Claytor2022}
{Claytor}, Z.~R., {van Saders}, J.~L., {Llama}, J., {et~al.} 2022, \apj, 927, 219, \dodoi{10.3847/1538-4357/ac498f}

\bibitem[{{Cosentino} {et~al.}(2012){Cosentino}, {Lovis}, {Pepe}, {Collier Cameron}, {Latham}, {Molinari}, {Udry}, {Bezawada}, {Black}, {Born}, {Buchschacher}, {Charbonneau}, {Figueira}, {Fleury}, {Galli}, {Gallie}, {Gao}, {Ghedina}, {Gonzalez}, {Gonzalez}, {Guerra}, {Henry}, {Horne}, {Hughes}, {Kelly}, {Lodi}, {Lunney}, {Maire}, {Mayor}, {Micela}, {Ordway}, {Peacock}, {Phillips}, {Piotto}, {Pollacco}, {Queloz}, {Rice}, {Riverol}, {Riverol}, {San Juan}, {Sasselov}, {Segransan}, {Sozzetti}, {Sosnowska}, {Stobie}, {Szentgyorgyi}, {Vick}, \& {Weber}}]{Cosentino2012}
{Cosentino}, R., {Lovis}, C., {Pepe}, F., {et~al.} 2012, in Society of Photo-Optical Instrumentation Engineers (SPIE) Conference Series, Vol. 8446, Ground-based and Airborne Instrumentation for Astronomy IV, ed. I.~S. {McLean}, S.~K. {Ramsay}, \& H.~{Takami}, 84461V, \dodoi{10.1117/12.925738}

\bibitem[{{Cutri} {et~al.}(2003){Cutri}, {Skrutskie}, {van Dyk}, {Beichman}, {Carpenter}, {Chester}, {Cambresy}, {Evans}, {Fowler}, {Gizis}, {Howard}, {Huchra}, {Jarrett}, {Kopan}, {Kirkpatrick}, {Light}, {Marsh}, {McCallon}, {Schneider}, {Stiening}, {Sykes}, {Weinberg}, {Wheaton}, {Wheelock}, \& {Zacarias}}]{Cutri2003}
{Cutri}, R.~M., {Skrutskie}, M.~F., {van Dyk}, S., {et~al.} 2003, VizieR Online Data Catalog, II/246

\bibitem[{{da Silva} {et~al.}(2006){da Silva}, {Udry}, {Bouchy}, {Mayor}, {Moutou}, {Pont}, {Queloz}, {Santos}, {S{\'e}gransan}, \& {Zucker}}]{Silva2006}
{da Silva}, R., {Udry}, S., {Bouchy}, F., {et~al.} 2006, \aap, 446, 717, \dodoi{10.1051/0004-6361:20054116}

\bibitem[{{Damasso} {et~al.}(2020){Damasso}, {Sozzetti}, {Lovis}, {Barros}, {Sousa}, {Demangeon}, {Faria}, {Lillo-Box}, {Cristiani}, {Pepe}, {Rebolo}, {Santos}, {Zapatero Osorio}, {Gonz{\'a}lez Hern{\'a}ndez}, {Amate}, {Pasquini}, {Zerbi}, {Adibekyan}, {Abreu}, {Affolter}, {Alibert}, {Aliverti}, {Allart}, {Allende Prieto}, {{\'A}lvarez}, {Alves}, {Avila}, {Baldini}, {Bandy}, {Benz}, {Bianco}, {Borsa}, {Bossini}, {Bourrier}, {Bouchy}, {Broeg}, {Cabral}, {Calderone}, {Cirami}, {Coelho}, {Conconi}, {Coretti}, {Cumani}, {Cupani}, {D'Odorico}, {Deiries}, {Dekker}, {Delabre}, {Di Marcantonio}, {Dumusque}, {Ehrenreich}, {Figueira}, {Fragoso}, {Genolet}, {Genoni}, {G{\'e}nova Santos}, {Hughes}, {Iwert}, {Kerber}, {Knudstrup}, {Landoni}, {Lavie}, {Lizon}, {Lo Curto}, {Maire}, {Martins}, {M{\'e}gevand}, {Mehner}, {Micela}, {Modigliani}, {Molaro}, {Monteiro}, {Monteiro}, {Moschetti}, {Mueller}, {Murphy}, {Nunes}, {Oggioni}, {Oliveira}, {Oshagh}, {Pall{\'e}}, {Pariani}, {Poretti}, {Rasilla}, {Rebord{\~a}o}, {Redaelli},
  {Riva}, {Santana Tschudi}, {Santin}, {Santos}, {S{\'e}gransan}, {Schmidt}, {Segovia}, {Sosnowska}, {Span{\`o}}, {Su{\'a}rez Mascare{\~n}o}, {Tabernero}, {Tenegi}, {Udry}, \& {Zanutta}}]{Damasso2020}
{Damasso}, M., {Sozzetti}, A., {Lovis}, C., {et~al.} 2020, \aap, 642, A31, \dodoi{10.1051/0004-6361/202038416}

\bibitem[{{Dawson} \& {Johnson}(2018)}]{Dawson2018}
{Dawson}, R.~I., \& {Johnson}, J.~A. 2018, \araa, 56, 175, \dodoi{10.1146/annurev-astro-081817-051853}

\bibitem[{{Dawson} {et~al.}(2014){Dawson}, {Johnson}, {Fabrycky}, {Foreman-Mackey}, {Murray-Clay}, {Buchhave}, {Cargile}, {Clubb}, {Fulton}, {Hebb}, {Howard}, {Huber}, {Shporer}, \& {Valenti}}]{Dawson2014}
{Dawson}, R.~I., {Johnson}, J.~A., {Fabrycky}, D.~C., {et~al.} 2014, \apj, 791, 89, \dodoi{10.1088/0004-637X/791/2/89}

\bibitem[{{de Beurs} {et~al.}(2023){de Beurs}, {de Wit}, {Venner}, {Berardo}, {Bryan}, {Winn}, {Fulton}, \& {Howard}}]{deBeurs2023}
{de Beurs}, Z.~L., {de Wit}, J., {Venner}, A., {et~al.} 2023, \aj, 166, 136, \dodoi{10.3847/1538-3881/acedf1}

\bibitem[{{De Rosa} {et~al.}(2020){De Rosa}, {Dawson}, \& {Nielsen}}]{DeRosa2020}
{De Rosa}, R.~J., {Dawson}, R., \& {Nielsen}, E.~L. 2020, \aap, 640, A73, \dodoi{10.1051/0004-6361/202038496}

\bibitem[{{Debras} {et~al.}(2021){Debras}, {Baruteau}, \& {Donati}}]{Debras2021}
{Debras}, F., {Baruteau}, C., \& {Donati}, J.-F. 2021, \mnras, 500, 1621, \dodoi{10.1093/mnras/staa3397}

\bibitem[{{Duffell} \& {Chiang}(2015)}]{Duffell2015}
{Duffell}, P.~C., \& {Chiang}, E. 2015, \apj, 812, 94, \dodoi{10.1088/0004-637X/812/2/94}

\bibitem[{{Endl} {et~al.}(2014){Endl}, {Caldwell}, {Barclay}, {Huber}, {Isaacson}, {Buchhave}, {Brugamyer}, {Robertson}, {Cochran}, {MacQueen}, {Havel}, {Lucas}, {Howell}, {Fischer}, {Quintana}, \& {Ciardi}}]{Endl2014}
{Endl}, M., {Caldwell}, D.~A., {Barclay}, T., {et~al.} 2014, \apj, 795, 151, \dodoi{10.1088/0004-637X/795/2/151}

\bibitem[{{Epstein} \& {Pinsonneault}(2014)}]{Epstein2014}
{Epstein}, C.~R., \& {Pinsonneault}, M.~H. 2014, \apj, 780, 159, \dodoi{10.1088/0004-637X/780/2/159}

\bibitem[{{Espinoza-Retamal} {et~al.}(2023{\natexlab{a}}){Espinoza-Retamal}, {Zhu}, \& {Petrovich}}]{EspinozaRetamal2023}
{Espinoza-Retamal}, J.~I., {Zhu}, W., \& {Petrovich}, C. 2023{\natexlab{a}}, \aj, 166, 231, \dodoi{10.3847/1538-3881/ad00b9}

\bibitem[{{Espinoza-Retamal} {et~al.}(2023{\natexlab{b}}){Espinoza-Retamal}, {Brahm}, {Petrovich}, {Jord{\'a}n}, {Stef{\'a}nsson}, {Sedaghati}, {Hobson}, {Mu{\~n}oz}, {Boyle}, {Leiva}, \& {Suc}}]{ER2023}
{Espinoza-Retamal}, J.~I., {Brahm}, R., {Petrovich}, C., {et~al.} 2023{\natexlab{b}}, \apjl, 958, L20, \dodoi{10.3847/2041-8213/ad096d}

\bibitem[{{Fabrycky} \& {Tremaine}(2007)}]{Fabry2007}
{Fabrycky}, D., \& {Tremaine}, S. 2007, \apj, 669, 1298, \dodoi{10.1086/521702}

\bibitem[{{Fabrycky} \& {Winn}(2009)}]{Fabrycky2009}
{Fabrycky}, D.~C., \& {Winn}, J.~N. 2009, \apj, 696, 1230, \dodoi{10.1088/0004-637X/696/2/1230}

\bibitem[{{Foreman-Mackey} {et~al.}(2017){Foreman-Mackey}, {Agol}, {Ambikasaran}, \& {Angus}}]{Foreman-Mackey2017}
{Foreman-Mackey}, D., {Agol}, E., {Ambikasaran}, S., \& {Angus}, R. 2017, \aj, 154, 220, \dodoi{10.3847/1538-3881/aa9332}

\bibitem[{{Foreman-Mackey} {et~al.}(2013){Foreman-Mackey}, {Hogg}, {Lang}, \& {Goodman}}]{FM2013}
{Foreman-Mackey}, D., {Hogg}, D.~W., {Lang}, D., \& {Goodman}, J. 2013, \pasp, 125, 306, \dodoi{10.1086/670067}

\bibitem[{{Fulton} {et~al.}(2018){Fulton}, {Petigura}, {Blunt}, \& {Sinukoff}}]{Fulton2018}
{Fulton}, B.~J., {Petigura}, E.~A., {Blunt}, S., \& {Sinukoff}, E. 2018, \pasp, 130, 044504, \dodoi{10.1088/1538-3873/aaaaa8}

\bibitem[{{Gaia Collaboration} {et~al.}(2022){Gaia Collaboration}, {Klioner}, {Lindegren}, {Mignard}, {Hern{\'a}ndez}, {Ramos-Lerate}, {Bastian}, {Biermann}, {Bombrun}, {de Torres}, {Gerlach}, {Geyer}, {Hilger}, {Hobbs}, {Lammers}, {McMillan}, {Steidelm{\"u}ller}, {Teyssier}, {Raiteri}, {Bartolom{\'e}}, {Bernet}, {Casta{\~n}eda}, {Clotet}, {Davidson}, {Fabricius}, {Garralda Torres}, {Gonz{\'a}lez-Vidal}, {Portell}, {Rowell}, {Torra}, {Torra}, {Brown}, {Vallenari}, {Prusti}, {de Bruijne}, {Arenou}, {Babusiaux}, {Creevey}, {Ducourant}, {Evans}, {Eyer}, {Guerra}, {Hutton}, {Jordi}, {Luri}, {Panem}, {Pourbaix}, {Randich}, {Sartoretti}, {Soubiran}, {Tanga}, {Walton}, {Bailer-Jones}, {Drimmel}, {Jansen}, {Katz}, {Lattanzi}, {van Leeuwen}, {Bakker}, {Cacciari}, {De Angeli}, {Fouesneau}, {Fr{\'e}mat}, {Galluccio}, {Guerrier}, {Heiter}, {Masana}, {Messineo}, {Mowlavi}, {Nicolas}, {Nienartowicz}, {Pailler}, {Panuzzo}, {Riclet}, {Roux}, {Seabroke}, {Sordo}, {Th{\'e}venin}, {Gracia-Abril}, {Altmann}, {Andrae}, {Audard},
  {Bellas-Velidis}, {Benson}, {Berthier}, {Blomme}, {Burgess}, {Busonero}, {Busso}, {C{\'a}novas}, {Carry}, {Cellino}, {Cheek}, {Clementini}, {Damerdji}, {de Teodoro}, {Nu{\~n}ez Campos}, {Delchambre}, {Dell'Oro}, {Esquej}, {Fern{\'a}ndez-Hern{\'a}ndez}, {Fraile}, {Garabato}, {Garc{\'\i}a-Lario}, {Gosset}, {Haigron}, {Halbwachs}, {Hambly}, {Harrison}, {Hestroffer}, {Hodgkin}, {Holl}, {Jan{\ss}en}, {Jevardat de Fombelle}, {Jordan}, {Krone-Martins}, {Lanzafame}, {L{\"o}ffler}, {Marchal}, {Marrese}, {Moitinho}, {Muinonen}, {Osborne}, {Pancino}, {Pauwels}, {Recio-Blanco}, {Reyl{\'e}}, {Riello}, {Rimoldini}, {Roegiers}, {Rybizki}, {Sarro}, {Siopis}, {Smith}, {Sozzetti}, {Utrilla}, {van Leeuwen}, {Abbas}, {{\'A}brah{\'a}m}, {Abreu Aramburu}, {Aerts}, {Aguado}, {Ajaj}, {Aldea-Montero}, {Altavilla}, {{\'A}lvarez}, {Alves}, {Anderson}, {Anglada Varela}, {Antoja}, {Baines}, {Baker}, {Balaguer-N{\'u}{\~n}ez}, {Balbinot}, {Balog}, {Barache}, {Barbato}, {Barros}, {Barstow}, {Bassilana}, {Bauchet}, {Becciani},
  {Bellazzini}, {Berihuete}, {Bertone}, {Bianchi}, {Binnenfeld}, {Blanco-Cuaresma}, {Boch}, {Bossini}, {Bouquillon}, {Bragaglia}, {Bramante}, {Breedt}, {Bressan}, {Brouillet}, {Brugaletta}, {Bucciarelli}, {Burlacu}, {Butkevich}, {Buzzi}, {Caffau}, {Cancelliere}, {Cantat-Gaudin}, {Carballo}, {Carlucci}, {Carnerero}, {Carrasco}, {Casamiquela}, {Castellani}, {Castro-Ginard}, {Chaoul}, {Charlot}, {Chemin}, {Chiaramida}, {Chiavassa}, {Chornay}, {Comoretto}, {Contursi}, {Cooper}, {Cornez}, {Cowell}, {Crifo}, {Cropper}, {Crosta}, {Crowley}, {Dafonte}, {Dapergolas}, {David}, {de Laverny}, {De Luise}, {De March}, {De Ridder}, {de Souza}, {del Peloso}, {del Pozo}, {Delbo}, {Delgado}, {Delisle}, {Demouchy}, {Dharmawardena}, {Diakite}, {Diener}, {Distefano}, {Dolding}, {Enke}, {Fabre}, {Fabrizio}, {Faigler}, {Fedorets}, {Fernique}, {Fienga}, {Figueras}, {Fournier}, {Fouron}, {Fragkoudi}, {Gai}, {Garcia-Gutierrez}, {Garcia-Reinaldos}, {Garc{\'\i}a-Torres}, {Garofalo}, {Gavel}, {Gavras}, {Giacobbe}, {Gilmore}, {Girona},
  {Giuffrida}, {Gomel}, {Gomez}, {Gonz{\'a}lez-N{\'u}{\~n}ez}, {Gonz{\'a}lez-Santamar{\'\i}a}, {Granvik}, {Guillout}, {Guiraud}, {Guti{\'e}rrez-S{\'a}nchez}, {Guy}, {Hatzidimitriou}, {Hauser}, {Haywood}, {Helmer}, {Helmi}, {Sarmiento}, {Hidalgo}, {H{\l}adczuk}, {Holland}, {Huckle}, {Jardine}, {Jasniewicz}, {Jean-Antoine Piccolo}, {Jim{\'e}nez-Arranz}, {Juaristi Campillo}, {Julbe}, {Karbevska}, {Kervella}, {Khanna}, {Kordopatis}, {Korn}, {K{\'o}sp{\'a}l}, {Kostrzewa-Rutkowska}, {Kruszy{\'n}ska}, {Kun}, {Laizeau}, {Lambert}, {Lanza}, {Lasne}, {Le Campion}, {Lebreton}, {Lebzelter}, {Leccia}, {Leclerc}, {Lecoeur-Taibi}, {Liao}, {Licata}, {Lindstr{\o}m}, {Lister}, {Livanou}, {Lobel}, {Lorca}, {Loup}, {Madrero Pardo}, {Magdaleno Romeo}, {Managau}, {Mann}, {Manteiga}, {Marchant}, {Marconi}, {Marcos}, {Marcos Santos}, {Mar{\'\i}n Pina}, {Marinoni}, {Marocco}, {Marshall}, {Polo}, {Mart{\'\i}n-Fleitas}, {Marton}, {Mary}, {Masip}, {Massari}, {Mastrobuono-Battisti}, {Mazeh}, {Messina}, {Michalik}, {Millar}, {Mints},
  {Molina}, {Molinaro}, {Moln{\'a}r}, {Monari}, {Mongui{\'o}}, {Montegriffo}, {Montero}, {Mor}, {Mora}, {Morbidelli}, {Morel}, {Morris}, {Muraveva}, {Murphy}, {Musella}, {Nagy}, {Noval}, {Oca{\~n}a}, {Ogden}, {Ordenovic}, {Osinde}, {Pagani}, {Pagano}, {Palaversa}, {Palicio}, {Pallas-Quintela}, {Panahi}, {Payne-Wardenaar}, {Pe{\~n}alosa Esteller}, {Penttil{\"a}}, {Pichon}, {Piersimoni}, {Pineau}, {Plachy}, {Plum}, {Poggio}, {Pr{\v{s}}a}, {Pulone}, {Racero}, {Ragaini}, {Rainer}, {Rambaux}, {Ramos}, {Re Fiorentin}, {Regibo}, {Richards}, {Rios Diaz}, {Ripepi}, {Riva}, {Rix}, {Rixon}, {Robichon}, {Robin}, {Robin}, {Roelens}, {Rogues}, {Rohrbasser}, {Romero-G{\'o}mez}, {Royer}, {Ruz Mieres}, {Rybicki}, {Sadowski}, {S{\'a}ez N{\'u}{\~n}ez}, {Sagrist{\`a} Sell{\'e}s}, {Sahlmann}, {Salguero}, {Samaras}, {Sanchez Gimenez}, {Sanna}, {Santove{\~n}a}, {Sarasso}, {Schultheis}, {Sciacca}, {Segol}, {Segovia}, {S{\'e}gransan}, {Semeux}, {Shahaf}, {Siddiqui}, {Siebert}, {Siltala}, {Silvelo}, {Slezak}, {Slezak}, {Smart},
  {Snaith}, {Solano}, {Solitro}, {Souami}, {Souchay}, {Spagna}, {Spina}, {Spoto}, {Steele}, {Stephenson}, {S{\"u}veges}, {Surdej}, {Szabados}, {Szegedi-Elek}, {Taris}, {Taylor}, {Teixeira}, {Tolomei}, {Tonello}, {Torralba Elipe}, {Trabucchi}, {Tsounis}, {Turon}, {Ulla}, {Unger}, {Vaillant}, {van Dillen}, {van Reeven}, {Vanel}, {Vecchiato}, {Viala}, {Vicente}, {Voutsinas}, {Weiler}, {Wevers}, {Wyrzykowski}, {Yoldas}, {Yvard}, {Zhao}, {Zorec}, {Zucker}, \& {Zwitter}}]{Gaia}
{Gaia Collaboration}, {Klioner}, S.~A., {Lindegren}, L., {et~al.} 2022, arXiv e-prints, arXiv:2204.12574.
\newblock \doarXiv{2204.12574}

\bibitem[{{Gandolfi} {et~al.}(2018){Gandolfi}, {Barrag{\'a}n}, {Livingston}, {Fridlund}, {Justesen}, {Redfield}, {Fossati}, {Mathur}, {Grziwa}, {Cabrera}, {Garc{\'\i}a}, {Persson}, {Van Eylen}, {Hatzes}, {Hidalgo}, {Albrecht}, {Bugnet}, {Cochran}, {Csizmadia}, {Deeg}, {Eigm{\"u}ller}, {Endl}, {Erikson}, {Esposito}, {Guenther}, {Korth}, {Luque}, {Monta{\~n}es Rodr{\'\i}guez}, {Nespral}, {Nowak}, {P{\"a}tzold}, \& {Prieto-Arranz}}]{Gandolfi2018}
{Gandolfi}, D., {Barrag{\'a}n}, O., {Livingston}, J.~H., {et~al.} 2018, \aap, 619, L10, \dodoi{10.1051/0004-6361/201834289}

\bibitem[{{Gibson} {et~al.}(2016){Gibson}, {Howard}, {Marcy}, {Edelstein}, {Wishnow}, \& {Poppett}}]{Gibson2016}
{Gibson}, S.~R., {Howard}, A.~W., {Marcy}, G.~W., {et~al.} 2016, in Society of Photo-Optical Instrumentation Engineers (SPIE) Conference Series, Vol. 9908, Ground-based and Airborne Instrumentation for Astronomy VI, ed. C.~J. {Evans}, L.~{Simard}, \& H.~{Takami}, 990870, \dodoi{10.1117/12.2233334}

\bibitem[{{Gibson} {et~al.}(2018){Gibson}, {Howard}, {Roy}, {Smith}, {Halverson}, {Edelstein}, {Kassis}, {Wishnow}, {Raffanti}, {Allen}, {Chin}, {Coutts}, {Cowley}, {Curtis}, {Deich}, {Feger}, {Finstad}, {Gurevich}, {Ishikawa}, {James}, {Jhoti}, {Lanclos}, {Lilley}, {Miller}, {Milner}, {Payne}, {Rider}, {Rockosi}, {Sandford}, {Schwab}, {Seifahrt}, {Sirk}, {Smith}, {Stuermer}, {Weisfeiler}, {Wilcox}, {Vandenberg}, \& {Wizinowich}}]{Gibson2018}
{Gibson}, S.~R., {Howard}, A.~W., {Roy}, A., {et~al.} 2018, in Society of Photo-Optical Instrumentation Engineers (SPIE) Conference Series, Vol. 10702, Ground-based and Airborne Instrumentation for Astronomy VII, ed. C.~J. {Evans}, L.~{Simard}, \& H.~{Takami}, 107025X, \dodoi{10.1117/12.2311565}

\bibitem[{{Gibson} {et~al.}(2020){Gibson}, {Howard}, {Rider}, {Roy}, {Edelstein}, {Kassis}, {Grillo}, {Halverson}, {Sirk}, {Smith}, {Allen}, {Baker}, {Beichman}, {Berriman}, {Brown}, {Casey}, {Chin}, {Coutts}, {Cowley}, {Deich}, {Feger}, {Fulton}, {Gers}, {Gurevich}, {Ishikawa}, {James}, {Jelinsky}, {Kaye}, {Lanclos}, {Li}, {Lilley}, {McCarney}, {Miller}, {Milner}, {O'Hanlon}, {Pember}, {Raffanti}, {Rockosi}, {Rubenzahl}, {Rumph}, {Sandford}, {Savage}, {Schwab}, {Seifahrt}, {Shaum}, {Smith}, {Stuermer}, {Thorne}, {Vandenberg}, {Von Boeckmann}, {Wang}, {Wang}, {Weisfeiler}, {Wilcox}, {Wishnow}, {Wizinowich}, {Wold}, \& {Wolfenberger}}]{Gibson2020}
{Gibson}, S.~R., {Howard}, A.~W., {Rider}, K., {et~al.} 2020, in Society of Photo-Optical Instrumentation Engineers (SPIE) Conference Series, Vol. 11447, Ground-based and Airborne Instrumentation for Astronomy VIII, ed. C.~J. {Evans}, J.~J. {Bryant}, \& K.~{Motohara}, 1144742, \dodoi{10.1117/12.2561783}

\bibitem[{{Goldreich} \& {Sari}(2003)}]{Goldreich2003}
{Goldreich}, P., \& {Sari}, R. 2003, \apj, 585, 1024, \dodoi{10.1086/346202}

\bibitem[{{Goldreich} \& {Tremaine}(1980)}]{Goldreich1980}
{Goldreich}, P., \& {Tremaine}, S. 1980, \apj, 241, 425, \dodoi{10.1086/158356}

\bibitem[{{G{\"u}nther} \& {Daylan}(2019)}]{allesfitter-code}
{G{\"u}nther}, M.~N., \& {Daylan}, T. 2019, {Allesfitter: Flexible Star and Exoplanet Inference From Photometry and Radial Velocity}, Astrophysics Source Code Library.
\newblock \doeprint{1903.003}

\bibitem[{{G{\"u}nther} \& {Daylan}(2021)}]{allesfitter-paper}
---. 2021, \apjs, 254, 13, \dodoi{10.3847/1538-4365/abe70e}

\bibitem[{{Hansen} \& {Murray}(2013)}]{Hansen2013}
{Hansen}, B. M.~S., \& {Murray}, N. 2013, \apj, 775, 53, \dodoi{10.1088/0004-637X/775/1/53}

\bibitem[{{H{\'e}brard} {et~al.}(2020){H{\'e}brard}, {D{\'\i}az}, {Correia}, {Collier Cameron}, {Laskar}, {Pollacco}, {Almenara}, {Anderson}, {Barros}, {Boisse}, {Bonomo}, {Bouchy}, {Bou{\'e}}, {Boumis}, {Brown}, {Dalal}, {Deleuil}, {Demangeon}, {Doyle}, {Haswell}, {Hellier}, {Osborn}, {Kiefer}, {Kolb}, {Lam}, {Lecavelier des {\'E}tangs}, {Lopez}, {Martin-Lagarde}, {Maxted}, {McCormac}, {Nielsen}, {Pall{\'e}}, {Prieto-Arranz}, {Queloz}, {Santerne}, {Smalley}, {Turner}, {Udry}, {Verilhac}, {West}, {Wheatley}, \& {Wilson}}]{Hernard2020}
{H{\'e}brard}, G., {D{\'\i}az}, R.~F., {Correia}, A.~C.~M., {et~al.} 2020, \aap, 640, A32, \dodoi{10.1051/0004-6361/202038296}

\bibitem[{{Hedges} {et~al.}(2020){Hedges}, {Angus}, {Barentsen}, {Saunders}, {Montet}, \& {Gully-Santiago}}]{Hedges2020}
{Hedges}, C., {Angus}, R., {Barentsen}, G., {et~al.} 2020, Research Notes of the American Astronomical Society, 4, 220, \dodoi{10.3847/2515-5172/abd106}

\bibitem[{{Hirano} {et~al.}(2011){Hirano}, {Narita}, {Shporer}, {Sato}, {Aoki}, \& {Tamura}}]{Hirano2011}
{Hirano}, T., {Narita}, N., {Shporer}, A., {et~al.} 2011, \pasj, 63, 531, \dodoi{10.1093/pasj/63.sp2.S531}

\bibitem[{{Howard} \& {Fulton}(2016)}]{Howard2016}
{Howard}, A.~W., \& {Fulton}, B.~J. 2016, \pasp, 128, 114401, \dodoi{10.1088/1538-3873/128/969/114401}

\bibitem[{{Howard} {et~al.}(2010){Howard}, {Johnson}, {Marcy}, {Fischer}, {Wright}, {Bernat}, {Henry}, {Peek}, {Isaacson}, {Apps}, {Endl}, {Cochran}, {Valenti}, {Anderson}, \& {Piskunov}}]{Howard2010}
{Howard}, A.~W., {Johnson}, J.~A., {Marcy}, G.~W., {et~al.} 2010, \apj, 721, 1467, \dodoi{10.1088/0004-637X/721/2/1467}

\bibitem[{{Hu} {et~al.}(2024){Hu}, {Rice}, {Wang}, {Wang}, {Shporer}, {Teske}, {Yee}, {Butler}, {Shectman}, {Crane}, {Collins}, \& {Collins}}]{Hu2024}
{Hu}, Q., {Rice}, M., {Wang}, X.-Y., {et~al.} 2024, \aj, 167, 175, \dodoi{10.3847/1538-3881/ad2855}

\bibitem[{{Huang} {et~al.}(2018){Huang}, {Burt}, {Vanderburg}, {G{\"u}nther}, {Shporer}, {Dittmann}, {Winn}, {Wittenmyer}, {Sha}, {Kane}, {Ricker}, {Vanderspek}, {Latham}, {Seager}, {Jenkins}, {Caldwell}, {Collins}, {Guerrero}, {Smith}, {Quinn}, {Udry}, {Pepe}, {Bouchy}, {S{\'e}gransan}, {Lovis}, {Ehrenreich}, {Marmier}, {Mayor}, {Wohler}, {Haworth}, {Morgan}, {Fausnaugh}, {Ciardi}, {Christiansen}, {Charbonneau}, {Dragomir}, {Deming}, {Glidden}, {Levine}, {McCullough}, {Yu}, {Narita}, {Nguyen}, {Morton}, {Pepper}, {P{\'a}l}, {Rodriguez}, {Stassun}, {Torres}, {Sozzetti}, {Doty}, {Christensen-Dalsgaard}, {Laughlin}, {Clampin}, {Bean}, {Buchhave}, {Bakos}, {Sato}, {Ida}, {Kaltenegger}, {Palle}, {Sasselov}, {Butler}, {Lissauer}, {Ge}, \& {Rinehart}}]{Huang2018}
{Huang}, C.~X., {Burt}, J., {Vanderburg}, A., {et~al.} 2018, \apjl, 868, L39, \dodoi{10.3847/2041-8213/aaef91}

\bibitem[{{Huber} {et~al.}(2013){Huber}, {Carter}, {Barbieri}, {Miglio}, {Deck}, {Fabrycky}, {Montet}, {Buchhave}, {Chaplin}, {Hekker}, {Montalb{\'a}n}, {Sanchis-Ojeda}, {Basu}, {Bedding}, {Campante}, {Christensen-Dalsgaard}, {Elsworth}, {Stello}, {Arentoft}, {Ford}, {Gilliland}, {Handberg}, {Howard}, {Isaacson}, {Johnson}, {Karoff}, {Kawaler}, {Kjeldsen}, {Latham}, {Lund}, {Lundkvist}, {Marcy}, {Metcalfe}, {Silva Aguirre}, \& {Winn}}]{huber2013}
{Huber}, D., {Carter}, J.~A., {Barbieri}, M., {et~al.} 2013, Science, 342, 331, \dodoi{10.1126/science.1242066}

\bibitem[{{Huber} {et~al.}(2017){Huber}, {Zinn}, {Bojsen-Hansen}, {Pinsonneault}, {Sahlholdt}, {Serenelli}, {Silva Aguirre}, {Stassun}, {Stello}, {Tayar}, {Bastien}, {Bedding}, {Buchhave}, {Chaplin}, {Davies}, {Garc{\'\i}a}, {Latham}, {Mathur}, {Mosser}, \& {Sharma}}]{huber2017}
{Huber}, D., {Zinn}, J., {Bojsen-Hansen}, M., {et~al.} 2017, \apj, 844, 102, \dodoi{10.3847/1538-4357/aa75ca}

\bibitem[{{Ida} \& {Lin}(2008)}]{Ida2008}
{Ida}, S., \& {Lin}, D.~N.~C. 2008, \apj, 673, 487, \dodoi{10.1086/523754}

\bibitem[{{Ito} \& {Ohtsuka}(2019)}]{Ito2019}
{Ito}, T., \& {Ohtsuka}, K. 2019, Monographs on Environment, Earth and Planets, 7, 1, \dodoi{10.5047/meep.2019.00701.0001}

\bibitem[{{Jenkins} {et~al.}(2016){Jenkins}, {Twicken}, {McCauliff}, {Campbell}, {Sanderfer}, {Lung}, {Mansouri-Samani}, {Girouard}, {Tenenbaum}, {Klaus}, {Smith}, {Caldwell}, {Chacon}, {Henze}, {Heiges}, {Latham}, {Morgan}, {Swade}, {Rinehart}, \& {Vanderspek}}]{Jenkins2016}
{Jenkins}, J.~M., {Twicken}, J.~D., {McCauliff}, S., {et~al.} 2016, in Society of Photo-Optical Instrumentation Engineers (SPIE) Conference Series, Vol. 9913, Software and Cyberinfrastructure for Astronomy IV, ed. G.~{Chiozzi} \& J.~C. {Guzman}, 99133E, \dodoi{10.1117/12.2233418}

\bibitem[{{Jones} {et~al.}(2002){Jones}, {Paul Butler}, {Tinney}, {Marcy}, {Penny}, {McCarthy}, {Carter}, \& {Pourbaix}}]{Jones2002}
{Jones}, H. R.~A., {Paul Butler}, R., {Tinney}, C.~G., {et~al.} 2002, \mnras, 333, 871, \dodoi{10.1046/j.1365-8711.2002.05459.x}

\bibitem[{{Kervella} {et~al.}(2019){Kervella}, {Arenou}, {Mignard}, \& {Th{\'e}venin}}]{Kervella2019}
{Kervella}, P., {Arenou}, F., {Mignard}, F., \& {Th{\'e}venin}, F. 2019, \aap, 623, A72, \dodoi{10.1051/0004-6361/201834371}

\bibitem[{{Knutson} {et~al.}(2014){Knutson}, {Fulton}, {Montet}, {Kao}, {Ngo}, {Howard}, {Crepp}, {Hinkley}, {Bakos}, {Batygin}, {Johnson}, {Morton}, \& {Muirhead}}]{Knutson2014}
{Knutson}, H.~A., {Fulton}, B.~J., {Montet}, B.~T., {et~al.} 2014, \apj, 785, 126, \dodoi{10.1088/0004-637X/785/2/126}

\bibitem[{{Kozai}(1962)}]{Kozai1962}
{Kozai}, Y. 1962, \aj, 67, 591, \dodoi{10.1086/108790}

\bibitem[{{Kraft}(1967)}]{kraft1967}
{Kraft}, R.~P. 1967, \apj, 150, 551, \dodoi{10.1086/149359}

\bibitem[{{Kunovac Hod{\v{z}}i{\'c}} {et~al.}(2021){Kunovac Hod{\v{z}}i{\'c}}, {Triaud}, {Cegla}, {Chaplin}, \& {Davies}}]{Kunovac2021}
{Kunovac Hod{\v{z}}i{\'c}}, V., {Triaud}, A. H.~M.~J., {Cegla}, H.~M., {Chaplin}, W.~J., \& {Davies}, G.~R. 2021, \mnras, 502, 2893, \dodoi{10.1093/mnras/stab237}

\bibitem[{{Lai}(2012)}]{lai2012}
{Lai}, D. 2012, \mnras, 423, 486, \dodoi{10.1111/j.1365-2966.2012.20893.x}

\bibitem[{{Lai} {et~al.}(2018){Lai}, {Anderson}, \& {Pu}}]{Lai2018}
{Lai}, D., {Anderson}, K.~R., \& {Pu}, B. 2018, \mnras, 475, 5231, \dodoi{10.1093/mnras/sty133}

\bibitem[{{Lee} \& {Peale}(2002)}]{Lee2002}
{Lee}, M.~H., \& {Peale}, S.~J. 2002, \apj, 567, 596, \dodoi{10.1086/338504}

\bibitem[{{Li}(2021)}]{Li2021}
{Li}, G. 2021, \apjl, 915, L2, \dodoi{10.3847/2041-8213/ac0620}

\bibitem[{{Li} {et~al.}(2014){Li}, {Naoz}, {Kocsis}, \& {Loeb}}]{Li2014b}
{Li}, G., {Naoz}, S., {Kocsis}, B., \& {Loeb}, A. 2014, \apj, 785, 116, \dodoi{10.1088/0004-637X/785/2/116}

\bibitem[{{Li} \& {Lai}(2023)}]{Li2023}
{Li}, J., \& {Lai}, D. 2023, \apj, 956, 17, \dodoi{10.3847/1538-4357/aced89}

\bibitem[{{Lidov}(1962)}]{Lidov1962}
{Lidov}, M.~L. 1962, \planss, 9, 719, \dodoi{10.1016/0032-0633(62)90129-0}

\bibitem[{{Lightkurve Collaboration} {et~al.}(2018){Lightkurve Collaboration}, {Cardoso}, {Hedges}, {Gully-Santiago}, {Saunders}, {Cody}, {Barclay}, {Hall}, {Sagear}, {Turtelboom}, {Zhang}, {Tzanidakis}, {Mighell}, {Coughlin}, {Bell}, {Berta-Thompson}, {Williams}, {Dotson}, \& {Barentsen}}]{LK2018}
{Lightkurve Collaboration}, {Cardoso}, J.~V.~d.~M., {Hedges}, C., {et~al.} 2018, {Lightkurve: Kepler and TESS time series analysis in Python}, Astrophysics Source Code Library.
\newblock \doeprint{1812.013}

\bibitem[{{Lin} {et~al.}(1996){Lin}, {Bodenheimer}, \& {Richardson}}]{Lin1996}
{Lin}, D.~N.~C., {Bodenheimer}, P., \& {Richardson}, D.~C. 1996, \nat, 380, 606, \dodoi{10.1038/380606a0}

\bibitem[{{Lu} {et~al.}(2024){Lu}, {An}, {Li}, {Millholland}, {Brandt}, \& {Brandt}}]{Lu2024}
{Lu}, T., {An}, Q., {Li}, G., {et~al.} 2024, arXiv e-prints, arXiv:2405.19511, \dodoi{10.48550/arXiv.2405.19511}

\bibitem[{{Luck}(2017)}]{Luck2017}
{Luck}, R.~E. 2017, \aj, 153, 21, \dodoi{10.3847/1538-3881/153/1/21}

\bibitem[{{Lund} {et~al.}(2017){Lund}, {Silva Aguirre}, {Davies}, {Chaplin}, {Christensen-Dalsgaard}, {Houdek}, {White}, {Bedding}, {Ball}, {Huber}, {Antia}, {Lebreton}, {Latham}, {Handberg}, {Verma}, {Basu}, {Casagrande}, {Justesen}, {Kjeldsen}, \& {Mosumgaard}}]{Lund2017}
{Lund}, M.~N., {Silva Aguirre}, V., {Davies}, G.~R., {et~al.} 2017, \apj, 835, 172, \dodoi{10.3847/1538-4357/835/2/172}

\bibitem[{{Maciejewski} {et~al.}(2024){Maciejewski}, {Niedzielski}, {Gozdziewski}, {Wolszczan}, {Villaver}, {Fernandez}, {Adamow}, \& {Sierzputowska}}]{Maciejewski2024}
{Maciejewski}, G., {Niedzielski}, A., {Gozdziewski}, K., {et~al.} 2024, arXiv e-prints, arXiv:2407.11706, \dodoi{10.48550/arXiv.2407.11706}

\bibitem[{{Malhotra}(1993)}]{Malhotra1993}
{Malhotra}, R. 1993, \nat, 365, 819, \dodoi{10.1038/365819a0}

\bibitem[{{Masuda} \& {Winn}(2020)}]{MandW2020M}
{Masuda}, K., \& {Winn}, J.~N. 2020, \aj, 159, 81, \dodoi{10.3847/1538-3881/ab65be}

\bibitem[{{McLaughlin}(1924)}]{McLaughlin1924}
{McLaughlin}, D.~B. 1924, \apj, 60, 22, \dodoi{10.1086/142826}

\bibitem[{{Murray} \& {Dermott}(1999)}]{Murray1999}
{Murray}, C.~D., \& {Dermott}, S.~F. 1999, {Solar System Dynamics}, \dodoi{10.1017/CBO9781139174817}

\bibitem[{{Naoz} {et~al.}(2011){Naoz}, {Farr}, {Lithwick}, {Rasio}, \& {Teyssandier}}]{Naoz2011}
{Naoz}, S., {Farr}, W.~M., {Lithwick}, Y., {Rasio}, F.~A., \& {Teyssandier}, J. 2011, \nat, 473, 187, \dodoi{10.1038/nature10076}

\bibitem[{{Naoz} {et~al.}(2013){Naoz}, {Kocsis}, {Loeb}, \& {Yunes}}]{Naoz2013}
{Naoz}, S., {Kocsis}, B., {Loeb}, A., \& {Yunes}, N. 2013, \apj, 773, 187, \dodoi{10.1088/0004-637X/773/2/187}

\bibitem[{{Ngo} {et~al.}(2016){Ngo}, {Knutson}, {Hinkley}, {Bryan}, {Crepp}, {Batygin}, {Crossfield}, {Hansen}, {Howard}, {Johnson}, {Mawet}, {Morton}, {Muirhead}, \& {Wang}}]{Ngo2016}
{Ngo}, H., {Knutson}, H.~A., {Hinkley}, S., {et~al.} 2016, \apj, 827, 8, \dodoi{10.3847/0004-637X/827/1/8}

\bibitem[{{Pepper} {et~al.}(2020){Pepper}, {Kane}, {Rodriguez}, {Hinkel}, {Eastman}, {Daylan}, {Mocnik}, {Dalba}, {Gaudi}, {Fetherolf}, {Stassun}, {Campante}, {Vanderburg}, {Huber}, {Bossini}, {Crossfield}, {Howell}, {Stephens}, {Furlan}, {Ricker}, {Vanderspek}, {Latham}, {Seager}, {Winn}, {Jenkins}, {Twicken}, {Rose}, {Smith}, {Glidden}, {Levine}, {Rinehart}, {Collins}, {Mann}, {Burt}, {James}, {Siverd}, \& {G{\"u}nther}}]{pepper2020}
{Pepper}, J., {Kane}, S.~R., {Rodriguez}, J.~E., {et~al.} 2020, \aj, 159, 243, \dodoi{10.3847/1538-3881/ab84f2}

\bibitem[{{Petigura}(2015)}]{Petigura_thesis}
{Petigura}, E.~A. 2015, PhD thesis, University of California, Berkeley

\bibitem[{{Petrovich}(2015)}]{Petrovich2015}
{Petrovich}, C. 2015, \apj, 805, 75, \dodoi{10.1088/0004-637X/805/1/75}

\bibitem[{{Petrovich} {et~al.}(2014){Petrovich}, {Tremaine}, \& {Rafikov}}]{Petrovich2014}
{Petrovich}, C., {Tremaine}, S., \& {Rafikov}, R. 2014, \apj, 786, 101, \dodoi{10.1088/0004-637X/786/2/101}

\bibitem[{{Press} \& {Rybicki}(1989)}]{Press1989}
{Press}, W.~H., \& {Rybicki}, G.~B. 1989, \apj, 338, 277, \dodoi{10.1086/167197}

\bibitem[{{Pu} \& {Lai}(2019)}]{Pu2019}
{Pu}, B., \& {Lai}, D. 2019, \mnras, 488, 3568, \dodoi{10.1093/mnras/stz1817}

\bibitem[{{Rafikov}(2005)}]{Rafikov2005}
{Rafikov}, R.~R. 2005, \apjl, 621, L69, \dodoi{10.1086/428899}

\bibitem[{{Rafikov}(2006)}]{Rafikov2006}
---. 2006, \apj, 648, 666, \dodoi{10.1086/505695}

\bibitem[{{Ramsey} {et~al.}(1998){Ramsey}, {Adams}, {Barnes}, {Booth}, {Cornell}, {Fowler}, {Gaffney}, {Glaspey}, {Good}, {Hill}, {Kelton}, {Krabbendam}, {Long}, {MacQueen}, {Ray}, {Ricklefs}, {Sage}, {Sebring}, {Spiesman}, \& {Steiner}}]{Ramsey1998}
{Ramsey}, L.~W., {Adams}, M.~T., {Barnes}, T.~G., {et~al.} 1998, in Society of Photo-Optical Instrumentation Engineers (SPIE) Conference Series, Vol. 3352, Advanced Technology Optical/IR Telescopes VI, ed. L.~M. {Stepp}, 34--42, \dodoi{10.1117/12.319287}

\bibitem[{{Rasio} \& {Ford}(1996)}]{Rasio1996}
{Rasio}, F.~A., \& {Ford}, E.~B. 1996, Science, 274, 954, \dodoi{10.1126/science.274.5289.954}

\bibitem[{{Rice} {et~al.}(2022){Rice}, {Wang}, \& {Laughlin}}]{Rice2022}
{Rice}, M., {Wang}, S., \& {Laughlin}, G. 2022, \apjl, 926, L17, \dodoi{10.3847/2041-8213/ac502d}

\bibitem[{{Romanova} {et~al.}(2024){Romanova}, {Koldoba}, {Ustyugova}, {Espaillat}, \& {Lovelace}}]{Romanova2024MNRAS}
{Romanova}, M.~M., {Koldoba}, A.~V., {Ustyugova}, G.~V., {Espaillat}, C., \& {Lovelace}, R.~V.~E. 2024, \mnras, 532, 3509, \dodoi{10.1093/mnras/stae1658}

\bibitem[{{Romanova} {et~al.}(2023){Romanova}, {Koldoba}, {Ustyugova}, {Lai}, \& {Lovelace}}]{Romanova2023}
{Romanova}, M.~M., {Koldoba}, A.~V., {Ustyugova}, G.~V., {Lai}, D., \& {Lovelace}, R.~V.~E. 2023, \mnras, 523, 2832, \dodoi{10.1093/mnras/stad987}

\bibitem[{{Rosenthal} {et~al.}(2021){Rosenthal}, {Fulton}, {Hirsch}, {Isaacson}, {Howard}, {Dedrick}, {Sherstyuk}, {Blunt}, {Petigura}, {Knutson}, {Behmard}, {Chontos}, {Crepp}, {Crossfield}, {Dalba}, {Fischer}, {Henry}, {Kane}, {Kosiarek}, {Marcy}, {Rubenzahl}, {Weiss}, \& {Wright}}]{Rosenthal2021}
{Rosenthal}, L.~J., {Fulton}, B.~J., {Hirsch}, L.~A., {et~al.} 2021, \apjs, 255, 8, \dodoi{10.3847/1538-4365/abe23c}

\bibitem[{{Rossiter}(1924)}]{Rossiter1924}
{Rossiter}, R.~A. 1924, \apj, 60, 15, \dodoi{10.1086/142825}

\bibitem[{Rubenzahl {et~al.}(2023)Rubenzahl, Halverson, Walawender, Hill, Howard, Brown, Ida, Tehero, Fulton, Gibson, Kassis, Smith, Wold, \& Payne}]{KPFSoCal}
Rubenzahl, R.~A., Halverson, S., Walawender, J., {et~al.} 2023, Publications of the Astronomical Society of the Pacific, 135, 125002, \dodoi{10.1088/1538-3873/ad0b30}

\bibitem[{{Rubenzahl} {et~al.}(2024){Rubenzahl}, {Howard}, {Halverson}, {Petrovich}, {Angelo}, {Stef{\'a}nsson}, {Dai}, {Householder}, {Fulton}, {Gibson}, {Roy}, {Shaum}, {Isaacson}, {Brodheim}, {Deich}, {Hill}, {Holden}, {Huber}, {Laher}, {Lanclos}, {Payne}, {Petigura}, {Schwab}, {Walawender}, {Wang}, {Weiss}, {Winn}, \& {Wright}}]{Rubenzahl2024}
{Rubenzahl}, R.~A., {Howard}, A.~W., {Halverson}, S., {et~al.} 2024, arXiv e-prints, arXiv:2407.21188.
\newblock \doarXiv{2407.21188}

\bibitem[{{Saunders} {et~al.}(2024){Saunders}, {Grunblatt}, {Chontos}, {Dai}, {Huber}, {Zhang}, {Stef{\'a}nsson}, {van Saders}, {Winn}, {Hey}, {Howard}, {Fulton}, {Isaacson}, {Beard}, {Giacalone}, {Van Zandt}, {Murphey}, {Rice}, {Blunt}, {Turtelboom}, {Dalba}, {Lubin}, {Brinkman}, {Louden}, {Page}, {Watkins}, {Collins}, {Stockdale}, {Tan}, {Schwarz}, {Massey}, {Howell}, {Vanderburg}, {Ricker}, {Jenkins}, {Seager}, {Christiansen}, {Daylan}, {Falk}, {Brodheim}, {Gibson}, {Hill}, {Holden}, {Householder}, {Kaye}, {Laher}, {Lanclos}, {Petigura}, {Roy}, {Rubenzahl}, {Schwab}, {Shaum}, {Sirk}, {Smith}, {Walawender}, \& {Yeh}}]{Saunders2024}
{Saunders}, N., {Grunblatt}, S.~K., {Chontos}, A., {et~al.} 2024, \aj, 168, 81, \dodoi{10.3847/1538-3881/ad543b}

\bibitem[{{Schlaufman}(2010)}]{Schlaufman2010}
{Schlaufman}, K.~C. 2010, \apj, 719, 602, \dodoi{10.1088/0004-637X/719/1/602}

\bibitem[{{Schofield} {et~al.}(2019){Schofield}, {Chaplin}, {Huber}, {Campante}, {Davies}, {Miglio}, {Ball}, {Appourchaux}, {Basu}, {Bedding}, {Christensen-Dalsgaard}, {Creevey}, {Garc{\'\i}a}, {Handberg}, {Kawaler}, {Kjeldsen}, {Latham}, {Lund}, {Metcalfe}, {Ricker}, {Serenelli}, {Silva Aguirre}, {Stello}, \& {Vanderspek}}]{Schofield2019}
{Schofield}, M., {Chaplin}, W.~J., {Huber}, D., {et~al.} 2019, \apjs, 241, 12, \dodoi{10.3847/1538-4365/ab04f5}

\bibitem[{{Smith} {et~al.}(2012){Smith}, {Stumpe}, {Van Cleve}, {Jenkins}, {Barclay}, {Fanelli}, {Girouard}, {Kolodziejczak}, {McCauliff}, {Morris}, \& {Twicken}}]{Smith2012}
{Smith}, J.~C., {Stumpe}, M.~C., {Van Cleve}, J.~E., {et~al.} 2012, \pasp, 124, 1000, \dodoi{10.1086/667697}

\bibitem[{{Speagle}(2020)}]{dynest2020}
{Speagle}, J.~S. 2020, \mnras, 493, 3132, \dodoi{10.1093/mnras/staa278}

\bibitem[{{Stumpe} {et~al.}(2014){Stumpe}, {Smith}, {Catanzarite}, {Van Cleve}, {Jenkins}, {Twicken}, \& {Girouard}}]{Stumpe2014}
{Stumpe}, M.~C., {Smith}, J.~C., {Catanzarite}, J.~H., {et~al.} 2014, \pasp, 126, 100, \dodoi{10.1086/674989}

\bibitem[{{Stumpe} {et~al.}(2012){Stumpe}, {Smith}, {Van Cleve}, {Twicken}, {Barclay}, {Fanelli}, {Girouard}, {Jenkins}, {Kolodziejczak}, {McCauliff}, \& {Morris}}]{Stumpe2012}
{Stumpe}, M.~C., {Smith}, J.~C., {Van Cleve}, J.~E., {et~al.} 2012, \pasp, 124, 985, \dodoi{10.1086/667698}

\bibitem[{{Teyssandier} {et~al.}(2019){Teyssandier}, {Lai}, \& {Vick}}]{Teyssandier2019}
{Teyssandier}, J., {Lai}, D., \& {Vick}, M. 2019, \mnras, 486, 2265, \dodoi{10.1093/mnras/stz1011}

\bibitem[{{Tull}(1998)}]{Tull1998}
{Tull}, R.~G. 1998, in Society of Photo-Optical Instrumentation Engineers (SPIE) Conference Series, Vol. 3355, Optical Astronomical Instrumentation, ed. S.~{D'Odorico}, 387--398, \dodoi{10.1117/12.316774}

\bibitem[{{van Leeuwen} {et~al.}(1997){van Leeuwen}, {Evans}, {Grenon}, {Grossmann}, {Mignard}, \& {Perryman}}]{vanLeeuwen1997}
{van Leeuwen}, F., {Evans}, D.~W., {Grenon}, M., {et~al.} 1997, \aap, 323, L61

\bibitem[{{Vogt} {et~al.}(1994){Vogt}, {Allen}, {Bigelow}, {Bresee}, {Brown}, {Cantrall}, {Conrad}, {Couture}, {Delaney}, {Epps}, {Hilyard}, {Hilyard}, {Horn}, {Jern}, {Kanto}, {Keane}, {Kibrick}, {Lewis}, {Osborne}, {Pardeilhan}, {Pfister}, {Ricketts}, {Robinson}, {Stover}, {Tucker}, {Ward}, \& {Wei}}]{Vogt1994}
{Vogt}, S.~S., {Allen}, S.~L., {Bigelow}, B.~C., {et~al.} 1994, Society of Photo-Optical Instrumentation Engineers (SPIE) Conference Series, Vol. 2198, {HIRES: the high-resolution echelle spectrometer on the Keck 10-m Telescope}, 362, \dodoi{10.1117/12.176725}

\bibitem[{{von Zeipel}(1910)}]{1910AN....183..345V}
{von Zeipel}, H. 1910, Astronomische Nachrichten, 183, 345, \dodoi{10.1002/asna.19091832202}

\bibitem[{{Wang} {et~al.}(2022){Wang}, {Rice}, {Wang}, {Pu}, {Stef{\'a}nsson}, {Mahadevan}, {Radzom}, {Giacalone}, {Wu}, {Esposito}, {Dalba}, {Avsar}, {Holden}, {Skiff}, {Polakis}, {Voeller}, {Logsdon}, {Klusmeyer}, {Schweiker}, {Wu}, {Beard}, {Dai}, {Lubin}, {Weiss}, {Bender}, {Blake}, {Dressing}, {Halverson}, {Hearty}, {Howard}, {Huber}, {Isaacson}, {Jackman}, {Llama}, {McElwain}, {Rajagopal}, {Roy}, {Robertson}, {Schwab}, {Shkolnik}, {Wright}, \& {Laughlin}}]{Wang2022}
{Wang}, X.-Y., {Rice}, M., {Wang}, S., {et~al.} 2022, \apjl, 926, L8, \dodoi{10.3847/2041-8213/ac4f44}

\bibitem[{{Winn} {et~al.}(2010{\natexlab{a}}){Winn}, {Fabrycky}, {Albrecht}, \& {Johnson}}]{Winn2010}
{Winn}, J.~N., {Fabrycky}, D., {Albrecht}, S., \& {Johnson}, J.~A. 2010{\natexlab{a}}, \apjl, 718, L145, \dodoi{10.1088/2041-8205/718/2/L145}

\bibitem[{{Winn} {et~al.}(2007){Winn}, {Johnson}, {Peek}, {Marcy}, {Bakos}, {Enya}, {Narita}, {Suto}, {Turner}, \& {Vogt}}]{Winn2007}
{Winn}, J.~N., {Johnson}, J.~A., {Peek}, K. M.~G., {et~al.} 2007, \apjl, 665, L167, \dodoi{10.1086/521362}

\bibitem[{{Winn} {et~al.}(2010{\natexlab{b}}){Winn}, {Johnson}, {Howard}, {Marcy}, {Isaacson}, {Shporer}, {Bakos}, {Hartman}, \& {Albrecht}}]{Winn2010b}
{Winn}, J.~N., {Johnson}, J.~A., {Howard}, A.~W., {et~al.} 2010{\natexlab{b}}, \apjl, 723, L223, \dodoi{10.1088/2041-8205/723/2/L223}

\bibitem[{{Wu} \& {Lithwick}(2011)}]{Wu2011}
{Wu}, Y., \& {Lithwick}, Y. 2011, \apj, 735, 109, \dodoi{10.1088/0004-637X/735/2/109}

\bibitem[{{Xuan} \& {Wyatt}(2020)}]{xw2020}
{Xuan}, J.~W., \& {Wyatt}, M.~C. 2020, \mnras, 497, 2096, \dodoi{10.1093/mnras/staa2033}

\bibitem[{{Yee} {et~al.}(2018){Yee}, {Petigura}, {Fulton}, {Knutson}, {Batygin}, {Bakos}, {Hartman}, {Hirsch}, {Howard}, {Isaacson}, {Kosiarek}, {Sinukoff}, \& {Weiss}}]{Yee2018}
{Yee}, S.~W., {Petigura}, E.~A., {Fulton}, B.~J., {et~al.} 2018, \aj, 155, 255, \dodoi{10.3847/1538-3881/aabfec}

\bibitem[{{Zanazzi} {et~al.}(2024){Zanazzi}, {Dewberry}, \& {Chiang}}]{Zanazzi2024}
{Zanazzi}, J.~J., {Dewberry}, J., \& {Chiang}, E. 2024, \apjl, 967, L29, \dodoi{10.3847/2041-8213/ad4644}

\bibitem[{{Zechmeister} {et~al.}(2018){Zechmeister}, {Reiners}, {Amado}, {Azzaro}, {Bauer}, {B{\'e}jar}, {Caballero}, {Guenther}, {Hagen}, {Jeffers}, {Kaminski}, {K{\"u}rster}, {Launhardt}, {Montes}, {Morales}, {Quirrenbach}, {Reffert}, {Ribas}, {Seifert}, {Tal-Or}, \& {Wolthoff}}]{SERVAL}
{Zechmeister}, M., {Reiners}, A., {Amado}, P.~J., {et~al.} 2018, \aap, 609, A12, \dodoi{10.1051/0004-6361/201731483}

\bibitem[{{Zhang} {et~al.}(2021){Zhang}, {Weiss}, {Huber}, {Blunt}, {Chontos}, {Fulton}, {Grunblatt}, {Howard}, {Isaacson}, {Kosiarek}, {Petigura}, {Rosenthal}, \& {Rubenzahl}}]{Zhang2021}
{Zhang}, J., {Weiss}, L.~M., {Huber}, D., {et~al.} 2021, \aj, 162, 89, \dodoi{10.3847/1538-3881/ac0634}

\bibitem[{{Zhang} {et~al.}(2024){Zhang}, {Weiss}, {Huber}, {Jensen}, {Brandt}, {Collins}, {Conti}, {Isaacson}, {Lewin}, {Marino}, {Massey}, {Murgas}, {Palle}, {Radford}, {Relles}, {Srdoc}, {Stockdale}, {Tan}, \& {Wang}}]{zhang2023}
---. 2024, \aj, 167, 89, \dodoi{10.3847/1538-3881/ad1189}

\bibitem[{{Zink} \& {Howard}(2023)}]{zink2023}
{Zink}, J.~K., \& {Howard}, A.~W. 2023, \apjl, 956, L29, \dodoi{10.3847/2041-8213/acfdab}

\end{thebibliography}
